\documentclass[11pt, oneside]{Thesis} 

\graphicspath{{Pictures/}} 
\usepackage[square, numbers, comma, sort&compress]{natbib} 
\hypersetup{urlcolor=blue, colorlinks=true} 
\title{\ttitle} 
\usepackage{relsize}
\usepackage{amsmath}
\usepackage{adjustbox}
\usepackage{leftidx}

\newcommand*{\defeq}{\mathrel{\vcenter{\baselineskip0.5ex \lineskiplimit0pt
                     \hbox{\scriptsize.}\hbox{\scriptsize.}}}%
                     =}

\newcommand{\veq}{\mathrel{\rotatebox{90}{$=$}}}

\renewcommand{\bibname}{References}

\begin{document}

\frontmatter 

\setstretch{1.25} 

\fancyhead{} 
\rhead{\thepage} 
\lhead{} 

\pagestyle{fancy} 

\newcommand{\HRule}{\rule{\linewidth}{0.5mm}} 

\hypersetup{pdftitle={\ttitle}}
\hypersetup{pdfsubject=\subjectname}
\hypersetup{pdfauthor=\authornames}
\hypersetup{pdfkeywords=\keywordnames}


\begin{titlepage}
\begin{center}

\textsc{\LARGE \univname}\\[1.5cm] 

\HRule \\[0.4cm] 
{\Large \bfseries \ttitle}\\[0.4cm] 
\HRule \\[1.5cm] 
 
\begin{minipage}{0.4\textwidth}
\begin{flushleft} \large
\emph{Author:}\\
\href{mailto:anon@cam.ac.uk}{\authornames} 
\end{flushleft}
\end{minipage}
\begin{minipage}{0.4\textwidth}
\begin{flushright} \large
\emph{Supervisors:} \\
{\supname} 
\end{flushright}
\end{minipage}\\[1cm]

\vspace{2cm}

\begin{figure}[h]
   \centering
   \includegraphics[scale=0.08]{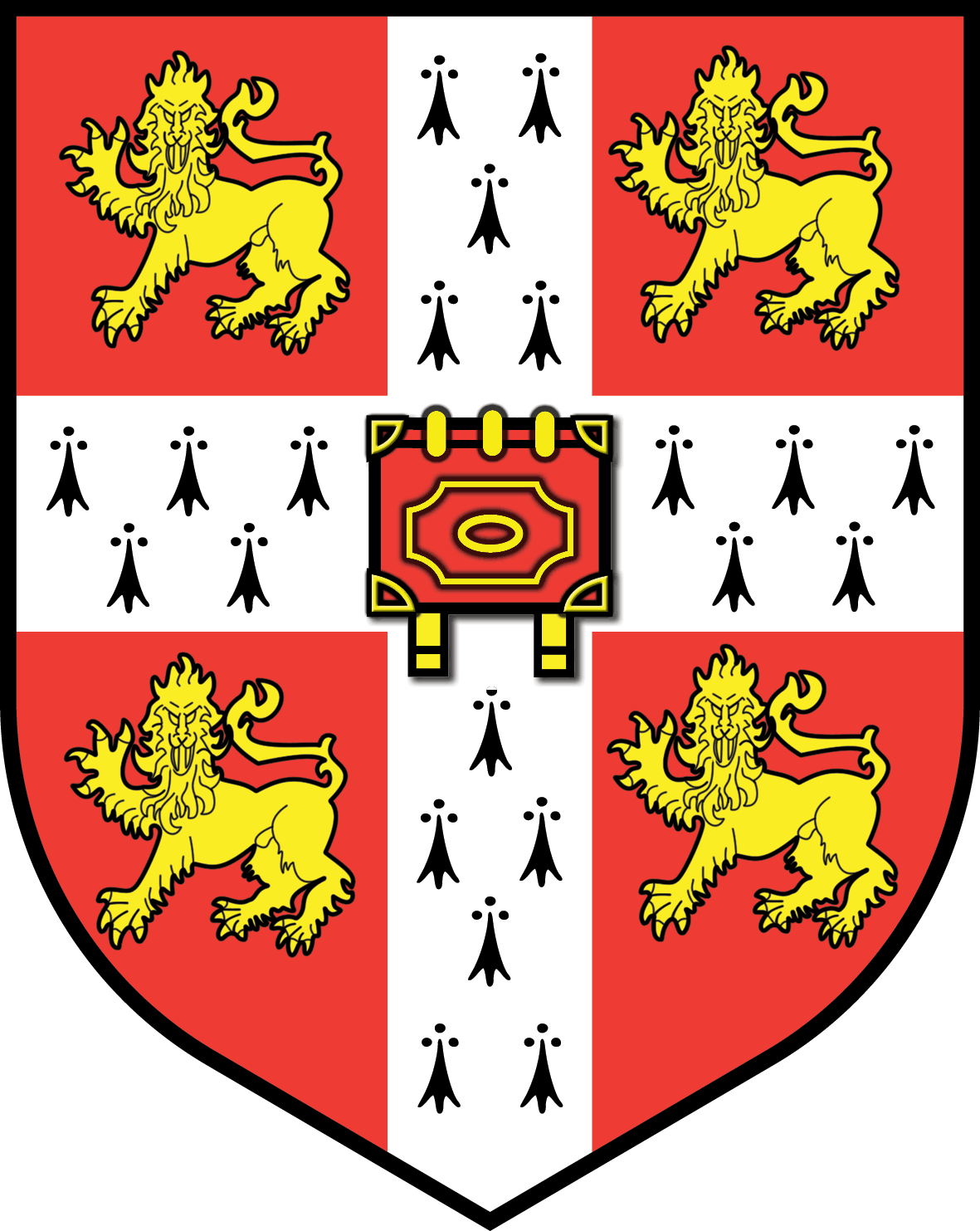}
\end{figure}

\vspace{3cm}

\large \textit{An essay submitted in fulfilment of the requirements\\ for the degree of \degreename}\\[0.3cm] 
\textit{in the}\\[0.2cm]
\deptname \\ 
 
\vspace{1cm}

{\large \today} 

\vfill
\end{center}

\end{titlepage}








\setstretch{1.3} 

\acknowledgements{\addtocontents{toc}{\vspace{1em}} 
The author thanks his supervisors, Prof$.$ Paul Shellard and Dr$.$ James Fergusson, for guidance during the year, and invaluable advice on the first draft; and Sarah Bosman, for many fruitful discussions and innumerable insights throughout the work.

}
\clearpage 

\begin{minipage}[0]{15cm}
\vspace{10cm} \hspace{6cm} \\
\begin{center}
To Nan
\end{center}
\end{minipage}

\clearpage

%

\pagestyle{fancy} 

\lhead{\emph{Contents}} 
\tableofcontents 


\mainmatter 

\pagestyle{fancy} 


\makeatletter
\def\@makechapterhead#1{%
  \vspace*{50\p@}%
  {\parindent \z@ \raggedright \normalfont
    \interlinepenalty\@M
    \Huge \bfseries #1\par\nobreak
    \vskip 40\p@
  }}
  \makeatother

\renewcommand{\chaptername}{}


\chapter{Introduction} 

\label{Chapter1} 

\lhead{1. \emph{Introduction}} 


\section{Overview}

The late twentieth century, and the following decades, have marked a pivotal period in the history of cosmology - a period which boasts the formulation, and precise parametrisation, of the accomplished $\Lambda$CDM cosmological model \citep{PlankLCDM}. This model, whilst not the full story, has seen extreme success in its ability to describe and predict the evolution of the Universe. The inclusion of cosmological inflation - a period of quasi-exponential expansion shortly after the Big Bang\footnote{Spanning approximately the first $10^{-36}$ - $10^{-34}$ seconds of the Universe.} - is paramount to solving a number of observational problems with the Big Bang theory\footnote{Namely, inflation solves the horizon and flatness problems.}. The detailed mechanics of this inflationary period form a contested yet integral part of modern cosmology. The concept of inflation was first proposed by Alan Guth in the early 1980s, and was motivated by attempting to understand the lack of observation of relic particles from the early Universe \citep{guth_inf}. Since then, it has become the most promising candidate for explaining the origin of structure in the Universe - a result of quantum fluctuations being stretched over classical distances due to accelerated expansion. These fluctuations act as the primordial seeds for generating density perturbations, which eventually undergo non-linear gravitational collapse to become stars and galaxies as seen today. Inflation provides a natural mechanism for the occurrence of density perturbations, and hence, observation of large scale structure (LSS) and the cosmic microwave background (CMB) can be used to constrain and differentiate between proposed inflationary models. Specifically, this essay plans to review the statistical imprints which these models predict are left on the CMB and LSS from primordial times.



\begin{figure}[h]
\centering
\includegraphics[scale=0.37]{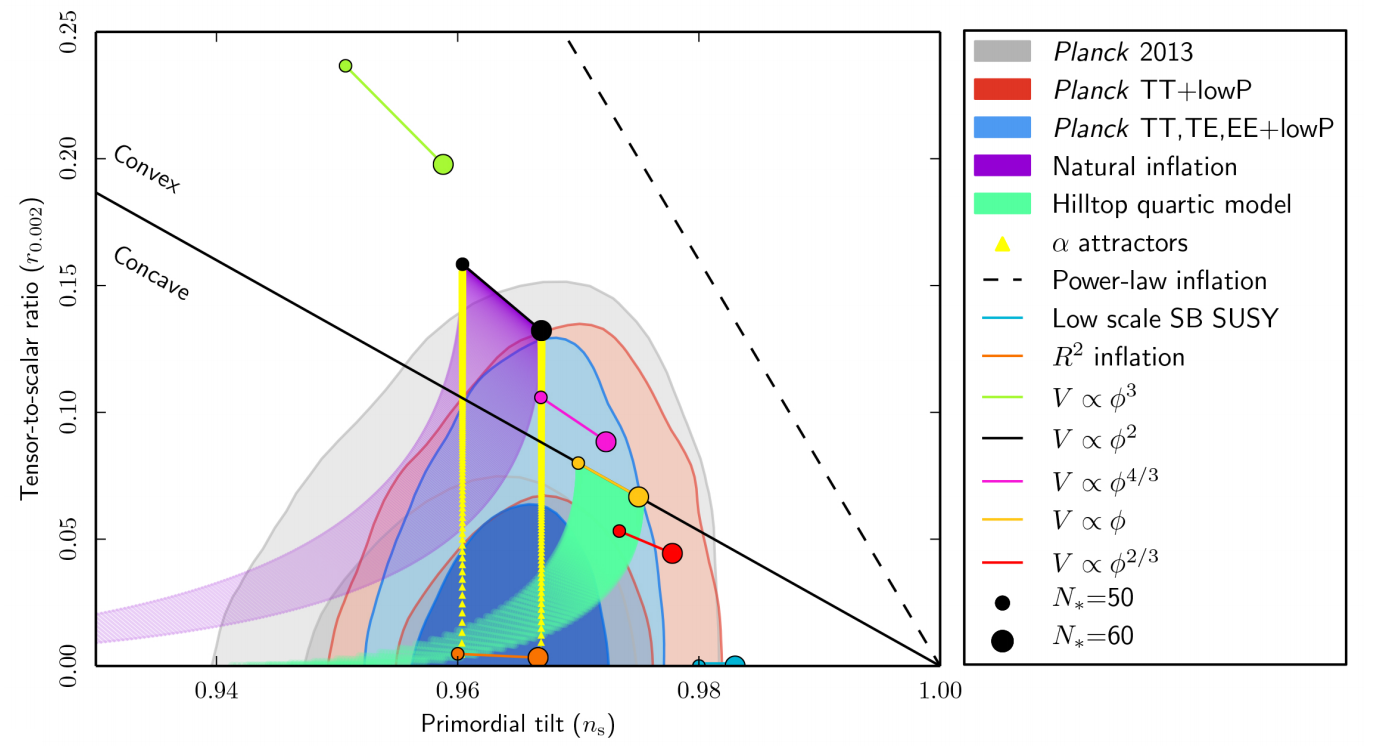}
\caption{\textit{Planck} 2015 constraints on the inflation parameters: $n_s$ and $r$. This data is compared to \textit{Planck} 2013 (grey) and various theoretical predictions of typical inflationary models \citep{plank2015}.}
\label{msa:r_vs_n}
\end{figure}

Historically, experiments regarding inflation have been mostly limited to the measurement of two parameters: the tensor to scalar ratio, $r$, and the scalar spectral index, $n_s$ \citep{WMAP}. The tensor to scalar ratio is defined as the ratio between the amplitude of tensor perturbations (gravitational waves) and scalar perturbations to the flat, Friedmann-Robertson-Walker (FRW) metric. Thus far, it remains consistent with zero, but is constrained to $r<0.11$ at $95\%$ confidence level (CL) \citep{plank2015}. Moreover, gravitational wave polarization signatures in the CMB - specifically, `B-mode' polarization - provide a further method of constraining (or detecting) primordial tensors with ground-based experiments such as BICEP2 \citep{BICEP2}. The scalar spectral index, $n_s$, is a measure of the deviation of the primordial curvature power spectrum from perfect scale-invariance. \textit{Planck} 2015 measures this scalar spectral index to be $n_s=0.968\pm0.006$ (68$\%$ CL), confirming the quasi-de Sitter nature of expansion during inflation. Fig$.$ \hspace{-0.25cm} \ref{msa:r_vs_n} depicts the latest constraints on these parameters provided by the \textit{Planck} mission in 2015 \citep{plank2015}; which improve upon earlier experiments such as \textit{Planck} 2013 \citep{plank2013} and WMAP \citep{WMAP}. The consequences these constraints have on selected inflationary models is also shown, notably disfavouring V($\phi$)$\propto\phi^2$ models of inflation. \\

\vspace{0.6cm}

Despite these observational successes, there still exists a large degeneracy of theoretical inflationary models which lie within the \textit{Planck} 2015 bounds on $r$ and $n_s$. However, a separate technique that is used to constrain, and distinguish between, models of inflation is measurements of \textit{non-Gaussianities} \citep{PlankNG}. This essay plans to provide a brief review of non-Gaussianities, specifically: what they are in the context of cosmological fields and observables, how they are generated in inflationary models, and how the \textit{Planck} 2015 data constrains such models.

\section{Essay Structure}


Section \ref{sec_2} begins with a comprehensive review of the statistical mathematics required to study non-Gaussianities. This will include, but is not limited to, a motivation for Gaussianity as a starting point, followed by a rigorous definition of the power spectrum and bispectrum from first principles. A brief introduction of relevant inflationary physics will then follow; focusing primarily on single field, slow roll inflation. This regime is explicitly detailed as it forms the basis of multiple ensuing formal calculations. Secondly, cosmological perturbation theory will be reviewed; the intention being to provide statements of key tools, which will later be referred to. For example, relevant gauges will be formally defined, along with the constant density curvature perturbation, $\zeta$. Thus, when the appropriate background physics and mathematics has been outlined, a discussion of phenomenological models of non-Gaussianity can begin. Firstly, the discussion will include a derivation of the local model of non-Gaussianity. This will introduce many fundamental concepts of the field of non-Gaussianities, such as a formal definition of a bispectrum \textit{amplitude} and \textit{shape function}. Furthermore, all three shape templates that \textit{Planck} explicitly constrains will be detailed - the local, equilateral, and orthogonal shapes. This will include an analysis of how non-Gaussianity is generated within inflationary models in each limit. Finally, an overview of observational considerations will be given. Specifically, the angular bispectrum of the CMB will be examined, along with a discussion of estimation techniques for amplitude of non-Gaussianity, and how two arbitrary shape functions can be correlated.

Section \ref{sec_23} will introduce the techniques required to \textit{calculate} non-Gaussianity given an inflationary model. A review of the quantum non-interacting theory will first be provided; outlining the quantisation procedure, and the resultant free field mode functions. Secondly, the `\textit{in-in}' formalism for calculating $n$-point quantum correlation functions with time-dependent interacting states will be motivated and summarised. Specifically, care will be taken to define the tree-level \textit{in-in} `master' formula that will be utilised in the subsequent sections.

Section \ref{sec_3} will be dedicated to detailing a seminal calculation made in Juan Maldacena's 2003 paper, \textit{Non-Gaussian features of primordial fluctuations in single field inflationary models} \citep{mald}. This calculation was used to determine that no observationally significant non-Gaussianity will be produced from single field, slow roll inflation.

Finally, Section \ref{sec_4} will investigate a class of inflationary models that \textit{can} produce an observationally significant amplitude of non-Gaussianity, namely, models whereby the initial state of inflation is \textit{non-Bunch-Davis}. The objective of this section will be to provide an introduction to the considerations that go into replacing the Bunch-Davis vacuum with excited initial states. As such, the discussion will be kept general, but will use the foundational paper, \textit{Enhanced Non-Gaussianity from Excited Initial States} by R. Holman and A. Tolley \citep{tol}, as a skeleton for the section. Toward the end, the work will also detail calculations of Ref$.$ \citep{meer} and Ref$.$ \citep{chennonbd}, in which various models are derived that are relevant to current experimental efforts. Particularly, the \textit{Planck} 2015 constraints for such models will be discussed and future prospects outlined. 

\section{Units and Conventions}

The metric signature $(-,+,+,+)$ will be used. 

Greek letters will denote spacetime indices, $\mu=0,1,2,3$; and Latin letters will denote spatial indices, $i=1,2,3$.

Natural units, $c=\hbar=1$, will be adopted throughout the work; where $M_{pl} \defeq (8 \pi G)^{-\frac{1}{2}} = 1$ unless explicitly stated otherwise.






\chapter{What is Non-Gaussianity?} 

\label{sec_2} 

\lhead{2. \emph{What is Non-Gaussianity?}} 


\section{Gaussian Random Fields}
\label{sec_21}

To understand what non-Gaussianity is in a cosmological setting, one must first begin with a discussion of the statistics of cosmological fields. Here, cosmological fields can refer to, for example, the temperature fluctuation field, $\delta T / \bar{T}$, which is used to probe anisotropies in the CMB. Most importantly for the following work, however, is the primordial scalar curvature perturbation on constant density space-like hypersurfaces, $\zeta$. With a convenient gauge choice, $\zeta$ will become a useful measure of quantum fluctuations during inflation. To begin with, Gaussian random fields (GRFs) will be quantitatively characterised, because initial perturbations from inflation are Gaussian random by the Central Limit Theorem (CLT). To see why this is, one can heuristically extend a statement of the CLT from random variables, $x_i$, to random fields, $f_i(\textbf{x})$. Foregoing the specific conditions under which the CLT is applicable, the statement is as follows: first, one must consider an arbitrary cosmological field, $f(\textbf{x})$. This field can now be split up into a sum of $n$ constituent fields which represent the \textit{same cosmological field}, but many different physical sources of the field,
\begin{equation}
\label{CLT1}
f(\textbf{x}) = \sum^n_{i=1} f_i(\textbf{x}).
\end{equation}
Each independently sourced constituent field, $f_i(\textbf{x})$, has its own arbitrary (not necessarily Gaussian) probability density functional (PDF), $P[f_i(\textbf{x})]$, pertaining to how the field is sourced\footnote{Note that a probability density functional for random fields will be more rigorously defined below.}. The CLT now states that, regardless of the underlying PDFs of the constituent fields, as $n\rightarrow\infty$, the PDF of the emergent field will be normally distributed, i.e. \textit{Gaussian},
\begin{equation}
\label{CLT1}
P[f(\textbf{x})] \propto \text{exp}[-f(\textbf{x})^2/\sigma^2].
\end{equation}
Hence, cosmological fields are said to be initially Gaussian random by the CLT, as no cosmological processes have yet had time to cause otherwise. Therefore, a departure from Gaussianity provides crucial insight into the physical processes that can drive such a deviation - hereafter referred to as \textit{non-linear} processes. A clear example of this is LSS formation in the late Universe, in which matter undergoes non-linear collapse due to Einsteinian gravity, and hence the distribution of matter in the Universe is highly non-Gaussian.

Given that Gaussianity has now been motivated as an appropriate starting point, it can be asked, how are Gaussian statistics treated for random fields, rather than variables? Particularly, we wish to arrive at `n-point correlators', which provide a measurable set of quantities that encode all the statistical information contained in a random field. Using the following convention, a random field $f$ can be converted between Fourier and real space as follows,
\begin{equation}
\label{fourier}
f(\textbf{k}) = \int d^3x \text{ } f(\textbf{x})e^{-i\textbf{k}.\textbf{x}},
\end{equation}
\begin{equation}
\label{fourier1}
f(\textbf{x}) = \int \frac{d^3k}{(2\pi)^3} f(\textbf{k})e^{i\textbf{k}.\textbf{x}},
\end{equation}
where the expansion
\begin{align}
\label{F_expansion}
f(\textbf{k})=a_{\textbf{k}} + ib_{\textbf{k}}
\end{align}
can be made without loss of generality. Constraints on these Fourier coefficients can be derived by enforcing the reality of $f(\textbf{x})$ in  (\ref{fourier}): $a_{\textbf{k}} = a_{-\textbf{k}}$, $b_{\textbf{k}}=-b_{-\textbf{k}}$. If one wishes to arrive at correlators for random fields, an expectation operation, $\langle \text{ } \rangle$, must be defined for fields. In the familiar case of random variables, an expectation value is defined as,
\begin{equation}
\label{rand_var}
\langle x\rangle = \int^{\infty}_{-\infty} dx \text{ } xP(x),
\end{equation}
where $P(x)$ is the probability density function of the random variable $x$. By analogy, one can now see that $P(x)$ from  (\ref{rand_var}) will be promoted to a \textit{probability density functional} $P[f(\textbf{k})]$, satisfying%
\begin{equation}
\label{exp_functional}
\langle F[f(\textbf{k})]\rangle = \int \mathcal{D}f\text{ }F[f(\textbf{k})]P[f(\textbf{k})].
\end{equation}
The functional, $F[f(\textbf{k})$], is simply any combination of the random field; and $\mathcal{D}f$, much like the path integral approach to quantum theory, denotes an integral over all possible field configurations. Therefore, using the expansion of $f(\textbf{k})$ in terms of Fourier coefficients, the expectation value of an arbitrary combination of GRFs satisfies
 \begin{equation}
\label{gauss_functional}
\langle F[f(\textbf{k})]\rangle = \Pi_{\textbf{k}}\int da_{\textbf{k}} \int db_{\textbf{k}}\text{ }F[f(\textbf{k})]P[f(\textbf{k})],
\end{equation}
\begin{equation}
\label{gauss_functional1}
P[f(\textbf{k})] = \frac{1}{\pi \sigma_\textbf{k}^2} \text{exp}[-(a_{\textbf{k}}^2 + b_{\textbf{k}}^2)/\sigma_\textbf{k}^2],
\end{equation}
for a given Fourier mode, \textbf{k}. Qualitatively, this means the random field has its Fourier coefficients drawn from the distribution in (\ref{gauss_functional1}) with variance $\sigma_{\textbf{k}}$ - such a field is known as \textit{Gaussian random}. Equipped with a well-defined expectation value operation for a combination of random fields, the n-point correlator can be defined as,
\begin{equation}
\label{n_pt_corr}
\langle f(\textbf{k}_1)f(\textbf{k}_2)\dots f(\textbf{k}_n)\rangle = \Pi_{\textbf{k}}\int da_{\textbf{k}} \int db_{\textbf{k}}\text{ }f(\textbf{k}_1)f(\textbf{k}_2)\dots f(\textbf{k}_n)P[f(\textbf{k})].
\end{equation}
Hence, n-point correlators measure the extent to which different Fourier modes are correlated for a given PDF - meaning the statistics of a field, $f$, are completely determined by the full hierarchy of correlators defined in  (\ref{n_pt_corr}). Interestingly, symmetries of the Gaussian PDF allow GRFs to be completely characterised by their 2-point correlator, or, the \textit{power spectrum}. To see why this is, the power spectrum must first be defined as follows: linearity of the expectation value operation allows the 2-point correlator to be expanded in terms of Fourier coefficients as such,
\begin{equation}
\label{2_pt_corr1}
\langle f(\textbf{k}_1)f(\textbf{k}_2)\rangle = \langle a_{\textbf{k}_1}a_{\textbf{k}_2} \rangle - \langle b_{\textbf{k}_1}b_{\textbf{k}_2} \rangle.
\end{equation}
Focusing on the first term in  (\ref{2_pt_corr1}), the explicit computation required is,
\begin{equation}
\label{2_pt_corr2}
\langle a_{\textbf{k}_1}a_{\textbf{k}_2} \rangle = \Pi_{\textbf{k}}\int da_{\textbf{k}} \int db_{\textbf{k}} \text{ } a_{\textbf{k}_1}a_{\textbf{k}_2} \frac{1}{\pi \sigma_\textbf{k}^2} \text{exp}[-(a_{\textbf{k}}^2 + b_{\textbf{k}}^2)/\sigma_\textbf{k}^2].
\end{equation}
One can immediately see that if $\textbf{k}_1 \neq \textbf{k}_2$, the above expression will always evaluate to zero. This is due to the product operator, $\Pi_\textbf{k}$, picking up zeros by unavoidably cycling through the first moment of the Gaussian PDF;
\begin{align}
\label{k1k2k3}
\textbf{k}_1 \neq \textbf{k}_2 = \textbf{k}: \langle a_{\textbf{k}_1}a_{\textbf{k}_2} \rangle = a_{\textbf{k}_1} \int da_{\textbf{k}} \int db_{\textbf{k}} \text{ } a_{\textbf{k}} \frac{1}{\pi \sigma_\textbf{k}^2} \text{exp}[-(a_{\textbf{k}}^2 + b_{\textbf{k}}^2)/\sigma_\textbf{k}^2] = 0,
\end{align}
which is a statement of zero mean for GRFs,
\begin{align}
\label{k1k2k3_2}
\langle f(\textbf{k}) \rangle = 0.
\end{align}
The remaining possibilities are exhausted by considering $\textbf{k}_1 = \textbf{k}_2$; whereby the resultant correlator picks up a factor of $\frac{1}{2}\sigma_{\textbf{k}}^2$ from the second moment of the Gaussian PDF ($\textbf{k}_1 = \textbf{k}_2 = \textbf{k}$), and unity from the zeroth moment ($\textbf{k}_1 = \textbf{k}_2 \neq \textbf{k}$), leaving the final expression as,
\begin{align}
\label{k1k2k3_3}
\langle a_{\textbf{k}_1}a_{\textbf{k}_2} \rangle = \delta(\textbf{k}_1 + \textbf{k}_2) \frac{1}{2}\sigma_{\textbf{k}}^2.
\end{align}
Symmetries between the Fourier coefficients result in,
\begin{align}
\langle a_{\textbf{k}_1}a_{\textbf{k}_2} \rangle = \langle b_{\textbf{k}_1}b_{-\textbf{k}_2} \rangle = -\langle b_{\textbf{k}_1}b_{\textbf{k}_2} \rangle,
\end{align}
which allows the full 2-point correlator of a GRF to be written as,
\begin{align}
\label{2_pt_grf}
\langle f(\textbf{k}_1)f(\textbf{k}_2)\rangle = \delta(\textbf{k}_1 + \textbf{k}_2) \sigma_{\textbf{k}}^2 \defeq  (2\pi)^3\delta(\textbf{k}_1 + \textbf{k}_2)P_f(\textbf{k}_1).
\end{align}
The power spectrum, $P_f(\textbf{k})$, has therefore been defined, and will later prove to be a crucial mathematical ingredient for studying non-Gaussianities.

The power spectrum can be simplified by noticing that the cosmological fields of interest are statistically isotropic, allowing $P_f(\textbf{k}) = P_f(|\textbf{k}|) = P_f(k)$. Moreover, the appearance of the Dirac delta function in  (\ref{2_pt_grf}) can be shown to be a result of the statistical homogeneity of $f$, and can hence be derived by enforcing translational invariance of the field\footnote{Which corresponds to transforming the field in Fourier space as $f(\textbf{k}) \rightarrow e^{-i\textbf{k}.\textbf{a}}f(\textbf{k})$.}. Given (\ref{2_pt_grf}), it is now possible to show that all statistical information about a GRF is contained within the power spectrum. First, via similar arguments as above, it can be shown that \textit{all} (2$n$+1)-point correlators vanish due to odd moments of the Gaussian PDF (which are zero by identity) becoming unavoidable in  (\ref{n_pt_corr}). This has the critical consequence that the 3-point correlator, or \textit{bispectrum}\footnote{This will be more rigorously defined, and motivated, in subsequent sections.}, vanishes for GRFs. Furthermore, all $2n$-point correlators of GRFs can be expressed in terms of the  power spectrum by a powerful contraction technique called \textit{Wick's theorem}\footnote{Also known as Isserlis' theorem. Wick proved this theorem in the context of quantum field theory, where it is used to decompose an arbitrary product of creation and annihilation operators.}. For brevity, Wick's theorem will not be proved, but simply stated as:
\begin{align}
\label{wick1}
\langle f(\textbf{k}_1)f(\textbf{k}_2)\dots f(\textbf{k}_{2n}) \rangle = \prod \sum \langle f(\textbf{k}_i)f(\textbf{k}_j)\rangle,
\end{align}
for $f$ Gaussian; where $\Pi \Sigma$ is introduced to denote the summation over $n$ products of all distinct pairs of $f(\textbf{k}_i)f(\textbf{k}_j)$. For example, the 4-point correlator is Wick contracted as so,
\begin{align}
\label{wick2}
\langle f(\textbf{k}_1)f(\textbf{k}_2)f(\textbf{k}_3)f(\textbf{k}_4) \rangle = \langle f(\textbf{k}_1)f(\textbf{k}_2) \rangle \langle f(\textbf{k}_3)f(\textbf{k}_4) \rangle&\\ \nonumber
+ \langle f(\textbf{k}_1)f(\textbf{k}_3) \rangle \langle f(\textbf{k}_2)f(\textbf{k}_4) \rangle&\\ \nonumber
+ \langle f(\textbf{k}_1)f(\textbf{k}_4) \rangle \langle f(\textbf{k}_2)f(\textbf{k}_3) \rangle&.
\end{align}
Hence, all statistical properties about GRFs are contained within the 2-point correlator,  (\ref{2_pt_grf}). It follows that Fourier modes in GRFs are \textit{uncorrelated} - this will not be the case in general for non-GRFs. Finally, the mathematical tools are now in place to define a quantity called the \textit{dimensionless power spectrum} - this will lead indirectly to how non-Gaussianity is quantified. The derivation of the dimensionless power spectrum involves the concept of \textit{scale invariance}, which is paramount to observational studies of inflationary models, and will hence be sketched here. One begins this derivation by enforcing that the statistics of the field, $f$, remain the same under a rescaling in real-space, $x \rightarrow \lambda x$:
\begin{align}
\label{scale1}
\langle f(\textbf{x}_1)f(\textbf{x}_2) \rangle = \langle f(\lambda\textbf{x}_1)f(\lambda\textbf{x}_2) \rangle.
\end{align}
Using the definition of $\langle \text{ } \rangle$, this condition is explicitly expressed as,
\begin{align}
\label{1234}
\int \frac{d^3 k_1}{(2\pi)^3} \int \frac{d^3 k_2}{(2\pi)^3} e^{i(\textbf{k}_1.\textbf{x}_1 + \textbf{k}_2.\textbf{x}_2)} &(2\pi)^3 \delta(\textbf{k}_1 + \textbf{k}_2)P(k_1)\\ \nonumber &= \int \frac{d^3 k_1}{(2\pi)^3} \int \frac{d^3 k_2}{(2\pi)^3} e^{i\lambda(\textbf{k}_1.\textbf{x}_1 + \textbf{k}_2.\textbf{x}_2)} (2\pi)^3 \delta(\textbf{k}_1 + \textbf{k}_2)P(k_1).
\end{align}
Thus, changing variables in the RHS of  (\ref{1234}) to $\textbf{k}_1' = \lambda\textbf{k}_1$ and $\textbf{k}_2' = \lambda\textbf{k}_2$, one finds,
\begin{align}
\label{2222}
\int \frac{d^3 k_1'}{\lambda^3(2\pi)^3} \int \frac{d^3 k_2'}{\lambda^3(2\pi)^3} e&^{i(\textbf{k}_1'.\textbf{x}_1 + \textbf{k}_2'.\textbf{x}_2)} (2\pi)^3 \delta(\frac{1}{\lambda}(\textbf{k}_1' + \textbf{k}_2'))P(k_1)\\ \nonumber
&= \frac{1}{\lambda^3}\int \frac{d^3 k_1'}{(2\pi)^3} \int \frac{d^3 k_2'}{(2\pi)^3} e^{i(\textbf{k}_1'.\textbf{x}_1 + \textbf{k}_2'.\textbf{x}_2)}(2\pi)^3 \delta(\textbf{k}_1' + \textbf{k}_2')P(k_1'),
\end{align}
by use of the three-dimensional Dirac delta identity,
\begin{align}
\delta(\frac{1}{\lambda}(\textbf{k}_1' + \textbf{k}_2')) = \lambda^3 \delta(\textbf{k}_1' + \textbf{k}_2').
\end{align}
Therefore,  (\ref{2222}) yields,
\begin{align}
\label{scale}
\langle f(\lambda\textbf{x}_1)f(\lambda\textbf{x}_2) \rangle = \frac{1}{\lambda^3}\langle f(\textbf{x}_1)f(\textbf{x}_2) \rangle,
\end{align}
which is clearly not scale invariant, unless the power spectrum scales as,
\begin{align}
\label{PS}
P(k) \propto \frac{1}{k^3}.
\end{align}
Under this assumption, the scaling of $\frac{1}{k^3}$ kills the factor of $\frac{1}{\lambda^3}$ in  (\ref{scale}) upon the above change of integration variables. Hence  (\ref{scale1}) is satisfied, and the statistics are scale invariant. Therefore, the dimensionless power spectrum, $\mathcal{P}(k)$, is defined by extracting the factor of $\frac{1}{k^3}$ as such:
\begin{align}
\label{S_I_PS}
\mathcal{P}(k) = k^3 P(k),
\end{align}
leading to the form of the power spectrum that is most often quoted,
\begin{align}
\langle f(\textbf{k}_1)f(\textbf{k}_2)\rangle = \frac{(2\pi)^3}{k^3}\delta(\textbf{k}_1 + \textbf{k}_2)\mathcal{P}_f(k_1).
\end{align}
As an aside, a key prediction of inflation is that the primordial power spectrum, $\mathcal{P}_{\zeta}(k)$, deviates slightly from perfect scale invariance, which is parametrised by the ansatz,
\begin{align}
\label{pps}
\mathcal{P}_{\zeta}(k) = A_s(k_{\star})\left(\frac{k}{k_{\star}}\right)^{n_s - 1}.
\end{align}
With a reference scale of $k_{\star}=0.05\text{Mpc}^{-1}$, \textit{Planck} 2015 has measured
\begin{align}
\label{ns}
n_s = 0.968 \pm 0.006
\end{align}
at a $68\%$ CL - which can be seen in Fig$.$ \hspace{-0.25cm} \ref{msa:r_vs_n}. Thus, a departure from perfect scale invariance, as predicted by the simplest inflationary models, has been confirmed. This result marks a triumph of modern observational and theoretical cosmology.






\section{Non-Gaussianities}

The statistics of GRFs have been rigorously defined in the previous section. We now wish to know, how are \textit{primordial} non-Gaussianities characterised in cosmology? This must first begin with a  discussion of what primordial means, which will be done by briefly reviewing relevant inflationary physics and cosmological perturbation theory. A simple, but illuminating parametrisation of non-Gaussianity will then be presented and used to demonstrate key theoretical and observational underpinnings.

\subsection{Inflation Review}

Inflation is defined as a period of accelerated expansion shortly after the Big Bang. This is equivalently expressed as a shrinking comoving Hubble radius,
\begin{align}
\label{inf}
\frac{d}{dt}(aH)^{-1} < 0,
\end{align}
where $a(t)$ is defined as the (cosmic) time-dependent scale factor which determines the expansion dynamics of the Universe via the flat FRW metric,
\begin{align}
\label{FRW}
ds^2 = -dt^2 + a(t)^2 d\textbf{x}^2.
\end{align}
Inflation is often mathematically stated in the above form because it stresses the notion that portions of the Universe fall out of causal contact with each other during inflation, thus solving the horizon problem \citep{Baumann}. Two parameters can now be defined which compactly capture the dynamics of inflation - the \textit{slow roll} parameters:
\begin{align}
\epsilon \defeq \frac{\dot{H}}{H^2},\\ \nonumber
\eta \defeq \frac{\dot{\epsilon}}{H\epsilon}. 
\end{align}
The \textit{slow roll conditions} follow by noticing that $\{\epsilon, |\eta|\}\ll1$ is traditionally required for inflation to begin and persist - this results in the potential term dominating over the kinetic term in simple scalar field models of inflation\footnote{In other words, the inflation scalar field, $\phi$, is slowly rolling down its potential V($\phi$) - hence `slow roll'.}. Such models form the basis of the work to follow, where quanta of the scalar field, $\phi$, are named \textit{inflatons}. The simplest action for a scalar field with a canonical kinetic term minimally coupled to gravity reads,
\begin{align}
S = \int d^4x \text{ } \sqrt{-g} \left[ \frac{M_{pl}^2}{2}\mathcal{R} - \frac{1}{2} g^{\mu \nu}\partial_{\mu}\phi \partial_{\nu} \phi - V(\phi) \right].
\end{align}
From such an action, classical dynamics of the inflationary model can be derived. By varying this action with respect to the inflaton field, a variant of the Klein-Gordon equation is found,
\begin{align}
\label{KG}
\ddot{\phi} + 3H\dot{\phi} + \partial_{\phi}V = 0.
\end{align}
Furthermore, expressions for $\text{P}_{\phi}$ and $\rho_{\phi}$ can be determined by varying the action with respect to the metric, yielding the stress-energy tensor, which can then be used to obtain the Friedmann equations,
\begin{align}
\label{fried}
H^2 = \frac{1}{3 M_{pl}^2}\left(\frac{1}{2}\dot{\phi}^2 + V(\phi)\right), \hspace{1.5cm} \dot{H} = -\frac{\dot{\phi}^2}{2 M_{pl}^2}.
\end{align}
The ratio of $\text{P}_{\phi}$ and $\rho_{\phi}$ form the equation of state of the inflaton field,
\begin{align}
\label{EoS}
w_{\phi} = \frac{\text{P}_{\phi}}{\rho_{\phi}} = \frac{\frac{1}{2}\dot{\phi}^2 - V(\phi)}{\frac{1}{2}\dot{\phi}^2 + V(\phi)}. 
\end{align}
Quasi-exponential expansion in fact requires $w_{\phi}<-\frac{1}{3}$ - thus violating the strong energy condition (SEC) \citep{Baumann}, but providing the negative pressure required to drive such an expansion. An important feature when considering SEC-violating fluids is the \textit{de Sitter limit}. This limit refers to a Universe in which the equation of state is $w=-1$, and hence de Sitter space is obtained in  (\ref{FRW}) by finding $a(t) \propto e^{\text{H}t}$, $H \approx \text{constant}$. Such dynamics refer to a Universe dominated by constant vacuum energy, or a \textit{cosmological constant} - a promising candidate for the nature of dark energy\footnote{As discussed, the inflationary period now has strong statistical evidence against perfect de Sitter dynamics (cosmological constant) by the running of the scalar spectral index.} \citep{darkenergy}.

Combined,  (\ref{KG}) and  (\ref{fried}) completely specify the scalar field and FRW dynamics respectively. This is called \textit{single field slow roll inflation}, and will be revisited in the context of primordial non-Gaussianities in Section \ref{sec_3}. Distinct models within the single field slow roll regime are then differentiated between by specifying a potential, $V(\phi)$. An example of a typical potential which adheres to the slow roll conditions is shown in Fig$.$ \ref{fig_SR} below.
\begin{figure}[h]
\centering
\includegraphics[scale=0.32]{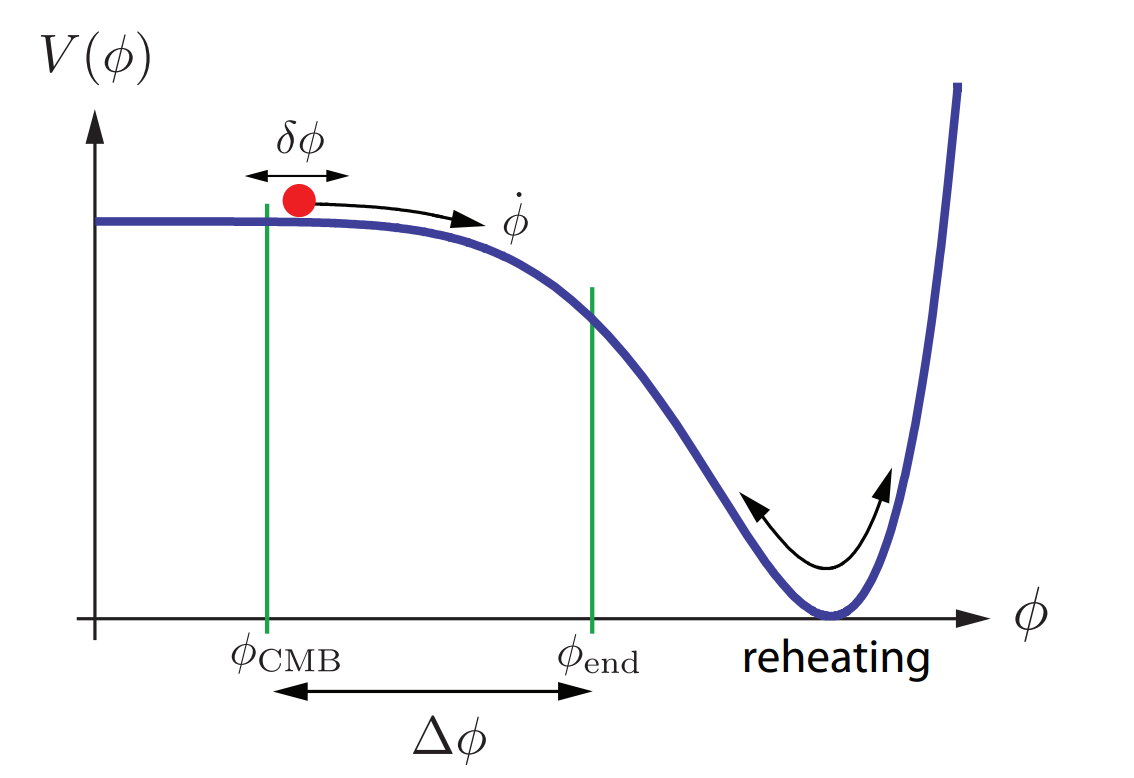}
\caption{The form of a typical inflation potential, where $\phi$ is `slowly rolling' down the potential toward $\phi_{\text{end}}$, marking the end of inflation. Decay of $\phi$ into the standard model occurs at the global minimum known as `reheating'. $\delta \phi$ denotes quantum fluctuations of the field around a classical background, $\bar{\phi}$. Image taken from D. Baumann's TASI Lectures on Inflation \citep{Baumann}.}
\label{fig_SR}
\end{figure}

Quantum fluctuations of the inflaton field are labelled as $\delta \phi$ on Fig$.$ \ref{fig_SR}, which modify the homogeneous classical background, $\bar{\phi}(t)$. Thus, the resultant field takes the form,
\begin{align}
\label{Q_fluc}
\phi(\textbf{x},t) = \bar{\phi}(t) + \delta \phi(\textbf{x}, t).
\end{align} 
It is these fluctuations, stretched over classical distances, that seed the formation of LSS as mentioned in previous sections. The origin of such fluctuations are quantum mechanical in nature, owing to a temporal uncertainty. Therefore, the inflaton field acts as a clock, counting toward the end of inflation, meaning local perturbations $\delta \phi(\textbf{x},t)$ can induce local differences in the density fields post-inflation. To elaborate, if $\delta \phi$ fluctuates up the potential in Fig$.$ \ref{fig_SR}, that region of space will inflate for longer, resulting in a region of lower density (and vice versa). These types of perturbation, which can be described by a local shift in time, are called \textit{adiabatic}. It is the goal of this work to review how these fluctuations could affect the statistics of cosmological observables. Hence, to treat them in a formal manner, cosmological perturbation theory will be required - the key mathematical components of which will now be briefly outlined.

\subsection{Cosmological Perturbation Theory}
\label{subsec_pert}

Cosmological perturbation theory, sparing no detail, can quickly become algebraically dense at a loss of transparency. Therefore, the intention of this section is to simply state and discuss the relevant components of cosmological inhomogeneity theory for later use in computing non-Gaussianities. This will first begin with an account of the mathematical setting in which non-linearity is most easily studied in cosmology - the 3+1 or ADM formalism of General Relativity \citep{ADM}. Ultimately, this discussion will include a definition of the primordial curvature scalar, $\zeta$, and will introduce the tools required to obtain a perturbed, order-$n$ action for fluctuations of the inflaton field. 

The 3+1 split of General Relativity is named as such because it separates the full, four-dimensional differentiable space-time manifold $\mathcal{M}$ into a set of constant $t$, space-like hypersurfaces. These hypersurfaces, $\Sigma_t$, are separated by a proper time defined by $N(t,x^i)dt$, and admit a change in spatial coordinates along a well-defined `normal' trajectory between the hypersurfaces of $x^i \rightarrow x^i + N^i dt$. This yields a metric of the form,
\begin{align}
\label{ADM}
ds^2 = -N^2 dt^2 + \leftidx{^{(3)}}g_{ij}(dx^i - N^idt)(dx^j - N^jdt),
\end{align}
where $N(t,x^i)$ is called the \textit{lapse function}, and $N^i(t,x^i)$ is called the \textit{shift vector}. Furthermore, a notion of curvature can be defined on $\Sigma_t$ and divided into two kinds - the \textit{intrinsic} and \textit{extrinsic} curvatures. The intrinsic curvature, R, is obtained in the usual way - with the Ricci scalar built out of Christoffel  symbols - under the replacement $g_{\mu \nu} \rightarrow \text{ }^{(3)}g_{ij}$. Perhaps more importantly the extrinsic curvature is defined by parallel propagation of a vector, $n^i$, normal to $\Sigma_t$ along an integral curve of a vector tangent to $\Sigma_t$. The resultant deviation of this normal vector is then defined as the extrinsic curvature,
\begin{align}
\label{ex_curv}
K_{ij} = \nabla_j n_i = - \frac{1}{2N} \left(\partial_0 \! \text{ } ^{(3)}g_{ij} + \mathcal{D}_j N_{i} + \mathcal{D}_i N_{j} \right),
\end{align}
where $\mathcal{D}$ denotes the covariant derivative on $\Sigma_t$. The general form of the Einstein-Hilbert action in this formalism therefore becomes,
\begin{align}
\label{action}
S = S_G + S_m = \frac{1}{16\pi G} \int dt d^3x \text{ } N \sqrt{^{(3)}g}\left[R - K^2 + K_{ij}K^{ij} \right] + \int dt d^3x \text{ } \mathcal{L}_m,
\end{align}
where $K$ is the trace of  (\ref{ex_curv}). Varying this action with respect to the fields it contains produces a host of equations specifying the dynamical behaviour of the theory and the various constraints that can be applied.

In order to solve the dynamics and constraint equations for perturbations introduced by varying  (\ref{action}), one must first obtain a set of scalar functions representing said perturbations for both geometry and matter. First, the metric can be perturbed to linear order about a FRW background. This background is redefined as,
\begin{align}
\label{FRW1}
ds^2 = -\bar{N}^2(t) dt^2 + a(t)^2 \delta_{ij}dx^i dx^j,
\end{align}
where $\bar{N} = 1$ refers to cosmic time, and $\bar{N} = a$ refers to conformal time. To perturb  (\ref{FRW1}), one must exhaust all possible combinations in which this metric can be modified using scalars, vectors, and tensors to first order. We will, however, discard vector and tensor perturbations for the remainder of this work, because they are decoupled from scalar perturbations - the focus of this essay. Performing a scalar-vector-tensor decomposition, to isolate all possible scalar contributions, results in the perturbed line element reading,
\begin{align}
ds^2 = -\bar{N}^2 (1+2\Psi)dt^2 + 2a(t)^2\partial_iBdtdx^i + a(t)^2\left[(1-2\Phi)\delta_{ij} + 2\partial_i \partial_j E \right]dx^idx^j.
\end{align}
Therefore, four scalar functions: $\Psi$, $\Phi$, $B$, and $E$ arise from perturbations to the flat FRW geometry. A further scalar, $\delta \phi$, is obtained by perturbing the inflaton field. It will later be convenient to shift some of these perturbations onto the trace and traceless parts of the extrinsic curvature defined in  (\ref{ex_curv}). In doing so, a compact set of linearised equations determining the evolution of both matter and metric perturbations can be found via the constraints derived from the Einstein equations. However, despite these constraints, there still exist spurious degrees of freedom in the metric due to the diffeomorphism invariance of General Relativity. This invariance reflects a redundancy in choice of coordinates - the gauge problem. There exist multiple techniques to deal with this redundancy. One such way is to \textit{fix} the gauge. Gauge fixing amounts to choosing a coordinate system in which scalar perturbations will be treated; ultimately arriving at an observable, which will be independent of the gauge choice. Different gauges have different attractions, ranging from mathematical compactness, to physical transparency - the two often overlapping. Two such gauge choices in particular are pertinent to this work: the \textit{comoving} gauge\footnote{A gauge in which the spatial slices move with the fluid, and hence is equivalent to setting the fluid velocity $u=0$.} ($B=E=0$) in which degrees of freedom will be forced onto the curvature scalar; and the spatially flat gauge ($\Phi=E=0$) in which degrees of freedom will be pushed onto the inflaton fluctuation scalar.

A further technique to circumvent the gauge problem is to find combinations of scalar perturbations which remain invariant under a gauge transformation\footnote{Formally computed by making an infinitesimal temporal and spatial coordinate transformation, and forcing the line element to remain the same.}. This has the advantage of dealing with manifestly physical degrees of freedom. Gauge invariant quantities were first introduced to cosmological perturbation theory by the seminal contributions of James Bardeen in the form of the Bardeen potentials: $\Psi_{gi}$, $\Phi_{gi}$ \citep{bardeen}. In this vein, the following gauge invariant quantity can be defined,
\begin{align}
\label{curv_gi}
\zeta = - \Phi + \frac{\delta \rho}{3(\bar{\rho} + \bar{P})}.
\end{align}
On slices of constant energy density, this expression reduces to the gravitational potential ($\Phi$) or the scalar curvature via the relation,
\begin{align}
R = \frac{4}{a^2}\nabla^2 \Phi.
\end{align}
Hence, $\zeta$ is formally given the name \textit{constant density curvature perturbation}. The construction of $\zeta$ was such that it is a convenient measure of \textit{adiabatic} perturbations (in our case, single field inflaton fluctuations\footnote{Multifield inflation models can, in fact, generate primordial fluctuations which are not adiabatic, but isocurvature perturbations.}) through its dependence on $\rho$. Furthermore, $\zeta$ has the exceptionally useful property of being \textit{conserved on superhorizon scales}\footnote{Superhorizon scales refers to when Fourier wavelengths are much greater than the Hubble horizon - $k \ll aH$.} - $\dot{\zeta}\approx0$. This is the defining feature of $\zeta$, because it will allow us to map features of the field during inflation to observables at later times, such as the CMB - thus bridging the gap between the primordial and observable Universe. A particularly illuminating depiction of this history of $\zeta$, from primordial times to the CMB, can be seen in Fig$.$ \ref{fig_zeta} below.
\begin{figure}[h]
\centering
\includegraphics[scale=0.24]{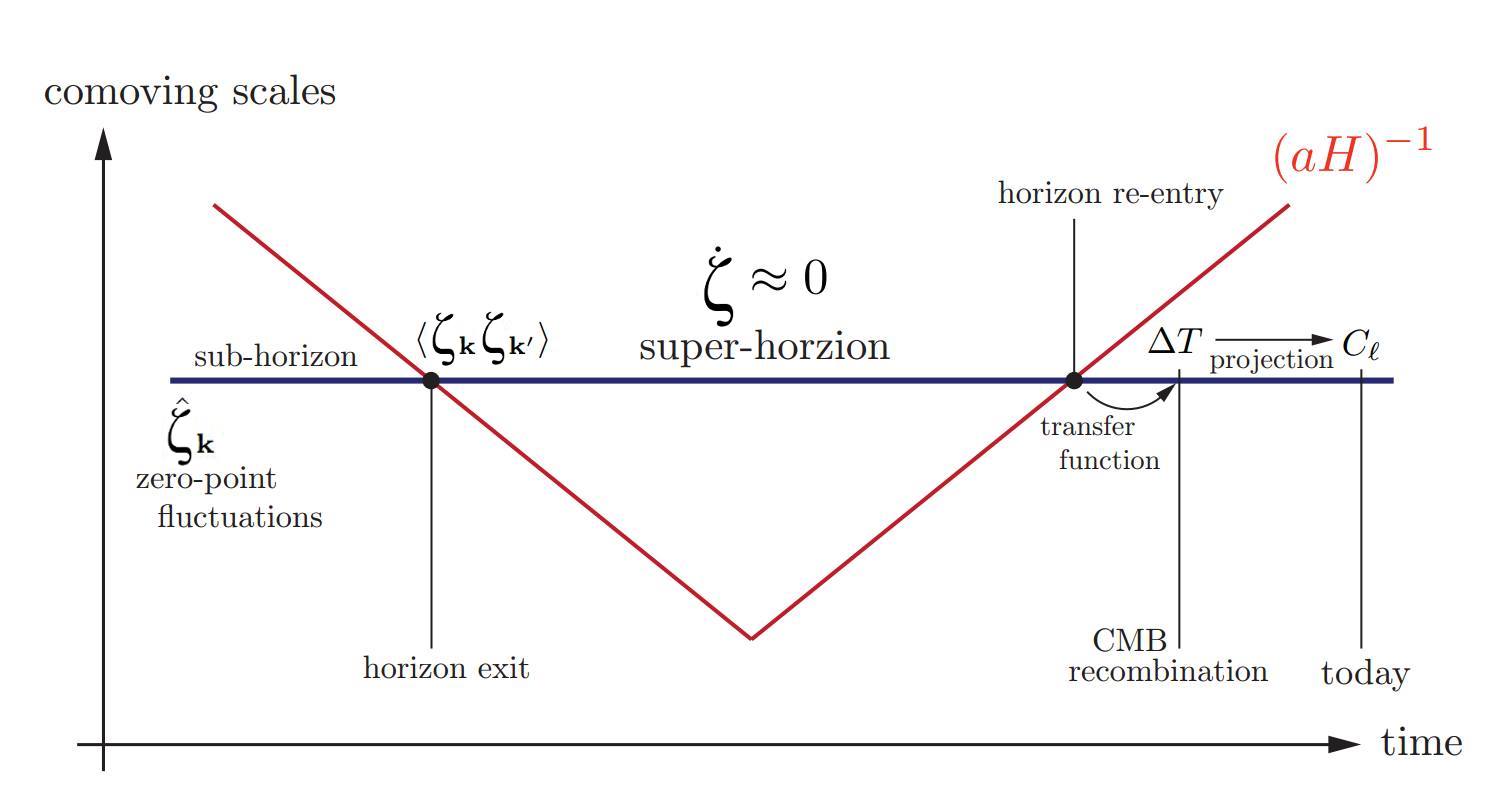}
\caption{Evolution of the curvature perturbation, $\zeta$; beginning at inflation, and ending at its imprint on the CMB as measured today. Source image taken (and edited) from D. Baumann's TASI Lectures on Inflation \citep{Baumann}.}
\label{fig_zeta}
\end{figure}
To elaborate on this history; $\zeta$ begins as a quantum field, $\hat{\zeta}$, which can be expanded as operators and time-dependent mode functions, satisfying a classical equation of motion. Due to interaction terms, finite zero-point fluctuations of the quantum field (with a well-defined vacuum) manifest in the 3-point expectation value. These fluctuations will freeze as they leave the horizon because of the aforementioned properties of $\zeta$. Hence, their behaviour can be measured when they re-enter the horizon at a later time, post-inflation. It can be seen in Fig$.$ \ref{fig_zeta} that a finite time elapses between the horizon re-entry of $\zeta_{\textbf{k}}$, and the creation of the CMB - where these fluctuations will eventually leave their imprint. Therefore, a calculation will be required to propagate the field statistics from horizon re-entry, to their effect on the angular bispectrum of the CMB measured at present day. Such a calculation is done computationally, using a transfer function. However, the physics involved in this process is nearly linear, and will hence not interfere with the non-Gaussian signature generated by inflation. 

The tools detailed in this section can now be used to obtain an order-$n$ action for the scalar degree of freedom during inflation. This will be done in Section \ref{sec_3} by considering a gauge-restricted non-linear line element of the form,
\begin{align}
\label{mald_line}
ds^2 = -\left[ (1+ \Psi)^2 - \partial_i B \partial^i B \right] dt^2 + 2a(t)^2 \partial_iB dt dx^i + a(t)^2 e^{2\zeta}\delta_{ij}dx^i dx^j.
\end{align}
This line element is expressed in the comoving gauge, where matter is unperturbed ($\delta \phi = 0$) and the spatial geometry is perturbed by a factor of $e^{2\zeta}$. Thus, there are three remaining scalar degrees of freedom: $\Psi$, $B$, and $\zeta$. The aforementioned gauge fixing process has removed two (of the original five) degrees of freedom by using the spatial and temporal re-parametrisations: $x^i \rightarrow x^i + \partial^i \alpha$ and $t \rightarrow t + \beta$. Moreover, energy and momentum constraint equations remove a further two degrees of freedom; meaning all perturbations are moved onto a singular degree of freedom: either $\zeta$ (geometry) or $\delta \phi$ (matter), depending on the gauge choice. From  (\ref{mald_line}), the lapse and shift can then be deduced, allowing for the calculation of intrinsic and extrinsic curvatures. These can  then be substituted into the general 3+1 Einstein-Hilbert action, defined in  (\ref{action}), to obtain contributions of arbitrary order in the perturbation field. The first non-trivial contributions (in which interactions are treated) appear at third order in inflaton fluctuations, and hence third order is what this action will be expanded to in Section \ref{sec_3}. From such an action, the quantum dynamical behaviour of these perturbations can be calculated using the \textit{`in-in' formalism}. For further details regarding the interim mathematics of cosmological perturbation theory that were omitted here, the reader is referred to A. Riotto's \textit{Inflation and the Theory of Cosmological Perturbations} \citep{Riotto}.\\

\subsection{Non-Gaussian Phenomenology}
\label{sec_223}

The mathematical tools have now been introduced which allow us to study phenomenological models of non-Gaussianity. However, \textit{calculating} the non-Gaussianity predicted by specific models of inflation will be left to Sections \ref{sec_3} and \ref{sec_4}. This is because the resultant bispectra from these calculations fall into broad, model-independent classes, which will be characterised and discussed below. To introduce and illustrate these classes, a simple ansatz can be made which involves a local correction term to a GRF ($\zeta_G$) proportional to the square of itself,
\begin{align}
\label{komat}
\zeta(\textbf{x}) = \zeta_G(\textbf{x}) + \frac{3}{5}f_{\text{NL}}^{\text{local}}\left[ \zeta_G(\textbf{x})^2 - \langle \zeta_G(\textbf{x})^2 \rangle \right],
\end{align}
hereafter referred to as the local model\footnote{The factor of $\frac{3}{5}$ is a historical one, as a result of the original work using the gravitational potential $\Phi$. On superhorizon scales, $\Phi = \frac{3}{5} \zeta$.}. This parametrisation of non-Gaussianity is attributed to Eiichiro Komatsu and David Spergel \citep{local}; where $f_{\text{NL}}^{\text{local}}$ is the parameter defining the amplitude of the non-Gaussianity\footnote{`NL' in $f_{\text{NL}}^{\text{local}}$ refers to non-linear, because the non-Gaussianity scales as the square of a GRF. `Local' refers to the correction being localised at \textbf{x}.}. Given this ansatz, the resulting bispectrum can now be calculated, thus arriving at measurable quantities that \textit{Planck} can constrain. The bispectrum is defined as the Fourier transform of the 3-point correlator, thus one begins by taking the Fourier transform of  (\ref{komat}). The non-linear terms transform non-trivially: firstly, the rightmost term is Fourier transformed into an expression involving the power spectrum,
\begin{align}
\mathcal{F}(\langle \zeta_G(\textbf{x})^2 \rangle) = \int d^3x\left(\int \frac{d^3k'}{(2\pi)^3} P_{\zeta}(k') \right)e^{i\textbf{k}.\textbf{x}},
\end{align} 
which used the definition,
\begin{align}
\langle \zeta_G(\textbf{x})^2 \rangle = \int \frac{d^3k'}{(2\pi)^3} P_{\zeta}(k').
\end{align}
Upon substitution of the Dirac delta identity, the first term transforms as,
\begin{align}
\mathcal{F}(\langle \zeta_G(\textbf{x})^2 \rangle) = (2 \pi)^3 \delta(\textbf{k})\left(\int \frac{d^3k'}{(2\pi)^3} P_{\zeta}(k') \right).
\end{align}
Secondly, the leftmost non-linear term is the Fourier transform of a product, thus requiring the convolution theorem,
\begin{align}
\mathcal{F}(\zeta_G(\textbf{x}) . \zeta_G^*(\textbf{x})) = \mathcal{F}(\zeta_G(\textbf{x})) \ast \mathcal{F}(\zeta_G^*(\textbf{x})),
\end{align}
where $\ast$ denotes a convolution. Hence, by the definition of a convolution,
\begin{align}
\zeta_G(\textbf{k}) \ast \zeta_G^*(\textbf{k}) = \int \frac{d^3 k'}{(2\pi)^3} \zeta_G(\textbf{k} + \textbf{k}')\zeta_G^*(\textbf{k}').
\end{align}
Combining these gives the full, Fourier transformed field,
\begin{align}
\zeta(\textbf{k}) &= \zeta_G(\textbf{k}) + \frac{3}{5}f_{\text{NL}}^{\text{local}}\left( \int \frac{d^3 k'}{(2\pi)^3} \zeta_G(\textbf{k} + \textbf{k}')\zeta_G^*(\textbf{k}') - (2 \pi)^3 \delta(\textbf{k})\left(\int \frac{d^3k'}{(2\pi)^3} P_{\zeta}(k') \right) \right)\\ \nonumber
&\defeq \zeta_G(\textbf{k}) + \zeta_{\text{NL}}(\textbf{k}).
\end{align}
The 3-point correlator in Fourier space is therefore,
\begin{align}
\label{3ptcor_local}
\langle \zeta(\textbf{k}_1)\zeta(\textbf{k}_2)\zeta(\textbf{k}_3) \rangle &= \langle (\zeta_G(\textbf{k}_1) + \zeta_{\text{NL}}(\textbf{k}_1))(\zeta_G(\textbf{k}_2) + \zeta_{\text{NL}}(\textbf{k}_2))(\zeta_G(\textbf{k}_3) + \zeta_{\text{NL}}(\textbf{k}_3)) \rangle \\ \nonumber
&= \langle \zeta_G(\textbf{k}_1)\zeta_G(\textbf{k}_2)\zeta_G(\textbf{k}_3) \rangle + \langle \zeta_G(\textbf{k}_1)\zeta_G(\textbf{k}_2)\zeta_{\text{NL}}(\textbf{k}_3) \rangle + \text{cyc. perms.},\\ \nonumber
& \hspace{2.2cm}  \veq \\[-0.3cm] \nonumber
&  \hspace{2.25cm} \vspace{-1.5cm} 0
\end{align}
where terms which are second order and above in the non-linear field have been ignored. It can now be seen that the non-zero contributions of  (\ref{3ptcor_local}) expand into a 4-point correlator of Gaussian fields, which can be expressed in terms of the power spectrum by Wick's theorem - detailed in Section \ref{sec_21}. The explicit computation required is,
\begin{align}
\label{1111111}
\langle \zeta_G(\textbf{k}_1)\zeta_G(\textbf{k}_2)\zeta_{\text{NL}}(\textbf{k}_3) \rangle = \frac{3}{5}f^{\text{local}}_{\text{NL}} \left< \zeta_G(\textbf{k}_1)\zeta_G(\textbf{k}_2)  \left( \int  \frac{d^3 k'}{(2\pi)^3} \zeta_G(\textbf{k}_3 + \textbf{k}')\zeta_G^*(\textbf{k}')\right. \right. \\ \nonumber
 - \left. \left.(2 \pi)^3 \delta(\textbf{k}_3)\left(\int \frac{d^3k'}{(2\pi)^3} P_{\zeta}(k') \right) \right) \right>.
\end{align}
The rightmost term can be factored out of the correlator, whereas the leftmost gets contracted into three terms, which will now be evaluated. The first term in the Wick contraction is simply the coefficient of the 2-point correlator in $\textbf{k}_1$ and $\textbf{k}_2$,
\begin{align}
\nonumber
\frac{3}{5}f^{\text{local}}_{\text{NL}} \langle \zeta_G(\textbf{k}_1)\zeta_G(\textbf{k}_2) \rangle \left( \int  \frac{d^3 k'}{(2\pi)^3}\langle  \zeta_G(\textbf{k}_3 + \textbf{k}')\zeta_G^*(\textbf{k}') \rangle - (2 \pi)^3 \delta(\textbf{k}_3)\left(\int \frac{d^3k'}{(2\pi)^3} P_{\zeta}(k') \right) \right),
\end{align}
which evaluates to zero by noticing that,
\begin{align}
\int  \frac{d^3 k'}{(2\pi)^3}\langle  \zeta_G(\textbf{k}_3 + \textbf{k}')\zeta_G^*(\textbf{k}') \rangle - &(2 \pi)^3 \delta(\textbf{k}_3)\left(\int \frac{d^3k'}{(2\pi)^3} P_{\zeta}(k') \right)\\ \nonumber
 &= \delta(\textbf{k}_3) \int  d^3 k' P_{\zeta}(k')  - (2 \pi)^3 \delta(\textbf{k}_3)\left(\int \frac{d^3k'}{(2\pi)^3} P_{\zeta}(k') \right) \\ \nonumber &= 0.
\end{align}
The second term of the Wick contraction is,
\begin{align}
\int \frac{d^3k'}{(2\pi)^3} \langle \zeta_G(\textbf{k}_1) \zeta_G(\textbf{k}_3 + \textbf{k}') \rangle  \langle \zeta_G(\textbf{k}_2) \zeta_G^*(\textbf{k}') \rangle &= (2 \pi)^3 \int d^3k' \delta(\textbf{k}_1 + \textbf{k}_3 + \textbf{k}')P_{\zeta}(k_1) \delta(\textbf{k}_2-\textbf{k}')P_{\zeta}(k')\\ \nonumber
 &= (2\pi)^3  \delta(\textbf{k}_1 + \textbf{k}_2 + \textbf{k}_3)P_{\zeta}(k_1)P_{\zeta}(k_2).
\end{align}
Finally, by symmetry, the third term equates to the second. Combining these three Wick contractions produces the following 3-point correlator,
\begin{align}
\langle \zeta_G(\textbf{k}_1)\zeta_G(\textbf{k}_2)\zeta_{\text{NL}}(\textbf{k}_3) \rangle = \frac{3}{5}f^{\text{local}}_{\text{NL}} 2 (2\pi)^3  \delta(\textbf{k}_1 + \textbf{k}_2 + \textbf{k}_3)P_{\zeta}(k_1)P_{\zeta}(k_2).
\end{align}
Furthermore, cyclic permutations of $\zeta_{\text{NL}}$ in  (\ref{1111111}) account for two additional contributions to the 3-point correlator of the field in  (\ref{komat}):
\begin{align}
\langle \zeta_G(\textbf{k}_1)\zeta_{\text{NL}}(\textbf{k}_2)\zeta_G(\textbf{k}_3) \rangle = \frac{3}{5}f^{\text{local}}_{\text{NL}}2(2\pi)^3  \delta(\textbf{k}_1 + \textbf{k}_2 + \textbf{k}_3)P_{\zeta}(k_1)P_{\zeta}(k_3), \\ \nonumber
\langle \zeta_{\text{NL}}(\textbf{k}_1)\zeta_G(\textbf{k}_2)\zeta_G(\textbf{k}_3) \rangle = \frac{3}{5}f^{\text{local}}_{\text{NL}}2(2\pi)^3  \delta(\textbf{k}_1 + \textbf{k}_2 + \textbf{k}_3)P_{\zeta}(k_2)P_{\zeta}(k_3).
\end{align}
Thus, a non-zero 3-point correlator has been obtained via the inclusion of a quadratic local term built out of an underlying GRF,
\begin{align}
\hspace{-0.5cm} \langle \zeta(\textbf{k}_1)\zeta(\textbf{k}_2)\zeta(\textbf{k}_3) \rangle = \frac{6}{5}f^{\text{local}}_{\text{NL}}(2\pi)^3 \delta(\textbf{k}_1 + \textbf{k}_2 + \textbf{k}_3) (P_{\zeta}(k_1)P_{\zeta}(k_2) + P_{\zeta}(k_1)P_{\zeta}(k_3) + P_{\zeta}(k_2)P_{\zeta}(k_3)).
\end{align}
A general bispectrum, $B$, is now defined by extending the definition of the power spectrum,
\begin{align}
\langle \zeta(\textbf{k}_1)\zeta(\textbf{k}_2)\zeta(\textbf{k}_3) \rangle = (2\pi)^3 \delta(\textbf{k}_1 + \textbf{k}_2 + \textbf{k}_3) B_{\zeta}(k_1, k_2, k_3) ,
\end{align}
which allows the local bispectrum to be deduced,
\begin{align}
B^{\text{local}}_{\zeta}(k_1, k_2, k_3) = \frac{6}{5}f^{\text{local}}_{\text{NL}}  (P_{\zeta}(k_1)P_{\zeta}(k_2) + P_{\zeta}(k_1)P_{\zeta}(k_3) + P_{\zeta}(k_2)P_{\zeta}(k_3)).
\end{align}
The amplitude parameter, $f_{\text{NL}}$, is often defined relative to the power spectrum via $f_{\text{NL}} \sim B_{\zeta}(k,k,k)/P^2_{\zeta}(k)$. Local non-Gaussianity is therefore characterised by a sum of products of power spectra, which makes for a relatively tractable calculation. If scale invariance is assumed, which has been measured to be approximately the case, the functional form of this bispectrum can be arranged as such: 
\begin{gather}
P_{\zeta}(k) \propto \frac{1}{k^3} \rightarrow P_{\zeta}(k) = \frac{\mathcal{P}_{\zeta}}{k^3}, \\ \nonumber
B^{\text{local}}_{\zeta}(k_1, k_2, k_3) = \frac{6}{5}f^{\text{local}}_{\text{NL}} \mathcal{P}^2_{\zeta} \left(\frac{1}{k_1^3 k_2^3} + \frac{1}{k_1^3 k_3^3} + \frac{1}{k_2^3 k_3^3}\right).
\end{gather} 
A few noteworthy properties of general bispectra can now be stated, before continuing with the local ansatz example. Firstly, it can be deduced that a scale invariant bispectrum will always scale as $\frac{1}{k^6}$, by the same method of enforcing real-space scale invariance in Section \ref{sec_21}. This property allows one to factorise out the implicit $k$-dependence, and obtain bispectra as a product of an amplitude and a \textit{shape function}. Secondly, the appearance of $\delta(\textbf{k}_1 + \textbf{k}_2 + \textbf{k}_3)$ in the 3-point correlator results in the Fourier wavevectors being constrained to forming a \textit{closed triangle} in $k$-space. This is, to some extent, a statement of the conservation of momentum. Finally, the local bispectrum defined above depends only on the magnitude of the three Fourier modes: $k_1$, $k_2$, and $k_3$. This will always be the case for an isotropic and homogeneous field. A heuristic argument\footnote{Without actually applying rotation and translation operators to the fields within the 3-point correlator.} as to why this is the case will now be provided. For a completely arbitrary field, the bispectrum contains a maximum of nine degrees of freedom: $\textbf{k}_1$, $\textbf{k}_2$, and $\textbf{k}_3$. Isotropy is a statement of rotational invariance. All scalar quantities that can be built out of the original nine degrees of freedom are manifestly invariant under a rotation - $k_1^2$, $k_2^2$, $k_3^2$, $\textbf{k}_1.\textbf{k}_2$, $\textbf{k}_2.\textbf{k}_3$, $\textbf{k}_1.\textbf{k}_3$. Homogeneity is then enforced via the aforementioned triangle condition, $\textbf{k}_1 + \textbf{k}_2 + \textbf{k}_3 = 0$. Thus, knowing two of these wavevectors fixes the third. Without loss of generality, $\textbf{k}_1$ and $\textbf{k}_2$ can be chosen to fix $\textbf{k}_3$, leaving three independent degrees of freedom: $k_1^2$, $k_2^2$, and $\textbf{k}_1.\textbf{k}_2 \sim k_3^2$. Therefore, the bispectrum of a homogeneous and isotropic field will depend on only three scalar degrees of freedom: $k_1$, $k_2$, and $k_3$. 

Returning to the local model, it will now prove useful to split this bispectrum up into a \textit{shape}, and an amplitude. This will illuminate the key underlying features of the model pertaining to how such a bispectrum could best be realised with observational data. Extracting the factor of $(k_1 k_2 k_3)^2$ implicit in scale invariant bispectra of this form, one finds,
\begin{align}
B^{\text{local}}_{\zeta}(k_1, k_2, k_3) = \frac{6}{5} \frac{f^{\text{local}}_{\text{NL}} \mathcal{P}^2_{\zeta}}{(k_1 k_2 k_3)^2} \left(\frac{k_3^2}{k_1 k_2} + \frac{k_2^2}{k_1 k_3} + \frac{k_1^2}{k_2 k_3}\right).
\end{align}
Therefore, a dimensionless shape function for such models is defined naturally as,
\begin{align}
\label{shape}
S(k_1, k_2, k_3) \defeq N (k_1 k_2 k_3)^2 B_{\zeta}(k_1, k_2, k_3),
\end{align} 
where $N$ is a normalisation factor. Thus, the local bispectrum, and in fact all scale invariant bispectra, have a shape defined by  (\ref{shape}) and an amplitude defined by $f_{\text{NL}}$. One therefore wishes to find the three Fourier modes which correspond to a peak of the shape function, thus allowing $f_{\text{NL}}^{\text{model}}$ to most easily be estimated. Clearly, the local shape function can be deduced as\footnote{The normalisation factor is ignored here, as it is somewhat arbitrary (often defined by enforcing $S(k,k,k)=1$) and does not effect the ensuing discussion and analysis.},
\begin{align}
\label{shape1}
S^{\text{local}}(k_1, k_2, k_3) = \frac{k_3^2}{k_1 k_2} + \frac{k_2^2}{k_1 k_3} + \frac{k_1^2}{k_2 k_3}.
\end{align} 
The wavevectors {$\textbf{k}_i$} are constrained to forming a closed triangle in Fourier space. The question now reduces to: which triangle configuration does the shape function in  (\ref{shape1}) peak at? Finding this configuration will allow one to choose the `correct' set of {$k_i$}'s which are most appropriate for measuring and constraining $f_{\text{NL}}^{\text{local}}$. In the case of CMB temperature fluctuations, this choice amounts to picking an appropriate set of multipoles, {$\ell_i$} (which also happen to form a triangle condition in $\ell$-space \citep{multipole}), with which to evaluate the 3-point correlator of $a_{\ell m}$\footnote{This will be discussed in a later section detailing experimental efforts.}. The shape function in  (\ref{shape1}) can be plotted and analysed, but first, it is worth noting some of the nuances that go into the plotting process. It is often the case that shape functions are $k$-scale invariant, and functions of this form can be factorised into two degrees of freedom (which is required for following plotting technique to span the full space of triangle configurations). The third degree of freedom is, of course, set at an arbitrary scale. Conventionally, these remaining two degrees of freedom are chosen to be the ratio of triangle sides: $x_1 = \frac{k_1}{k_2}$ and $x_3 = \frac{k_3}{k_2}$. In this coordinate system, the local shape function is re-expressed as,
\begin{align}
\label{shape2}
S^{\text{local}}(x_1, 1, x_3) = \frac{x_1^2}{x_3} + \frac{1}{x_1 x_3} + \frac{x_3^2}{x_1}.
\end{align}
It will be convenient to now make a separate transformation to the coordinates which lend themselves well to \textit{polar plots}\footnote{Note that most literature uses $x_1$ and $x_3$ as plot axes (assuming scale invariance). Moreover, an advanced technique which can account for possible scale dependence is to plot the full \textit{tetrahedron} in three-dimensional $k$($\ell$)-space \citep{fs222}; where sets of density contours are displayed within the tetrahedron denoting the amplitude of the bispectrum.}. All scale invariant shape functions hereafter will be plotted using the coordinates,
\begin{gather}
\label{coords}
r = \frac{k_1}{k_2} \\ \nonumber
\theta = \cos^{-1} \left( \frac{k_1^2 + k_2^2 - k_3^2}{2k_1 k_2} \right),
\end{gather}
where $\theta$ is seen on an arbitrary triangle in Fig$.$ \ref{fig_tri1} below.
\begin{figure}[h]
\centering
\includegraphics[scale=0.1]{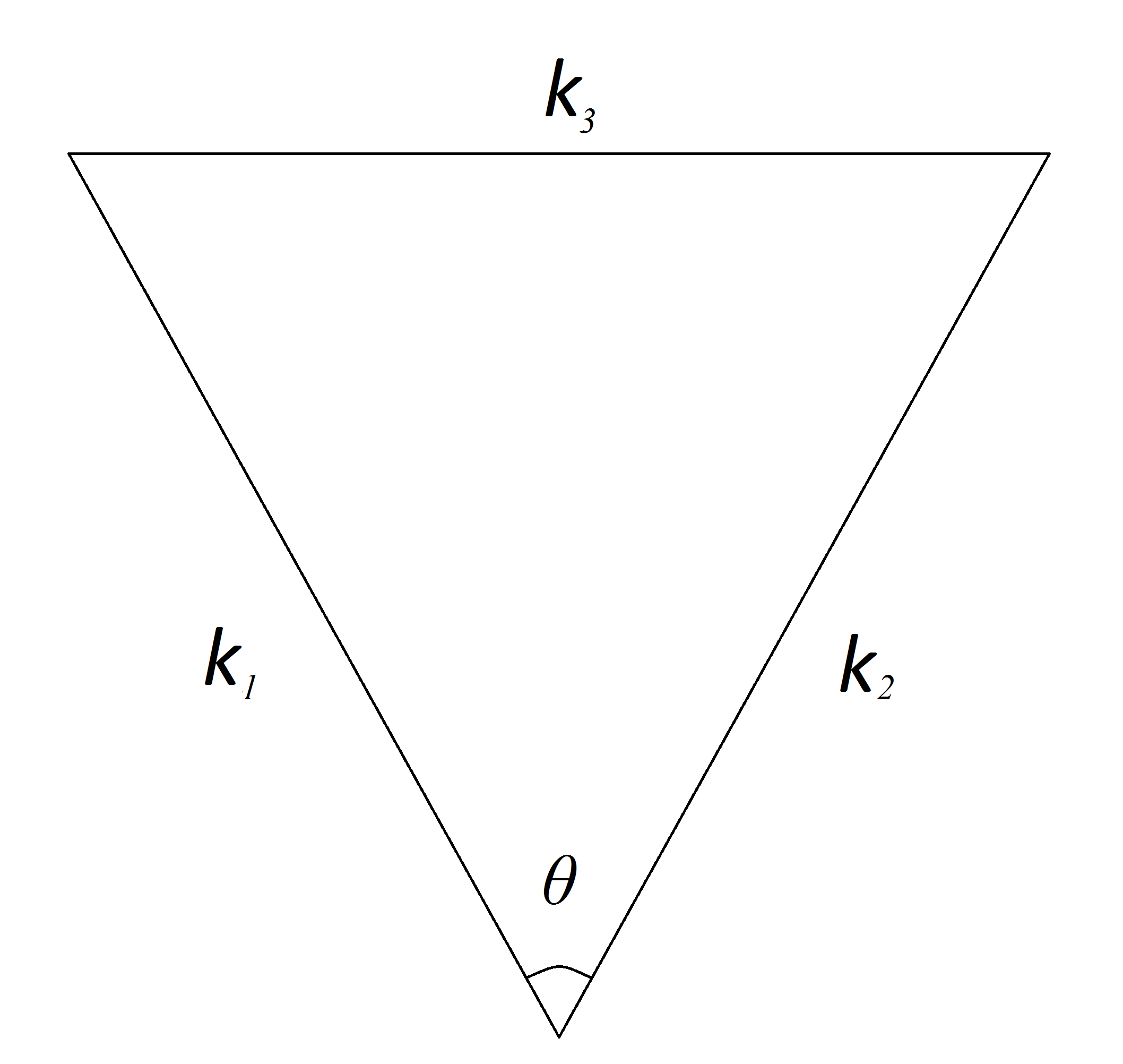}
\caption{An generic triangle shape to demonstrate the definition of $\theta$ and $r = \frac{k_1}{k_2}$.}
\label{fig_tri1}
\end{figure} 
\begin{figure}[h]
\centering
\includegraphics[scale=0.42]{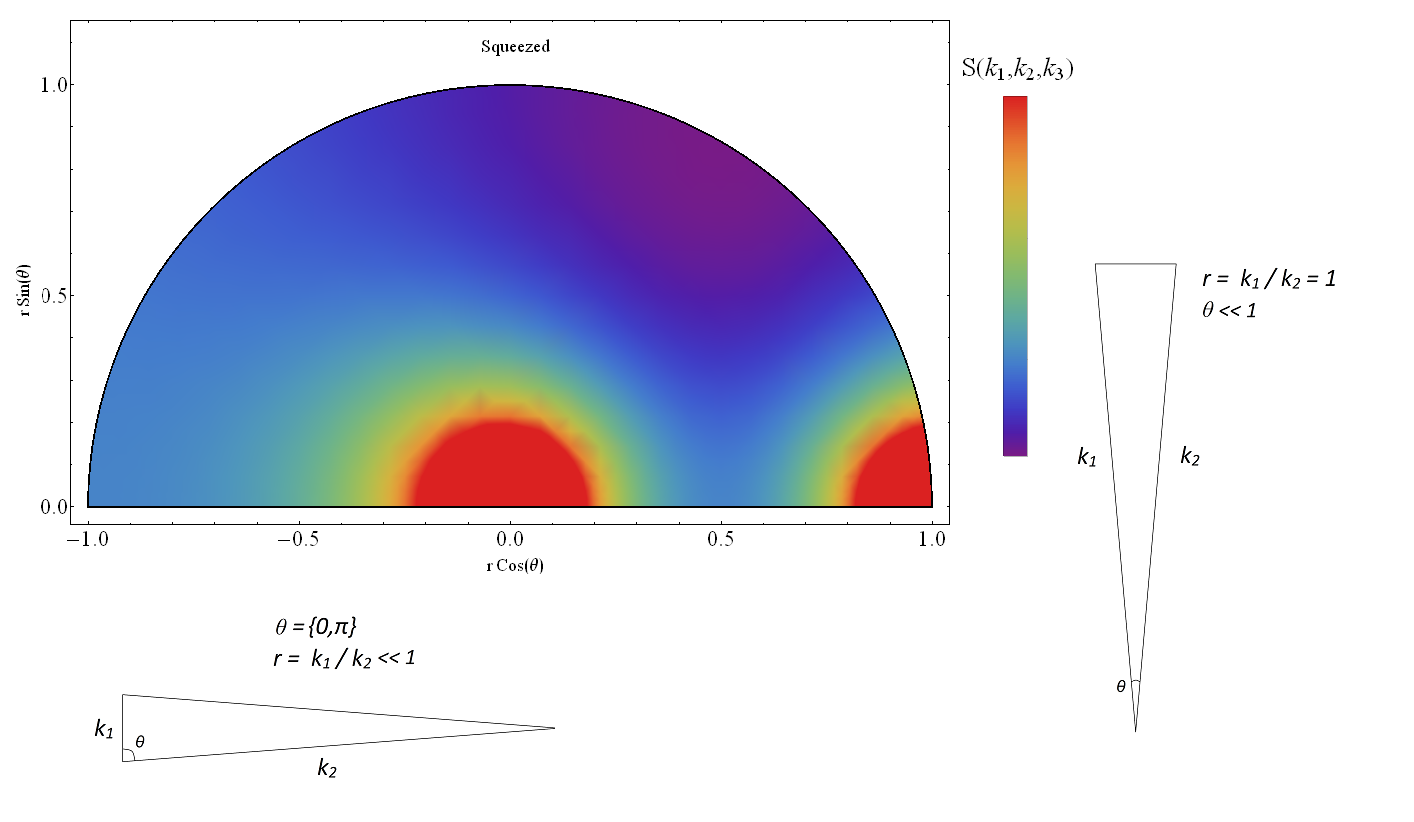}
\caption{A plot of the local shape function in polar coordinates, where the polar radius is $r$, and polar angle is $\theta$, running from 0 to $\pi$. Two peaks exist on this function: 1$.$ at $r \ll 1$  $\forall$ $\theta$ and 2. $\theta \ll 1$, $r=1$. The shapes of the triangles corresponding to these peaks are shown, and are hereafter referred to as squeezed triangles.}
\label{fig_sq1}
\end{figure}
The analytic expressions for $S(r, \theta)$ are often quite cumbersome and will therefore not be displayed. However, these coordinates are convenient for polar plots; where the polar angle is $\theta$ and the polar radius is $r$ - both as defined in  (\ref{coords}). The local shape function in polar coordinates with these definitions can be seen in Fig$.$ \ref{fig_sq1}. The entirety of the space of triangle configurations for $S^{\text{local}}$ is spanned within the ranges $r=( 0,1 )$ and $\theta=(0,\pi)$, due to the equivalence of $r < 1$ and $r > 1$ with respect to the shape of the triangles in these limits. It should also be noted that the function $S^{\text{local}}$ is pathological due to its unphysical behaviour at $r=k_1=0$. Therefore, a maximum `cut-off' value for $S^{\text{local}}$ was enforced to avoid the singularities in Fig$.$ \ref{fig_sq1}. There are two clear peaks in this local shape function, both occurring as $S(r,\theta)$ approaches a pole. These peaks correspond to triangles of the \textit{same shape}, but rotated. Such triangles are called \textit{squeezed} triangles, and are displayed, along with their corresponding conditions for existence, on Fig$.$ \ref{fig_sq1}. The formal definition of a squeezed triangle, which is most often quoted, is $k_1 \ll k_2 \approx k_3$, under an arbitrary ordering of $k_i$ which can be done without loss of generality. Therefore, it has been deduced that the local shape function peaks in the squeezed triangle limit. This is an exceptionally important result, because many models of inflation lead to a bispectrum which peaks in this limit, and thus have their shape functions well approximated by the local template. It will now be convenient to discuss the range of inflationary models for which this is the case, and how non-Gaussianity manifests in these models.

As discussed, it is non-linearity that gives rise to a departure from Gaussian statistics. Therefore, we wish to know how non-linearity is generated in models where one Fourier mode is much smaller than the other two. It is evident from Fig$.$ \ref{fig_zeta} that such long wavelength modes exit the Hubble horizon much before the short wavelength modes. The scalar curvature perturbation, however, freezes on superhorizon scales. Hence, non-Gaussianity will be generated by subhorizon modes, $\zeta_{\textbf{k}_2}$ and $\zeta_{\textbf{k}_3}$, dynamically evolving in the background of a single, frozen superhorizon mode, $\zeta_{\textbf{k}_1}$\footnote{The arbitrary ordering $k_1 \leq k_2 \leq k_3$ has been used here.}. This frozen mode will act as a perturbation of the background, thus altering the time of horizon crossing of the subhorizon modes. Such behaviour is present in simple, single field, slow roll inflation models. The seminal contribution by Juan Maldacenea, which will be detailed in Section \ref{sec_3}, was to explicitly calculate that no detectable non-Gaussianity exists in such a model of inflation \citep{mald}. In fact, this result was later shown to be a subset of a more general consistency relation that proves \textit{all} single field inflation models have a bispectrum which is suppressed by $(n_s-1)$ in the squeezed limit. That is to say, all models in which a single inflaton field acts as a `clock' counting toward the end of inflation, irrespective of the inflationary dynamics, is `slow roll suppressed' - assuming $n_s - 1 \approx \mathcal{O}(\epsilon)$ \citep{Baumann}. This theorem is attributed to Paolo Creminelli and Matias Zaldarriaga in their short 2004 letter, \textit{Single field consistency relation for the 3-point function}, which involves a relatively compact proof (see Ref$.$ \citep{allsf}). For a more detailed treatment, highlighting where assumptions, or lack thereof, have been made, the reader is referred to Ref$.$ \citep{cheung} by C. Cheung \textit{et al}. Heuristically, this result is due to the frozen superhorizon mode acting as a local rescaling of the background spatial coordinates. Given the ADM metric defined in  (\ref{ADM}), a gauge can be chosen in which matter is unperturbed and
\begin{align}
^{(3)}g_{ij} = a(t)^2 e^{2\zeta}\delta_{ij}.
\end{align}
On superhorizon scales, the lapse and shift become $N=1$ and $N^i=0$ respectively. Thus, the single scalar fluctuation reduces to a local rescaling of coordinates $x' = e^{\zeta(x)}x$,
\begin{align}
ds^2 = -dt^2 + a(t)^2 \delta_{ij} dx'^i dx'^j,
\end{align}
which is the unperturbed FRW background. Using this fact, the 3-point correlator can be computed in a two step process. Firstly, the 2-point correlator of $\zeta_{\textbf{k}_{2,3}}$ can be calculated in the presence of the slowly varying background mode - $\langle \zeta_{\textbf{k}_2} \zeta_{\textbf{k}_3} \rangle_{\zeta_{\textbf{k}_1}}$. Converting to real space, one finds the variation scale of the background, $\zeta_{\textbf{x}_3}$, is much larger than the real space separation of the large Fourier modes, $|\textbf{x}_2 - \textbf{x}_3|$. Thus, a Taylor expansion can be done in powers of the smaller background mode\footnote{One would not necessarily be able to perform this expansion if there were more than a single field driving inflation.}. To first order in the expansion variable, the resultant correlator includes the vacuum 2-point correlator, and a term including the log derivative of a small wavelength mode, $k_2$ or $k_3$. Finally, the 3-point correlator is obtained by re-correlating $\langle \zeta_{\textbf{k}_2} \zeta_{\textbf{k}_3} \rangle_{\zeta_{\textbf{k}_1}}$ with the superhorizon background mode, $\zeta_{\textbf{k}_3}$, to produce,
\begin{align}
\lim_{\textbf{k}_1 \rightarrow 0} \langle \zeta_{\textbf{k}_1}\zeta_{\textbf{k}_2}\zeta_{\textbf{k}_3} \rangle = \langle \zeta_{\textbf{k}_1} \langle \zeta_{\textbf{k}_2} \zeta_{\textbf{k}_3} \rangle_{\zeta_{\textbf{k}_1}} \rangle = (2\pi)^3 \delta(\textbf{k}_1 + \textbf{k}_2 + \textbf{k}_3) P(k_1)P(k_2) \frac{d \text{ln}(\mathcal{P}(k_2))}{d \text{ln}
(k_2)}.
\end{align}
Note that the rightmost derivative in this expression is the formal definition of the scalar spectral index,
\begin{align}
n_s \defeq \frac{d \text{ln}(\mathcal{P}(k_2))}{d \text{ln} (k_2)} + 1,
\end{align}
which has been measured to be approximately unity. Therefore, \textit{all} single field inflation models have a bispectrum which is slow roll suppressed in the squeezed limit\footnote{Technically, the non-Gaussianity is suppressed by the spectral tilt, however, in single field, slow roll inflation models $n_s - 1 \approx \mathcal{O}(\epsilon)$ \citep{mald}.}. This means the magnitude of the non-Gaussianity produced from such models would be $f_{\text{NL}} \approx \mathcal{O}(n_s - 1) \approx  \mathcal{O}(\epsilon)$ - well outside of experimental limits. For reference, \textit{Planck} 2015 is probing $f_{\text{NL}} \approx \mathcal{O}(1)$; where $f_{\text{NL}}$ here is appropriately normalised, often with respect to the power spectrum, for comparative purposes. Any detection of non-Gaussianity in the squeezed limit would therefore favour multifield models of inflation. 

Multifield models of inflation are not strictly limited to the squeezed limit, but much of the literature regarding these models predict a shape function comparable to the local one. The reason for this is, within such models, non-linearity is generated on superhorizon scales. Specifically, non-linear mechanisms are introduced via the transition of isocurvature perturbations (in the additional degrees of freedom) to adiabatic ones - i.e. total energy density perturbations, which we have only considered thus far. Isocurvature perturbations are generated by a causal connection between matter species, in which a stress force is exerted between them. Such perturbations have no net effect on the perturbed geometry, and it is the transition of these perturbations to curvature ones that induces non-Gaussianity in many multifield models \citep{PlankNG}. The amount of non-Gaussianity predicted by these models is typically $f^{\text{local}}_{\text{NL}} \approx \mathcal{O}(1)$, which is within feasible experimental bounds. Thus, multifield inflation forms an exciting prospect, on the cusp of potential detection. To name a specific example within the multifield inflationary regime, models in which the additional degree of freedom is the \textit{curvaton} have been the subject of much research \citep{curv1,curv2,curv3}. For further information on multifield models, the reader is referred to the review in Ref$.$ \citep{multi} by C. Byrnes \& K. Choi, and references therein. For completeness, it is worth mentioning that there also exist more exotic inflationary models by which an appreciable amount of non-Gaussianity is generated in the squeezed limit. Examples of these include: p-adic inflation, a non-local model based on string theory \citep{padic}; and ekpyrotic models, some of which are now strongly disfavoured due to their prediction of $f^{\text{local}}_{\text{NL}} \approx \mathcal{O}(100)$ \citep{ekp}.

\begin{figure}[h]
\centering
\includegraphics[scale=0.42]{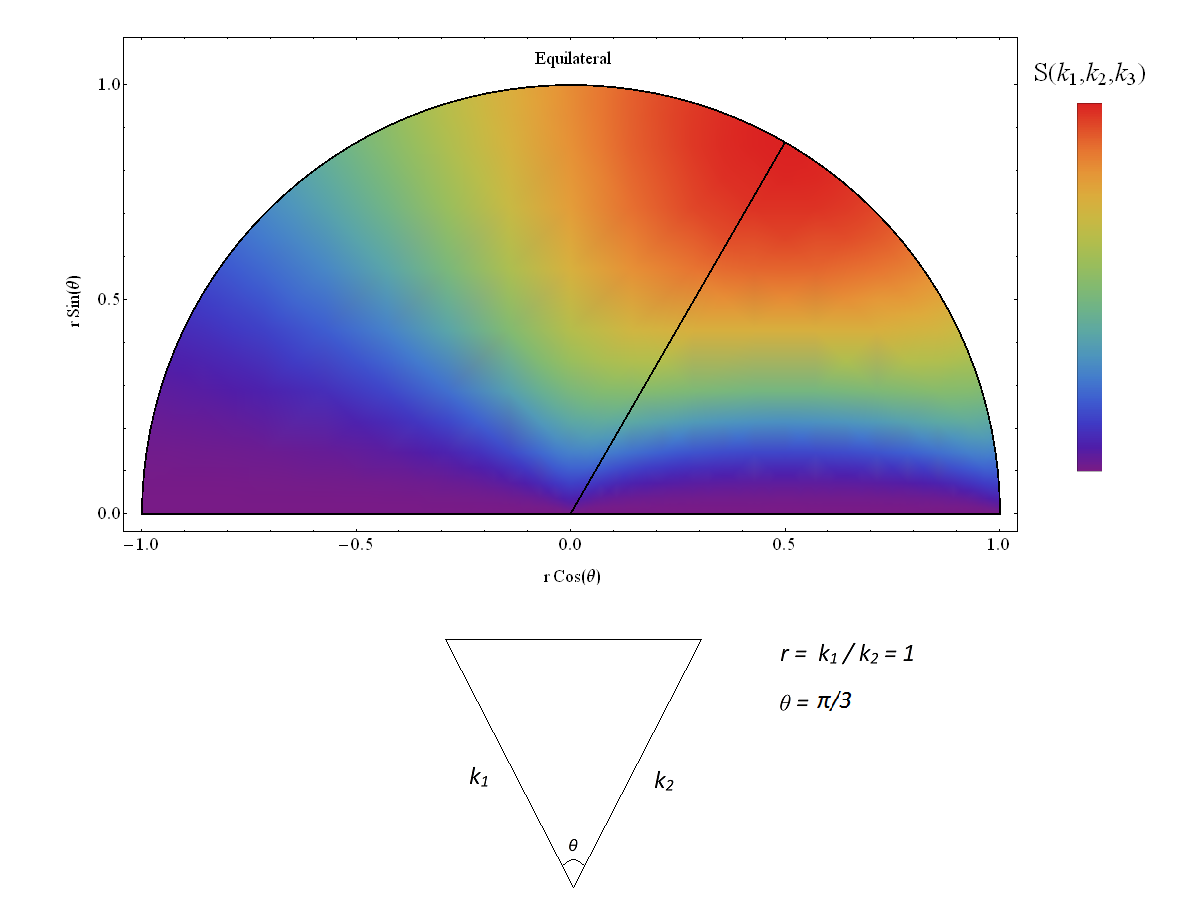}
\caption{A polar plot of the equilateral shape function. This function reveals a clear peak at $r \approx 1$ and $\theta \approx \frac{\pi}{3}$, corresponding to an equilateral triangle, which is shown below the heat map.}
\label{fig_eq}
\end{figure}

A second, perhaps equally important, shape of non-Gaussianity is one in which the shape function peaks for \textit{equilateral} triangle configurations of Fourier modes. The template for such a shape function is not derived, but is instead phenomenologically chosen as,
\begin{align}
S^{\text{equil}}(k_1, k_2, k_3) = \frac{(k_1 + k_2 - k_3)(k_1 - k_2 + k_3)(-k_1 + k_2 + k_3)}{k_1 k_2 k_3}.
\end{align}
It is not immediately obvious that this function peaks in the equilateral limit. However, plotting the function using polar coordinates defined in  (\ref{coords}) reveals this to be the case (Fig$.$ \ref{fig_eq}).
The equilateral condition, $k_1 \approx k_2 \approx k_3$, converts to $r \approx 1$ and $\theta \approx \frac{\pi}{3}$ in these coordinates. A line at $\theta \approx \frac{\pi}{3}$ is shown on Fig$.$ \ref{fig_eq}, thus clearly displaying a peak for equilateral triangle configurations.

By definition, inflation models which admit a shape function well approximated by the equilateral template generate non-Gaussianity on subhorizon scales. This is because \textit{all} Fourier modes in such models are approximately equal. Therefore, as a result of $\zeta$ being frozen in the superhorizon limit, non-Gaussianity must be generated on subhorizon scales - in contrast to models well approximated by the local template. More specifically, the bispectrum becomes suppressed when an individual mode is considerably outside the horizon. Hence, non-Gaussianity is expected to be at its largest when all three Fourier wavelengths are approximately the size of the Hubble horizon. This behaviour is most commonly displayed by so-called \textit{higher derivative} theories \citep{Baumann}. That is to say, models in which the kinetic term in the inflationary action is not canonical, but instead takes a more general form,
\begin{align}
\label{PXPHI}
S = \int d^4x \sqrt{-g} \left[ \frac{M_{pl}^2}{2}\mathcal{R} - P(X,\phi) \right],
\end{align} 
where $P(X,\phi)$ is an arbitrary function of the canonical kinetic term, $X \defeq \frac{1}{2} (\partial \phi)^2$. The slow roll limit is therefore recovered with $P(X,\phi) = X - V(\phi)$. Computing a third order action in such theories reveals that the size of the non-Gaussianity is controlled by the speed of propagation of the inflaton fluctuations, or the \textit{sound speed}. The sound speed, in general, appears as a factor of $\frac{1}{c_s^2}$ in the action, and is defined as,
\begin{align}
c_s^2 \defeq \frac{\partial_X P}{ \partial_X P + 2X \partial_X \partial_X P}.
\end{align}
Clearly, in the slow roll limit $c_s^2 =1$, and the third order action remains slow roll suppressed\footnote{This slow roll suppression of the cubic action will be explicitly shown in Section \ref{sec_3}.} by a factor of $\epsilon^2$. However, if one could construct a physically well motivated theory in which the kinetic term allows $c_s^2 \ll 1$; the factor of $\frac{1}{c_s^2}$ could effectively override this slow roll suppression, and produce an appreciable amplitude of non-Gaussianity. In fact, there is \textit{no} slow roll suppression for leading order terms in the non-canonical bispectrum when $c_s^2 \ll 1$ due to the introduction of $P(X, \phi)$ \citep{chen2}. This result will be explored in further detail in Section \ref{sec_4}. There exist multiple such models which fulfil this criterion. One example is Dirac-Born-Infeld (DBI) inflation \citep{DBI}. This model is string theoretic in nature, and its brane dynamics are governed by the DBI matter action,
\begin{align}
S_m = -T \int d^4x \sqrt{1- (\partial_{\mu}r)^2},
\end{align}
from which the non-canonical kinetic terms can be deduced. The shape function has been calculated in this regime, and is very well approximated by the equilateral template. Therefore, the amplitude of non-Gaussianity in such theories is constrained by \textit{Planck} to be $f_{\text{NL}}^{\text{equil}} = -4 \pm 43$ ($68\%$ CL), which places a lower limit on the DBI sound speed of $c_s > 0.087$ ($95\%$ CL) \citep{PlankNG}. Other ways in which such a sound speed has been achieved with non-canonical kinetic terms include: \textit{K-inflation}, whereby higher derivative terms in $X$ are explicitly added to $P(X,\phi)$ \citep{kinf}; and \textit{ghost inflation}, whereby the inflationary period is driven by a ghost scalar field \citep{ghost}.

Finally, the last phenomenologically relevant shape template which \textit{Planck} explicitly constrains is called the \textit{orthogonal} shape \citep{senatore}. This shape function is named as such because the \textit{correlation} (see Section \ref{obsss}) between the orthogonal template, and local and equilateral templates, is low (hence `orthogonal'). The functional form of this shape is given by, 
\begin{align}
\label{orthooo}
S^{\text{ortho}}(k_1, k_2, k_3) = -6\left( \frac{k_1^2}{k_2 k_3} +  \frac{k_2^2}{k_1 k_3} +  \frac{k_3^2}{k_1 k_2} \right) + 6 \left( \frac{k_1}{k_2} + \frac{k_1}{k_3} + \frac{k_2}{k_1} + \frac{k_2}{k_3} + \frac{k_3}{k_1} + \frac{k_3}{k_2}\right) - 18.
\end{align}  
which has a \textit{negative} peak for \textit{flattened} triangle configurations, defined as $k_3 \approx k_1 + k_2$. The flattened shape function is therefore $S^{\text{flat}} =  (S^{\text{equil}}-S^{\text{ortho}})/2$, which can be seen depicted in Fig$.$ \ref{fig_flat} below. There is a slight nuance to flattened triangles in polar coordinates, namely, that substitution of $k_3 \approx k_1 + k_2$ into the polar angle formula (\ref{coords}) yields peaks \textit{independent} of $r=\frac{k_1}{k_2}$, which appear at angles $\theta \approx \arccos(\pm 1) = 0,\pi$ (i.e. the behaviour in Fig$.$ \ref{fig_flat}).
\begin{figure}[h]
\centering
\includegraphics[scale=0.42]{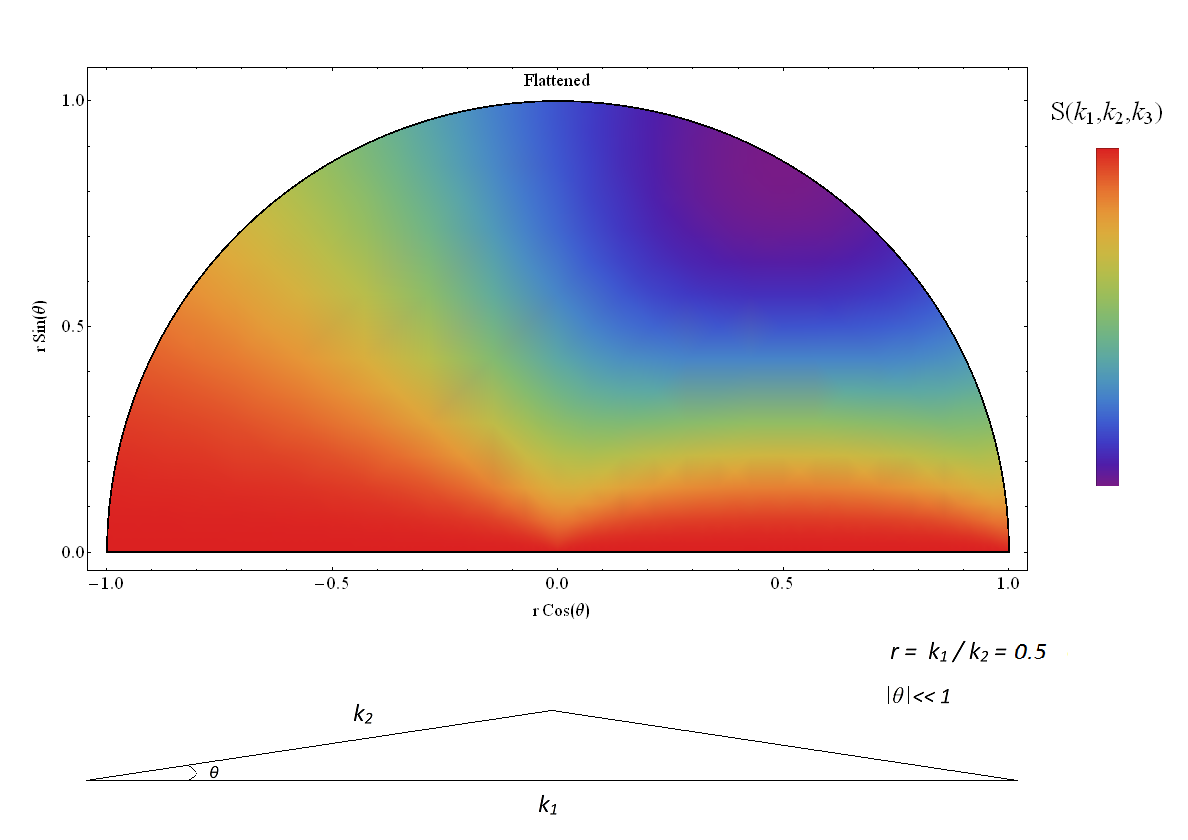}
\caption{A polar plot of the flattened shape function. This function reveals peaks along the regions $\theta \ll 1 \text{ } \forall \text{ } r$, and $\theta \approx \pi \text{ } \forall \text{ } r$, overlapping with flattened triangle configurations, an example of which is shown below the heat map.}
\label{fig_flat}
\end{figure}
This template is particularly relevant to the following work, because such behaviour manifests in models where the initial inflation vacuum is not Bunch-Davies \citep{PlankNG}. The Bunch-Davis vacuum is chosen by noticing that all cosmological modes are deep inside the Hubble horizon at asymptotic past infinity, and thus their behaviour is determined by Minkowski initial conditions - which will be detailed shortly. However, if a mechanism existed by which the Universe was in an excited initial state, a significant amount of non-Gaussianity could be generated. If this were the case, non-Gaussianity of this kind could provide information regarding the trans-Planckian physics at play during the Big Bang, and will be discussed in more detail in Section \ref{sec_4}.

The three bispectrum templates defined above were quoted without reference to normalisation. It is important, however, to have a consistent scheme of normalisation in order to properly define and compare the amplitude parameters, $f^{\text{model}}_{\text{NL}}$. To this end, an additional shape function is worth mentioning - the \textit{constant} model,
\begin{align}
S^{\text{constant}}(k_1, k_2, k_3) = 1.
\end{align}
This model defines a function which peaks equally for all triangle configurations. The usefulness of such a function lies within its simplicity, which allows for analytic calculations of the large angle reduced CMB bispectrum \citep{shapesferg}. Therefore, it is appropriate to define the constant shape as a standard by which all shape functions (such as the above) are normalised against so that the definitions of $f^{\text{model}}_{\text{NL}}$ can become regularised. There also exist separate, physical motivations for the investigation of the constant model, which are detailed in Ref$.$ \citep{bigreview}. 

To summarise; this section began by arguing that standard, slow roll inflation produces a negligible non-Gaussian signature with respect to current experimental limits\footnote{The calculation for which will be detailed in Section \ref{sec_3}.}. A plethora of models then followed, which broke one, or more, of the assumptions that underlaid this calculation. These assumptions can be collected into a `no-go' theorem, which states that, if any of the following conditions are violated, an appreciable non-Gaussian signature can be generated:

$\bullet$ Initial inflationary vacuum is Bunch-Davis

$\bullet$ A single-field driving inflation

$\bullet$ Slow roll conditions 

$\bullet$ A canonical kinetic term

$\bullet$ Einsteinian gravity

Selected models which break one, or more, of these conditions were then classified into groups, separated by the limit in which their bispectra peak: ranging from squeezed, to equilateral, to flattened triangle configurations of Fourier modes in $k$-space\footnote{There do exist triangle configurations in between these, such as isosceles triangles, $k_1 > k_2 = k_3$, but they will not be important for the remainder of this work.}. The next section will motivate why only three shape functions are considered observationally, and how general models can correlate their predictions with the constrains that \textit{Planck} derives for these three templates. 

\subsection{Observational Considerations}
\label{obsss}

The statistical nature of fluctuations from inflation are imprinted primarily on two cosmological observables: LSS and the CMB. Galactic surveys have been carried out with great accuracy \citep{des}. However, the matter distribution bispectrum bears a non-trivial relation to primordial fluctuations, because the formation of structure is inherently non-linear. Thus, to constrain predictions from inflation requires large $N$-body simulations to separate the gravitational contribution from the primordially sourced contribution in the matter bispectrum \citep{bigreview}. Temperature fluctuations in the CMB are, however, linearly related to primordial fluctuations. Therefore, observations of the CMB have provided some of the best constraints on inflation yet with the release of measurements from the \textit{Planck} 2015 experiment \citep{plank2015}.

Experiments such as \textit{Planck} constrain non-Gaussianity by determining limits for the amplitude parameter, $f_{\text{NL}}$. Ideally, one would do this by measuring the CMB bispectrum for individual multipoles $\ell$, and fitting the aforementioned shape templates. However, the signal to noise ratio within the data is too low to adopt this approach. Therefore, a statistical \textit{estimator} must be used, which finds the best fit value of $f_{\text{NL}}$ for a given bispectrum template by averaging over \textit{all} multipoles. This is done by first obtaining the temperature fluctuation bispectrum as the 3-point correlator of CMB multipoles,
\begin{align}
\langle a_{\ell_1 m_1} a_{\ell_2 m_2} a_{\ell_2 m_2} \rangle = \mathcal{G}^{\ell_1 \ell_2 \ell_3}_{m_1 m_2 m_3} b_{\ell_1 \ell_2 \ell_3},
\end{align}
where $\mathcal{G}$ is a known factor called the \textit{Gaunt integral} \citep{bigreview}, and $b$ is the reduced bispectrum. The multipoles, $a_{\ell m}$, are introduced as coefficients of the 2D projection of the CMB temperature fluctuations in the spherical harmonic basis, 
\begin{align}
\label{wowowo}
\frac{\Delta T}{T}(\hat{\textbf{n}}) = \sum_{\ell m} a_{\ell m} Y_{\ell m} (\hat{\textbf{n}}),
\end{align}
and are related to primordial fluctuations, $\zeta$, by
\begin{align}
a_{\ell m} = 4\pi i^{\ell} \int \frac{d^3k}{(2\pi)^3} \zeta(\textbf{k},t) \Delta_{\ell}(\textbf{k})Y^*_{\ell m}(\hat{\textbf{k}}),
\end{align}
where $\Delta_{\ell}(\textbf{k})$ are the previously mentioned radiation transfer functions. It can be shown that, for a given shape template, the so-called \textit{optimal estimator}\footnote{An estimator which gives the best possible constraints on the magnitude of $f_{\text{NL}}$. Optimal, in this case, means the estimator produces the smallest variance.} can be written as \citep{estimator},
\begin{align}
\label{estimator}
\mathcal{E} (a) = \frac{1}{N} \sum_{\ell_i m_i}& \left( \langle a_{\ell_1 m_1} a_{\ell_2 m_2} a_{\ell_2 m_2} \rangle C^{-1}_{\ell_1 m_1,\ell_4 m_4} C^{-1}_{\ell_2 m_2,\ell_5 m_5} C^{-1}_{\ell_3 m_3,\ell_6 m_6} a_{\ell_4 m_4} a_{\ell_5 m_5} a_{\ell_6 m_6} \right. \\ \nonumber &- 3 \left. \langle a_{\ell_1 m_1} a_{\ell_2 m_2} a_{\ell_2 m_2} \rangle C^{-1}_{\ell_1 m_1,\ell_2 m_2} C^{-1}_{\ell_3 m_3,\ell_4 m_4} \right),
\end{align}
where $C_{\ell m}$ is the covariance matrix, and $N$ the normalisation factor,
\begin{align}
N = \sum_{\ell_i m_i} \langle a_{\ell_1 m_1} a_{\ell_2 m_2} a_{\ell_2 m_2} \rangle   C^{-1}_{\ell_1 m_1,\ell_4 m_4} C^{-1}_{\ell_2 m_2,\ell_5 m_5} C^{-1}_{\ell_3 m_3,\ell_6 m_6} \langle a_{\ell_4 m_4} a_{\ell_5 m_5} a_{\ell_6 m_6} \rangle,
\end{align}
which is derived using statistical estimation theory. The expression in  (\ref{estimator}) contains a multidimensional integral over the bispectrum shape template, $S$, which is computationally infeasible in the general case unless simplifying assumptions are made. One such method of rendering this expression computationally tractable is to have a separable shape function,
\begin{align}
\label{sep}
S(k_1, k_2, k_3) = X(k_1)Y(k_2)Z(k_3) + 5\text{ perms.}.
\end{align} 
If the shape function is separable in this manner, calculating the reduced bispectrum, $b$, becomes a matter of evaluating three one-dimensional integrals. Hence, the computational cost of evaluating the full bispectrum estimator is drastically reduced. This is, amongst other complications, the reason why \textit{Planck} only constrains the three aforementioned shape templates. The separability condition,  (\ref{sep}), turns out to be very restrictive. Therefore, work has been done in attempting to arrive at a computationally feasible method for calculating bispectrum estimators with non-separable shape functions (see Ref$.$ \citep{shapesferg} for further details).

Using these techniques, the \textit{Planck} 2015 constraints on non-Gaussianity with both temperature and polarisation data can be summarised as,
\begin{align*}
f_{\text{NL}}^{\text{local}} = 0.8 \pm 5.0, \\
f_{\text{NL}}^{\text{equil}} = -4 \pm 43.0, \\
f_{\text{NL}}^{\text{ortho}} = -26 \pm 21,
\end{align*}
at $68 \%$ CL, for the bispectrum shape templates defined in Section \ref{sec_223}. Thus, no official detection of non-Gaussianity has been made at an appropriate degree of statistical significance. However, these new bounds are much improved over its predecessors, and can therefore rule out many of (or constrain parameters within) the myriad of inflationary models that exist. The implications of the \textit{Planck} 2015 data on a selected class of models - non-Bunch-Davis \textit{and} non-canonical - will be outlined in Section \ref{sec_4}.

In general, inflationary models \textit{do not} possess a shape function of the form of these three templates. Therefore, in order to draw conclusions about how well constrained an arbitrary shape function is, a systematic, quantitative measure of the difference between two shape functions must be defined. This is done with a \textit{bispectrum shape correlator} of the form,
\begin{align}
F[S, S'] = \int_{\mathcal{V}_{k}} S(k_1, k_2, k_3) S'(k_1, k_2, k_3) \omega (k_1, k_2, k_3) d\mathcal{V}_{k},
\end{align}
where $\omega$ is a weight function, which is appropriately approximated as \citep{bigreview},
\begin{align}
\omega (k_1, k_2, k_3) = \frac{1}{k_1 + k_2 + k_3},
\end{align}
and $\mathcal{V}_{k}$ is the domain of all triangle configurations up to a cut-off (which formally takes the shape of a tetrahedron). Thus, the normalised bispectrum shape correlator takes the form,
\begin{align}
C[S, S'] = \frac{F[S, S']}{\sqrt{F(S, S) F(S', S')}}.
\end{align}
Correlations between typical inflationary models, and the three shape templates that \textit{Planck} constrains, can be now be calculated, and are displayed in Table$.$ \ref{TABLE1} \citep{shapesferg}.

\begin{table}[h]
\centering
\setlength\belowcaptionskip{3pt}
\caption{Normalised fractional correlations, $C[S, S']$, between selected primordial shape functions. Data obtained from Ref$.$ \citep{shapesferg}.}
\begin{tabular}{ c  c  c  c  c  c  c  c }
  \hline
  \hline                       
  $ $ &\vline & Local & Equilateral & Flattened & DBI & Ghost & Single Field \\
  \hline 
  Local &\vline & 1.00 & 0.46 & 0.62 & 0.5 & 0.37 & 1.00 \\
  \hline 
  Equilateral &\vline & 0.46 & 1.00 & 0.30 & 0.99 & 0.98 & 0.46 \\
  \hline 
  Flattened &\vline & 0.62 & 0.30 & 1.00 & 0.39 & 0.15 & 0.62 \\
  \hline 
  \hline 
\end{tabular}
\label{TABLE1}
\end{table}
It can be seen that, to three significant figures, the local template is maximally correlated with the shape function from single field, slow roll inflation - which will be derived in Section \ref{sec_3}. Therefore, the \textit{Planck} constraints on the local template are directly applicable to single field, slow roll inflation. Moreover, as posited in this previous section, DBI and ghost inflation are both very well approximated by the equilateral template. In fact, the equilateral template was chosen as a separable ansatz for DBI inflation \citep{bigreview}. 

This section has therefore motivated, and explained, the use of three separable shape templates by \textit{Planck} and related experiments. A systematic technique was then stated which allows one to correlate any general shape function to these template shapes. Thus, one can obtain reasonable constraints for a given model of inflation based on the \textit{Planck} 2015 data (following, for example, the methodology of Ref$.$ \citep{fs222}). All subsequent sections will now be concerned with how one can \textit{calculate} such a bispectrum, given an inflationary model.
 

\chapter{The `\textit{in-in}' Formalism} 

\label{sec_23} 

\lhead{3. \emph{The `\textit{in-in}' Formalism}} 

The `\textit{in-in}' formalism, developed primarily by Maldacena \citep{mald} and Weinberg \citep{wein1} in a cosmological context, is a technique to calculate correlation functions of \textit{interacting} quantised fields. Hence, in this regime, one can compute the non-Gaussianity produced in primordial times as a result of zero-point fluctuations,
\begin{align}
\label{exp}
\langle Q(t) \rangle = \langle \textit{in} | Q(t) | \textit{in} \rangle,
\end{align}
where the observable $Q(t)$, in our case, will be the 3-point function of curvature perturbations, $\zeta \zeta \zeta$, at the end of inflation, $t$. Moreover, $|\textit{in}\rangle$ is the initial, time dependent interacting vacuum at a time far into the past, $t_i < t$. Crucially, at such a time, interactions are turned off, and the $|\textit{in}\rangle$ state reduces to the vacuum state of the non-interacting theory - often chosen to be the Bunch-Davis vacuum. Both states in (\ref{exp}) are `\textit{in}' states, therefore, this expression reduces to an average, or expectation value of the observable $Q$. It is not immediately obvious that it is appropriate to equate this quantum expectation value to the classical statistical expectation values detailed above. A case is made for why this is possible in Ref$.$ \citep{lim}, and involves noticing that the perturbation fields commute on superhorizon scales\footnote{Specifically, $[\dot{\zeta},\zeta] \rightarrow 0$ exponentially fast. Thus, superhorizon scales are said to be the classical limit of $\hat{\zeta}$.}; hence the mode functions of the non-interacting theory are identified with the classical variance in the statistical 2-point correlator. As an aside, because we are obtaining an expectation value and \textit{not} scattering amplitudes, no reference need be made to a final state\footnote{Where, in our case, a final state does not exist.}. This is in contrast with the `\textit{in-out}' formalism of standard quantum field theory applied to particle physics, where both `\textit{in}' and `\textit{out}' states are defined at asymptotic past and future times respectively. Without further ado, the key results of the `\textit{in-in}' formalism can be detailed, which will allow  (\ref{exp}) to be evaluated.

\section{Non-Interacting Theory}

\label{sec_331}

Firstly, it can be seen that the expectation value in  (\ref{exp}) requires a non-trivial evolution in time of $Q(t)$ back to $Q(t_i)$\footnote{The time-dependence in the operator $Q$ is introduced by the time-dependence of the scalar perturbation during inflation, $\delta \phi (t)$ or $\zeta (t)$.} - when the $|\textit{in} \rangle$ states are defined. Naively, this would be done using the interaction (third order and above) Hamiltonian, which involves complicated non-linear equations of motion. However, it is possible to avoid this by working in the \textit{interaction picture} of quantum mechanics. In this regime, the interaction picture fields, $\zeta^I$, have their time evolution determined by the non-interacting (up to second order) Hamiltonian. Interactions are then introduced perturbatively in powers of $H_{\text{int}}$ as correction terms. Hence, we will begin with a brief review of the non-interacting theory, which will allow us to define the relevant mode functions, and the Bunch-Davis vacuum. The second order action in the comoving gauge, following the process detailed in Section \ref{subsec_pert}, is,
\begin{align}
\label{2ndo}
S = \frac{1}{2} \int dt d^3x \text{ } a^3 \frac{\dot{\phi}^2}{H^2} \left[\dot{\zeta}^2 - \frac{1}{a^2} (\partial_i \zeta)^2 \right].
\end{align}
It will now be convenient to define the \textit{Mukhanov variable},
\begin{align}
v \defeq z \zeta,
\end{align}
where $z^2 = 2a^2 \epsilon$, and $\epsilon = \dot{\phi}^2/2H$. Thus, switching to conformal time, the second order action is re-expressed as,
\begin{align}
S = \frac{1}{2} \int d\tau d^3x \text{ } \left[ (v')^2 - (\partial_i v)^2 + \frac{z''}{z} v^2 \right],
\end{align}
where $'$ denotes a derivative with respect to conformal time, $d \tau = dt/a$. Upon variation, this action yields the classical Fourier space equation of motion,
\begin{align}
\label{eom1}
v_{k}'' + \left( k^2 - \frac{z''}{z} \right) v_{k} \defeq v_k'' + \omega^2_k(\tau) v_k = 0,
\end{align} 
which is called the \textit{Mukhanov-Sasaki equation}. Following the standard quantisation procedure, the field, $v$, and its momentum conjugate, $\pi$, are promoted to Heisenberg picture operators satisfying the equal-time commutation relations (ETCRs),
\begin{align}
[\hat{v}_k(\tau), \hat{\pi}_{k'}(\tau)] = i \delta(\textbf{k} + \textbf{k}').
\end{align}
Hence, a mode expansion of the quantised perturbation field can be made,
\begin{align}
\hat{v}_k(\tau) = v_k(\tau) \hat{a}_{\textbf{k}} + v^*_k(\tau) \hat{a}^{\dagger}_{\textbf{k}},
\end{align}
where $v_k$ and $v^*_k$ are the classical time-dependent mode functions separately satisfying the equation of motion in  (\ref{eom1}). Moreover, $\hat{a}^{\dagger}$ and $\hat{a}$ are creation and annihilation operators which are used to construct the Hilbert space with the (soon-to-be Bunch-Davis) vacuum $|0 \rangle$ satisfying,
\begin{align}
\label{annih}
\hat{a}_{\textbf{k}} |0 \rangle = 0.
\end{align}
Therefore, the problem now becomes \textit{fixing} these (appropriately normalised) mode functions, which will provide a unique definition of the vacuum state. The non-uniqueness of the vacuum defined in  (\ref{annih}) is best illustrated by noticing that, because $v_k$ and $v^*_k$ separately satisfy the Mukhanov-Sasaki equation, so do an arbitrary linear combination of these solutions,
\begin{align}
u_k(\tau) = \alpha_k v_k(\tau) + \beta_k v_k^* (\tau).
\end{align}
A separate but equally valid mode expansion can then be made,
\begin{align}
\hat{v}_k(\tau) = u_k(\tau) \hat{b}_{\textbf{k}} + u^*_k(\tau) \hat{b}^{\dagger}_{\textbf{k}},
\end{align}
meaning $\hat{a}$ and $\hat{b}$ are related by the coefficients $\alpha_k$ and $\beta_k$ via a \textit{Bogolyubov transformation} \citep{alph1}. Clearly, $\hat{b}_{\textbf{k}}$ will \textit{not} annihilate the vacuum state defined in  (\ref{annih}), but will instead satisfy,
\begin{align}
\hat{b}_{\textbf{k}} |0 \rangle_b = 0,
\end{align}
where $|0 \rangle \neq |0 \rangle_b$ - i.e. the vacuum state is not unique. Additional physical input is therefore required to determine a preferred set of mode functions to fix the vacuum, which is done as follows: firstly, the general solution to the Mukhanov-Sasaki equation is,
\begin{align}
v_k(\tau) = \alpha_k \frac{e^{-ik\tau}}{\sqrt{2k}} \left(1 - \frac{i}{k \tau} \right) + \beta_k \frac{e^{ik\tau}}{\sqrt{2k}} \left(1 + \frac{i}{k \tau} \right),
\end{align}
where de Sitter spacetime dynamics have been assumed\footnote{Note that, if de Sitter spacetime dynamics are not assumed, the general solution of the  Mukhanov-Sasaki equation involves \textit{Hankel functions}.}, $\omega^2_k(\tau) \approx k^2 - \frac{2}{\tau^2}$. Thus, initial conditions must be defined to determine the coefficients of integration, and, in doing so, the mode functions will become fixed. This is done by noticing that, at very early times (large negative conformal time), all modes are inside the Hubble horizon, and thus $\omega_k$ become time-independent as $\frac{2}{\tau ^2} \rightarrow 0$. The reduced, time-independent Mukhanov-Sasaki equation then yields the (positive frequency) solution,
\begin{align}
\label{IC1}
\lim_{\tau \rightarrow -\infty} v_k(\tau) = \frac{1}{\sqrt{2k}} e^{-ik \tau},
\end{align}
which, as an initial condition, determines the constants of integration to be $\alpha_k = 1$ and $\beta_k = 0$, resulting in the fixed mode functions:
\begin{align}
v_k(\tau) = \frac{1}{\sqrt{2k}} e^{-ik \tau} \left( 1 - \frac{i}{k \tau} \right),
\end{align}
and its complex conjugate. Hence, the Bunch-Davis vacuum is defined with Minkowski initial conditions in (\ref{IC1}). Here, `Minkowski initial conditions' refers to the fact that, as $\tau \rightarrow -\infty$, the comoving scales become arbitrarily short. Therefore, on these arbitrarily short scales, the behaviour of the theory is independent of space-time curvature - it looks locally flat. The time-dependent equations of motion for the non-interacting scalar perturbation during inflation have thus been determined, and will later be substituted in place of interaction picture fields.

\section{The \textit{in-in} `Master' Formula}

We now seek to derive the \textit{in-in} `master' formula,
\begin{align}
\label{inin}
\langle Q(t) \rangle = \langle 0 |  \bar{T} e^{ i \int^t_{-\infty(1-i\epsilon)}dt' \text{ } H^I_{\text{int}}(t')} \text{ } Q^I(t) \text{ } T e^{ - i \int^t_{-\infty(1+i\epsilon)}dt' \text{ } H^I_{\text{int}}(t')}  |0 \rangle,
\end{align}
which is rendered tractable by evaluating it at \textit{tree-level}\footnote{This is done by expanding the exponential to first order in $H_{\text{int}}$; which corresponds to a Feynman diagram with two vertices and no loops.},
\begin{align}
\label{inin1}
\langle Q(t) \rangle = \text{Re} \left[ \langle 0 | -2i Q^I(t) \int^t_{-\infty(1+i\epsilon)} dt' \text{ } H^I_{\text{int}} |0 \rangle \right].
\end{align}
All relevant terms will now be defined in a brief derivation of  (\ref{inin}), closely following Ref$.$ \citep{bau}. As usual, time evolution of the inflaton field, $\phi$, and its conjugate momentum, $\pi$, are determined using the Hamiltonian,
\begin{align}
\label{hamm}
H[\phi(t), \pi (t)] = \int d^3x \text{ }\mathcal{H}[\phi(t,\textbf{x}), \pi (t,\textbf{x})];
\end{align}
$\mathcal{H}$ denoting the Hamiltonian density. The fields, $\phi$ and $\pi$ (obeying the ETCRs), have Heisenberg equations of motion of,
\begin{align}
\dot{\phi} = i[H, \phi], \\ \nonumber
\dot{\pi} = i[H, \pi].
\end{align}
We are, however, only interested in quantising the perturbation field about a classical homogeneous background; which means making the substitution,
\begin{align}
\label{perts}
\phi(t, \textbf{x}) = \bar{\phi}(t) + \delta \phi (t, \textbf{x}), \\ \nonumber
\pi(t, \textbf{x}) = \bar{\pi}(t) + \delta \pi (t, \textbf{x}).
\end{align}
Therefore, from now on, the time-dependent background is treated classically with equations of motion,
\begin{align}
\dot{\bar{\phi}} = \frac{\partial \mathcal{H}}{\partial \bar{\pi}}, \\ \nonumber
\dot{\bar{\pi}} = \frac{\partial \mathcal{H}}{\partial \bar{\phi}},
\end{align}
and thus, only the quantised perturbations now obey the ETCRs and Heisenberg equations of motion. Upon substitution of  (\ref{perts}) into (\ref{hamm}), the Hamiltonian can be expanded as such,
\begin{align}
H[\phi, \pi; t] = \bar{H}[\bar{\phi}, \bar{\pi}] + H_0[\delta \phi, \delta \pi ;t] +  H_{\text{int}}[\delta \phi, \delta \pi ;t];
\end{align}
where $\bar{H}$ is the \textit{background} Hamiltonian, $H_0$ contains terms up to quadratic in the perturbation fields, and $H_{\text{int}}$ contains terms cubic and higher order. Therefore, in the interaction picture, the perturbation fields are obtained by solving the appropriate Heisenberg equations of motion, which have the familiar solutions,
\begin{align}
\label{intpic}
\delta \phi^I(t, \textbf{x}) = U_0^{-1}(t, t_i) \delta \phi (t_i, \textbf{x}) U_0 (t, t_i), \\ \nonumber
\delta \pi^I(t, \textbf{x}) = U_0^{-1}(t, t_i) \delta \pi (t_i, \textbf{x}) U_0 (t, t_i).
\end{align}
The unitary operator, $U_0$, which has the initial condition $U_0(t_i, t_i) \defeq 1$, satisfies,
\begin{align}
\label{uuu}
\frac{d}{dt} U_0 (t, t_i) = -i H_0[\delta \phi, \delta \pi ;t] U_0 (t, t_i).
\end{align} 
Returning to the quantum correlator in  (\ref{exp}), we now have the tools required to express this expectation value in terms of interaction picture fields, $Q^I(t)$. Firstly, we use the \textit{Heisenberg picture} unitary operator $U$\footnote{This unitary operator is defined by solving the Heisenberg equations of motion for the full, Heisenberg picture fields, and hence satisfies an equation similar to  (\ref{uuu}), but with a Hamiltonian defined by, $\tilde{H} \defeq H_0 + H_{\text{int}}$.}, to evolve $Q(t)$ back to $Q(t_i)$,
\begin{align}
\langle \textit{in} | Q(t) | \textit{in} \rangle = \langle \textit{in} | U^{-1} (t, t_i) Q(t_i) U (t, t_i)  | \textit{in} \rangle.
\end{align}
Substitution of the identity, $1 = U_0(t, t_i) U_0^{-1}(t, t_i)$, on both sides of $Q$ then yields,
\begin{align}
\langle Q(t) \rangle = \langle \textit{in} | &F^{-1}(t, t_i) U_0^{-1}(t,t_i) Q(t_i) U_0 (t, t_i) F(t, t_i) | \textit{in} \rangle, \\ \nonumber
&F(t, t_i) \defeq U_0^{-1} (t, t_i) U(t, t_i).
\end{align}
Therefore, using the definition of interaction picture fields,  (\ref{intpic}), we find, 
\begin{align}
U_0^{-1}(t,t_i) Q(t_i) U_0 (t, t_i) = Q^I(t),
\end{align}
which leads to,
\begin{align}
\label{1a2a}
\langle Q(t) \rangle =  \langle \textit{in} | &F^{-1}(t, t_i) Q^I(t) F(t, t_i) | \textit{in} \rangle.
\end{align}
In order to arrive at the final \textit{in-in} expression, we must now find the first-order differential equation (in $t$) that $F$ satisfies, solve it, and substitute it into  (\ref{1a2a}). The operator, $F$, is built out of $U_0$ and $U$, which satisfy  (\ref{uuu}) and
\begin{align}
\frac{d}{dt} U (t, t_i) = -i (H_0 + H_{\text{int}}) U (t, t_i)
\end{align}
respectively. Upon substitution of these equations into the definition of $F$, one finds,
\begin{align}
\frac{d}{dt} F(t, t_i) = -i H_{\text{int}} [ \delta \phi^I (t), \delta \pi^I (t);t] F(t,t_i),
\end{align}
which has the solution,
\begin{align}
\label{f1}
F(t, t_i) = T \text{exp} \left( -i \int_{t_i}^t dt \text{ } H_{\text{int}}(t) \right).
\end{align}
It is standard practice to include the time-ordering operator, $T$, which ensures $H$ and $Q$ always appear in the correct order with its commutative properties. The final step in this derivation is to fix $t_i$ at some time far into the past, when inflation begins, which is done by setting $t_i = -\infty (1 + i \epsilon)$. At this time, interactions are turned off (with the $i \epsilon$ prescription), and $| \textit{in} \rangle$ reduces to the Bunch-Davis vacuum, $| 0 \rangle$. Thus, substitution of  (\ref{f1}) into  (\ref{1a2a}) yields the \textit{in-in} `master' equation,  (\ref{inin}). There are many (QFT-derived) nuances involved within the $i \epsilon$ prescription used here. For example, the integration contour of the path integral does not close; the effect of which will become apparent in Section \ref{sec_3}. For further information regarding these nuances, the reader is referred to Ref$.$ \citep{lim}. To summarise, the tools have now been introduced which will allow us to calculate a quantum $n$-point correlator of scalar perturbations during inflation at tree-level. Thus, the seminal calculation  by Juan Maldacena \citep{mald} detailing the (lack of) non-Gaussianity produced by single field, slow roll inflation can now be reviewed in the following section.


\chapter{Non-Gaussianity in Single-Field, Slow-Roll Inflation Models} 

\label{sec_3} 

\lhead{4. \emph{Non-Gaussianity in Single-Field, Slow-Roll Inflation}} 


Using the tools defined in Sections \ref{sec_2} and \ref{sec_23}, we can now explicitly compute the amplitude, and $k$-dependence, of the primordial non-Gaussianity predicted by single field, slow roll inflation. This derivation will follow Ref$.$ \citep{wow} closely, with some notational deviations and additional comments. The calculation will begin by determining a suitable interaction Hamiltonian, which will then be substituted into the \textit{in-in} master formula. This Hamiltonian is found via the perturbation theory techniques outlined in Section \ref{subsec_pert}, and will be expressed in terms of the primordial curvature perturbation, $\zeta$, at third order. Finally, the interacting quantum 3-point correlator of $\zeta$ is then explicitly evaluated by Wick contracting into products of 2-point correlators. Thus, the time integral from within the tree-level \textit{in-in} equation (\ref{inin1}) is computed by substituting the free field mode functions in place of interaction picture fields. Therefore, the primordial bispectrum predicted by single field, slow roll inflation can be deduced.

\section{The Calculation}

\subsection{Laying the Groundwork}
\label{Part1}

As expected, we begin with the single field, slow roll inflation matter action,
\begin{align}
S_m = \int d^4x \text{ } \mathcal{L}_m = - \int d^4x \text{ } \frac{1}{2} (\partial_{\mu}\phi)^2 + V(\phi),
\end{align}
which is coupled to gravity in the ADM formalism via  (\ref{action}), hereafter setting $M_{pl} \defeq 1$. For convenience, the ADM metric and action are restated as,
\begin{align}
 ds^2 = -N^2 dt^2 + \leftidx{^{(3)}}g_{ij}(dx^i - N^idt)(dx^j - N^jdt), \label{1212} \\ 
S = \frac{1}{2} \int d^4x \text{ } N \sqrt{^{(3)}g}\left[ R - K^2 + K_{ij}K^{ij}  + \mathcal{L}_m \right]. \label{ADMAC}
\end{align}
The ultimate goal here is to express the ADM action (\ref{ADMAC}) in terms of the single, gauge invariant perturbation, $\zeta$, at third order (and to leading order in slow roll parameters). This is done by comparing  (\ref{1212}) with a gauge-restricted perturbed line element in order to deduce the lapse and shift (which are now acting as Lagrange multipliers). Hence, $R$, $K$, and $K_{ij}$ must be expressed in terms of the lapse and shift (and spatial 3-metric), so that the ADM action can be appropriately calculated from the perturbed line element. Furthermore, energy and momentum constraint equations will be stated which allow one to reduce the gauge fixed, perturbative degrees of freedom during inflation to a single function. 

\begin{center}
\textbf{\textit{Equation for $R$}}
\end{center}

First, the intrinsic curvature, $R$, is directly related to the 3-metric by the standard result of General Relativity,
\begin{align}
R = 2 \leftidx{^{(3)}}g^{jk} \left( \leftidx{^{(3)}}\Gamma^i_{j[k,i]} + \leftidx{^{(3)}}\Gamma^l_{j[k}\Gamma^i_{i]l} \right),
\end{align} 
where $\leftidx{^{(3)}}\Gamma^i_{jk}$ are the 3-Christoffel symbols defined as usual (for a Levi-Civita connection),
\begin{align}
\label{cristof}
\leftidx{^{(3)}}\Gamma^i_{jk} = \frac{1}{2} \leftidx{^{(3)}}g^{il} \left( \leftidx{^{(3)}}g_{jl,k} + \leftidx{^{(3)}}g_{lk,j} - \leftidx{^{(3)}}g_{jk,l} \right).
\end{align}

\begin{center}
\textbf{\textit{Equation for $K_{ij}$}}
\end{center}

Second, $K_{ij}$ has been defined in Section \ref{subsec_pert}, by parallel propagation of a normal vector, $n_i$, across the 3-geometry $\Sigma_t$,
\begin{align}
K_{ij} = n_{i;j} = - \frac{1}{2N} \left( \leftidx{^{(3)}}g_{ij,0} + N_{i|j} + N_{j|i} \right),
\end{align}
where the following notation has now been adopted: `$,$' for partial derivatives, `$;$' for covariant derivatives, and `$|$' for covariant derivatives on $\Sigma_t$.

\begin{center}
\textbf{\textit{Equation for $K$}}
\end{center}

Finally, $K$ is simply the contraction of the extrinsic curvature with the 3-metric,
\begin{align}
K = \leftidx{^{(3)}}g_{ij} K^{ij}.
\end{align}
Moreover, substitution of the lapse and shift (deduced from a gauge-fixed metric) into these equations still leaves us with three degrees of freedom remaining. However, this can be reduced to one by noticing that variation of the ADM action with respect to the metric yields two additional constraints: the energy and momentum constraint equations.

\begin{center}
\textbf{\textit{Constraint Equations}}
\end{center}

To linear order in perturbations, these equations are expressed in the gauge \textit{unrestricted} form,
\begin{gather}
\label{eng}
\Delta \Phi - H \left[  3 \left( \dot{\Phi} + H \Psi \right) - a^2 \Delta (B - \dot{E}) \right] = \frac{1}{2} \delta \phi, \\ \label{mom}
3 \left( \dot{\Phi} + H \Psi \right) = \frac{3}{2} \dot{\bar{\phi}} \delta \phi.
\end{gather}
where $\Delta = \partial^i \partial_i = \frac{\nabla^2}{a^2}$. 

\subsection{The Third Order Action}
\label{Part2}

We are now in the position to state a perturbed, gauge-restricted line element (as originally defined in Maldacena's calculation) of the form,
\begin{align}
ds^2 = -\left[ (1+ \Psi)^2 - \partial_i B \partial^i B \right] dt^2 + 2a(t)^2 \partial_iB dt dx^i + a(t)^2 e^{2\zeta}\delta_{ij}dx^i dx^j,
\end{align}
which is expressed in the comoving gauge, defined by,
\begin{align}
\label{gugu}
\delta \phi = 0, \hspace{2cm} \leftidx{^{(3)}}g_{ij} = a(t)^2 e^{2\zeta} \delta_{ij}.
\end{align}
Thus, this gauge leaves us with three metric degrees of freedom: $\Psi$, $B$, and $\zeta$. The lapse and shift can therefore be deduced as,
\begin{gather}
N = 1 + \Psi, \\
N_i = a^2 B_{,i}.
\end{gather}
Before substituting these into the three curvature equations defined in Section \ref{Part1}, it will now be convenient to solve the energy (\ref{eng}) and momentum (\ref{mom}) constraint equations. As we are working in the comoving gauge, the following substitutions can be made: $E=0$, $\delta \phi =0$, and $\Phi = - \zeta$. In doing so, the linearised constraint equations are re-expressed as,
\begin{align}
\label{eng1}
\Delta \zeta - 3 H \dot{\zeta} =& -3H^2 \Psi + H a^2 \Delta B, \\ \label{mom1}
&\dot{\zeta} = H \Psi.
\end{align}
These equations have the (comoving gauge) solutions,
\begin{gather}
\Psi = \frac{\dot{\zeta}}{H}, \\
B = - \frac{\zeta}{a^2 H} + \epsilon \nabla^{-2} \dot{\zeta},\label{bbbb}
\end{gather}
which reduces the lapse and shift to,
\begin{gather}
\label{1010}
N = 1 + \frac{\dot{\zeta}}{H}, \\\label{1011}
N_i = - a^2 \partial_i \left( \frac{\zeta}{a^2 H} - \epsilon \nabla^{-2} \dot{\zeta} \right).
\end{gather}
Crucially, $N$ and $N_i$ here have only been expanded to \textit{linear} order in perturbations. This is because higher order contributions are cleverly ignored by noticing that they appear as a pre-factor to the (appropriate order) equations of motion\footnote{Specifically, an order-$n$ action requires a lapse and shift expanded to order $n-2$. This is exemplified by considering $n=3$, where the non-linear terms will multiply only the \textit{background} equations of motion (which vanish).}: $\frac{\delta \mathcal{L}}{\delta N}$ and $\frac{\delta \mathcal{L}}{\delta N_i}$. These are, of course, set equal to zero (as solutions); hence, terms higher order than linear within $N$ and $N_i$ are ignored when considering a third order action. We can now obtain this third order action by substitution of the lapse (\ref{1010}) and shift (\ref{1011}) into the curvature equations. It should be noted, however, that at this point, literature often makes a gauge transformation to the spatially flat gauge \citep{mald}. In this gauge, all degrees of freedom are pushed onto the inflaton fluctuations, $\delta \phi$. On subhorizon scales, $\delta \phi$ becomes a more mathematically transparent quantity to work with. However, in this approach, gauge transformations are required to revert back to the comoving gauge on superhorizon scales (due to the constancy of $\zeta$ in this limit). Therefore, as a stylistic choice, this work will prefer the conceptually simpler route of dealing with $\zeta$ from start to finish. 


\begin{center}
\textbf{\textit{Determination of $R$}}
\end{center}
This task, and the following two, are algebraically very dense, and will hence be done primarily in Mathematica \citep{mma}. However, the key steps in each calculation will be identified. Firstly, we must determine the 3-Christoffel symbols (\ref{cristof}) given the comoving gauge 3-metric in  (\ref{gugu}). This substitution yields,
\begin{align}
\leftidx{^{(3)}}\Gamma^k_{ij} = \delta^{kl} \left( \delta_{ik} \partial_j \zeta + \delta_{jk} \partial_i \zeta - \delta_{ij} \partial_k \zeta \right),
\end{align}
where $R$ is appropriately built out of these, and takes the form,
\begin{align}
R = - \frac{2}{a^2 e^{2\zeta}} \left( 2 \nabla^2 \zeta + (\nabla \zeta)^2 \right).
\end{align}

\begin{center}
\textbf{\emph{\textit{Determination of $K_{ij}$}}}
\end{center}

This is perhaps the most algebraically intensive step thus far. Hence, an interim expression for $K_{ij}$ is first found inclusive of the lapse and shift,
\begin{align}
\label{pop}
K_{ij} = \frac{1}{N} \left[ a^2 e^{2\zeta} \delta_{ij} - \partial_{(i}N_{j)} + (2N_{(i}\partial_{j)} \zeta - \delta_{ij} N_k \partial^k \zeta) \right],
\end{align}
which, in the interest of calculating the $K^{ij}K_{ij}$ term in the ADM action, has the following inverse,
\begin{align}
K^{ij} = \frac{1}{a^4 e^{4\zeta}} \delta^{ik} \delta^{jl} K_{kl}.
\end{align}

\begin{center}
\textbf{\emph{\textit{Determination of $K$}}}
\end{center}

Finally, $K$ is determined (again in terms of the lapse and shift), by contracting  (\ref{pop}) with the 3-metric as follows,
\begin{align}
K = \leftidx{^{(3)}}g_{ij} K^{ij} = 3(H+ \dot{\zeta}) - \frac{1}{a^2 e^{2\zeta}} (\partial^k N_k + N_k \partial^k \zeta).
\end{align}

\begin{center}
\textbf{\emph{\textit{Substitution Into the Action}}}
\end{center}

The following fields: $K$, $K_{ij}$, $R$, and $\leftidx{^{(3)}}g_{ij}$, can now be substituted into the ADM (gravity) action to yield,
\begin{align}
S_G = \frac{1}{2} \int d^4x \text{ } N a^3 e^{3\zeta} & \left[  - \frac{2}{a^2 e^{2\zeta}} \left( 2 \nabla^2 \zeta + (\nabla \zeta)^2 \right) -  \left( 3(H+ \dot{\zeta}) - \frac{1}{a^2 e^{2\zeta}} (\partial^k N_k + N_k \partial^k \zeta) \right)^2 \right. \\ \nonumber
 &+ \left. \frac{1}{N^2} \frac{1}{a^4 e^{4\zeta}} \delta^{ik} \delta^{jl} \left[ a^2 e^{2\zeta} \delta_{ij} - \partial_{(i}N_{j)} + (2N_{(i}\partial_{j)} \zeta - \delta_{ij} N_k \partial^k \zeta) \right] \right. \\ \nonumber & \times \left[ a^2 e^{2\zeta} \delta_{kl} - \partial_{(k}N_{l)} + (2N_{(k}\partial_{l)} \zeta - \delta_{kl} N_i \partial^i \zeta) \right] \Bigg].
\end{align}
Substitution of the lapse and shift into this expression is now certainly a job for Mathematica. The `name of the game' is to isolate all terms leading order in slow roll parameters,
\begin{align*}
\epsilon = \frac{\dot{H}}{H^2} = \frac{\dot{\bar{\phi}}^2}{2H^2}, \hspace{2cm} \eta = -\frac{\dot{\epsilon}}{H\epsilon},
\end{align*}
and terms of cubic order in $\zeta$. This was done in \textit{detail} (in the comoving gauge) by Ref$.$ \citep{wow}. Thus, the third order action reads, 
\begin{align}
\label{thirdoac}
S_G[3] = \int d^4x \text{ } a \epsilon^2 \left[ a^2 \zeta \dot{\zeta}^2 - \zeta (\partial_i \zeta)^2 + 2a^2 \dot{\zeta} (\partial_i \zeta) \partial_i \left( \nabla^{-2} \dot{\zeta} \right) \right] + 2f(\zeta) \frac{\delta \mathcal{L}_2}{\delta \zeta},
\end{align}
where (following Maldacena's approach) all terms proportional to the first order equation of motion (\ref{eom1}) are collected into $f(\zeta)$. However, this term can be conveniently removed from the action via the field redefinition,
\begin{align}
\zeta = \zeta_n - f(\zeta_n).
\end{align}
Upon substitution, this redefinition kills all terms proportional to the first order equation of motion in (\ref{thirdoac}). Hence, $\zeta_n$ will now be the field of interest until the end of the calculation. Crucially, however, we wish to eventually arrive at the 3-point correlator of $\zeta$, and \textit{not} $\zeta_n$\footnote{$\zeta_n$ is, in fact, \textit{not} conserved on superhorizon scales. Hence, special care is needed when calculating the final correlator of $\zeta$.}. Therefore, the 3-point correlator of $\zeta$ is found to gain the following additional terms,
\begin{gather}
\label{cor1}
\langle \zeta_{\textbf{x}_1} \zeta_{\textbf{x}_2} \zeta_{\textbf{x}_3} \rangle = \langle \zeta_n (\textbf{x}_1) \zeta_n (\textbf{x}_2) \zeta_n (\textbf{x}_3) \rangle + \frac{\eta}{4} \left( \langle \zeta_n (\textbf{x}_1) \zeta_n (\textbf{x}_2) \rangle \langle \zeta_n (\textbf{x}_1) \zeta_n (\textbf{x}_3) \rangle + \text{sym}. \right), \\ \label{popop}
f(\zeta) \defeq \frac{\eta}{4} \zeta^2.
\end{gather}

It should be noted that $f(\zeta)$ here has been vastly reduced from its full form. This is because the correlators (\ref{cor1}) possess no contribution from subhorizon modes\footnote{As detailed in Maldacena's paper, the Minkowski space initial conditions for modes deep inside the horizon result in no contributions from expectation values of the rapidly oscillating fields.}. In fact, the correlators are evaluated at the end of inflation, as $\tau \rightarrow 0$, and hence all modes are superhorizon. In this limit, all terms with derivative operators acting on $\zeta$ can be ignored due to the constancy of $\zeta$ - leaving $f(\zeta)$ as defined in (\ref{popop}). For completion, the third order action that will be used to determine the interaction Hamiltonian is,
\begin{align}
\label{3rdo}
S_G[3] = \int d^4x \text{ } a \epsilon^2 \left[ a^2 \zeta_n \dot{\zeta}_n^2 + \zeta_n (\partial_i \zeta_n)^2 + 2a^2 \dot{\zeta}_n (\partial_i \zeta_n) \partial_i \left( \nabla^{-2} \dot{\zeta}_n \right) \right],
\end{align}
where the subscript $n$ will now be dropped for the remainder of this work. \\

\subsection{The Interaction Hamiltonian}
\label{Part3}

We now seek to determine the interaction Hamiltonian, $H_{\text{int}}$, that will be substituted into the tree-level \textit{in-in} formula. In Section \ref{sec_23}, the following Hamiltonian density was defined,
\begin{align}
\label{known}
\tilde{\mathcal{H}} = \mathcal{H}_0 + \mathcal{H}_{\text{int}},
\end{align}
which contains terms second order and above in perturbations. To find $\mathcal{H}_{\text{int}}$, the second (\ref{2ndo}) and third (\ref{3rdo}) order actions in $\zeta$ are collected as such,
\begin{align}
S[2] + S[3] = \int d^4x \text{ } \tilde{\mathcal{L}}.
\end{align}
The Hamiltonian density of this theory is therefore,
\begin{gather}
\tilde{\mathcal{H}} = \pi \dot{\zeta} - \tilde{\mathcal{L}},  \\ \label{cm}
\pi = \frac{\partial \tilde{\mathcal{L}}}{\partial \dot{\zeta}}.
\end{gather}
However, $\mathcal{H}_0$ in (\ref{known}) is known - the free field Hamiltonian density (also determined in Section \ref{sec_23}). Thus, in order to find $\mathcal{H}_{\text{int}}$, the conjugate momentum (\ref{cm}) must be calculated, and then rearranged for $\dot{\zeta}$. One can then substitute this resultant expression into $\tilde{\mathcal{H}}$, extract the known $\mathcal{H}_0$, and be left with $\mathcal{H}_{\text{int}}(\tilde{\mathcal{L}})$. There are a number of nuances which prevent this method from being a generally straightforward process. For example, the conjugate momentum is not always invertible for Lagrangians which contain higher order terms in $\dot{\zeta}$. This complication, along with others, are detailed in Ref$.$ \citep{lim}. Hence, following this process, the interaction Hamiltonian is determined as,
\begin{gather}
\mathcal{H}_{\text{int}} = - \mathcal{L}_{\text{int}} + \mathcal{O}(\zeta^4), \\
H_{\text{int}} = - \int d^3x \text{ }\mathcal{L}_{\text{int}} = - \int d^3x \text{ } a \epsilon^2 \left[ a^2 \zeta \dot{\zeta}^2 + \zeta (\partial_i \zeta)^2 + 2a^2 \dot{\zeta} (\partial_i \zeta) \partial_i \left( \nabla^{-2} \dot{\zeta} \right) \right].
\end{gather}\\

\subsection{Evaluating the \textit{in-in} Formula}
\label{Part4}

All the components have now been explicitly introduced which will allow us to evaluate,
\begin{align}
\langle Q(t) \rangle = \text{Re} \left[ \langle 0 | -2i Q^I(t) \int^t_{-\infty(1+i\epsilon)} dt' \text{ } H^I_{\text{int}} |0 \rangle \right].
\end{align}
The first step is to note that the induced non-Gaussianity should be calculated at the \textit{end} of inflation, $\tau \rightarrow 0$. Assuming de Sitter space, $a \approx -1/(H\tau)$, the limit $\tau \rightarrow 0$ results in all modes existing far outside the horizon, $k\tau \ll 0$. Hence, the observable becomes,
\begin{align}
Q(0) = \zeta(\textbf{k}_1, 0) \zeta(\textbf{k}_2, 0) \zeta(\textbf{k}_3, 0),
\end{align}
with the corresponding quantum correlator,
\begin{align}
\label{6pt}
\hspace{-1cm} \langle \zeta(\textbf{k}_1, 0) \zeta(\textbf{k}_2, 0) \zeta(\textbf{k}_3, 0) \rangle = \text{Re} \left[ \langle 0 | -2i \zeta^I(\textbf{k}_1, 0) \zeta^I(\textbf{k}_2, 0) \zeta^I(\textbf{k}_3, 0) \int^{t}_{-\infty(1+i\epsilon)} dt' \int d^3x \text{ } \mathcal{L}_{\text{int}}(\zeta^I(\textbf{x}, \tau')) | 0 \rangle \right].
\end{align}
The superscript $I$, denoting interaction picture fields, will be ignored hereafter for notational convenience. Converting the fields within the $\mathcal{L}_{\text{int}}$ to Fourier space yields\footnote{Where, because these are interaction picture fields, they have been expanded in terms of the mode functions derived for the non-interacting theory.},
\begin{align}
\zeta(\textbf{x}, \tau) = \int \frac{d^3k}{(2\pi)^3} \text{ } \left[ a_{\textbf{k}} v_{k}(\tau) e^{i\textbf{k}.\textbf{x}} + a^{\dagger}_{\textbf{k}} v^*_{k}(\tau) e^{-i\textbf{k}.\textbf{x}}  \right],
\end{align}
where the following redefinition of the Mukhanov variable has been made, $v_k \defeq z v_k$\footnote{This redefinition is made because, as detailed in Section \ref{sec_23}, $\zeta = z v_k$; where $z = \sqrt{2a^2\epsilon}$.}. It can now be seen that, because $\mathcal{L}_{\text{int}}$ is constructed such that it only contains terms $\mathcal{O}(\zeta^3)$, (\ref{6pt}) becomes a 6-point correlator. This 6-point correlator of interaction picture fields can be contracted into products of 2-point correlators via Wick's theorem. Thus, the 2-point correlator in the non-interacting limit is required, and takes the most general form,
\begin{align}
\langle 0 | \zeta_{\textbf{k}_1} (\tau_1) \zeta_{\textbf{k}_2} (\tau_2) | 0 \rangle & = v_{k_1}(\tau_1) v^*_{k_2}(\tau_2) \langle 0 | [a_{\textbf{k}_1} a^{\dagger}_{\textbf{k}_2}] | 0 \rangle \\ \nonumber
& = v_{k_1}(\tau_1) v^*_{k_2}(\tau_2) (2\pi)^3 \delta(\textbf{k}_1 - \textbf{k}_2),
\end{align}
which will be substituted appropriately when the Wick contractions are made\footnote{Note that these 2-point correlators are proportional to the absolute magnitudes of the individual mode functions, and are thus manifestly real quantities (also known as \textit{Wightman functions}).}. Therefore, the calculation of (\ref{6pt}) can begin, first by noticing that $\mathcal{L}_{\text{int}}$ contributes the following three terms,
\begin{gather}
\mathcal{L}_{\text{int}} = a \epsilon^2 \left[ a^2 \zeta  \dot{\zeta}^2 + \zeta ( \partial_i \zeta)^2 + 2a^2 \dot{\zeta} (\partial_i \zeta ) \partial_i \left( \nabla^{-2} \dot{\zeta} \right) \right], \\ \nonumber
\hspace{0.8cm} 1 \hspace{1.3cm} 2 \hspace{2.8cm} 3
\end{gather}

which will now be evaluated separately. \newpage

\begin{center}
\textbf{\textit{Term 1}}
\end{center}

Substituting term 1 of $\mathcal{L}_{\text{int}}$ into (\ref{6pt}) produces,
\begin{gather}
\label{22222}
\hspace{-3cm} \langle \zeta(\textbf{k}_1, 0) \zeta(\textbf{k}_2, 0) \zeta(\textbf{k}_3, 0) \rangle_1 = \text{Re} \Bigg[  \langle 0 | -2i \zeta(\textbf{k}_1, 0) \zeta(\textbf{k}_2, 0) \zeta(\textbf{k}_3, 0) \int^{\tau}_{-\infty(1+i\epsilon)} d\tau' \int d^3x  \\ \nonumber \hspace{4cm}
 \left. \times \int \frac{d^3p_1}{(2\pi)^3} \frac{d^3p_2}{(2\pi)^3} \frac{d^3p_3}{(2\pi)^3} \text{ } e^{-i(\textbf{p}_1 + \textbf{p}_2 + \textbf{p}_3).\textbf{x}} a^2 \epsilon^2 \zeta(\textbf{p}_1, \tau') \zeta'(\textbf{p}_2, \tau') \zeta'(\textbf{p}_3, \tau') | 0 \rangle \right],
\end{gather}
where a switch to conformal time has been made. This expression is vastly simplified by noticing that only three out of a possible fifteen contractions survive\footnote{The contractions that `don't survive' are due to the creation and annihilation vacuum identities: $\hat{a}_{\textbf{k}} |0 \rangle = 0$, $ \langle 0 | \hat{a}^{\dagger}_{\textbf{k}}= 0.$}, which are,
\begin{align}
\hspace{-0.25cm}\langle \zeta(\textbf{k}_1, 0) \zeta(\textbf{p}_1, \tau') \rangle& \langle \zeta(\textbf{k}_2, 0) \zeta'(\textbf{p}_2, \tau') \rangle \langle \zeta(\textbf{k}_3, 0) \zeta'(\textbf{p}_3, \tau') \rangle \\ \nonumber
&= (2\pi)^9 \delta(\textbf{k}_1 - \textbf{p}_1)\delta(\textbf{k}_2 - \textbf{p}_2)\delta(\textbf{k}_3 - \textbf{p}_3) v_{k_1}(0)v^*_{p_1}(\tau') v_{k_2}(0) v'^{*}_{p_2}(\tau') v_{k_3}(0)v'^{*}_{p_3}(\tau'),
\end{align}
and, by symmetry, two additional contributions under $1 \leftrightarrow 2$ and $1 \leftrightarrow 3$. Moreover, (\ref{22222}) is simplified further by substitution of the identity,
\begin{align}
\int d^3x \text{ } e^{-i(\textbf{p}_1 + \textbf{p}_2 + \textbf{p}_3).\textbf{x}} = (2\pi)^3 \delta(\textbf{p}_1 + \textbf{p}_2 + \textbf{p}_3),
\end{align}
which leaves,
\begin{align}
\label{label}
\hspace{-2.5cm} \langle \zeta(\textbf{k}_1, 0) \zeta(\textbf{k}_2, 0) \zeta(\textbf{k}_3, 0) \rangle_1  = \text{Re} \Bigg[ &-4i \int^{\tau}_{-\infty(1+i\epsilon)} d\tau' \int d^3 p_1 d^3 p_2 d^3 p_3 \text{ } a^2 \epsilon^2 (2\pi)^3 \delta(\textbf{p}_1 + \textbf{p}_2 + \textbf{p}_3) \\ \nonumber
& \times v_{k_1}(0) v_{k_2}(0) v_{k_3}(0)v^*_{p_1}(\tau')  v'^{*}_{p_2}(\tau')  v'^{*}_{p_3}(\tau') \delta(\textbf{k}_1 - \textbf{p}_1)\delta(\textbf{k}_2 - \textbf{p}_2)\delta(\textbf{k}_3 - \textbf{p}_3) + \text{sym.} \Bigg],
\end{align}
where `sym.' denotes the additional two terms from $1 \leftrightarrow 2$ and $1 \leftrightarrow 3$. The relevant mode functions (derived from Section \ref{sec_23}) now take the following form,
\begin{gather}
v_k(\tau) = \frac{H}{\sqrt{4 \epsilon k^3}} e^{-ik \tau} \left( 1 + ik \tau \right), \\ \label{consttt}
\lim_{\tau \rightarrow 0} v_k(\tau) \defeq  v_k(0) =  \frac{H}{\sqrt{4 \epsilon k^3}}, \\ 
v'_k(\tau) = \frac{H}{\sqrt{4 \epsilon k^3}} k^2 \tau e^{-ik \tau}.
\end{gather}
Substituting these into (\ref{label}), and integrating out the delta functions, yields,
\begin{align}
\label{mommm}
\hspace{-1.6cm} \langle \zeta(\textbf{k}_1, 0) \zeta(\textbf{k}_2, 0) \zeta(\textbf{k}_3, 0) \rangle_1 = \text{Re}  \Bigg[ & -4i \frac{H^6}{(4\epsilon)^3} \frac{1}{(k_1 k_2 k_3)^3} \int^{\tau}_{-\infty(1+i\epsilon)} d\tau' \text{ } \frac{\epsilon^2}{(H \tau')^2} \\ \nonumber
\times & (2\pi)^3 \delta(\textbf{k}_1 + \textbf{k}_2 + \textbf{k}_3) e^{iK\tau'} (k_2 k_3)^2 \tau'^2 (1+ik_1 \tau') + \text{sym}. \Bigg],
\end{align}
where $K = k_1 + k_2 + k_3$. The final step now is to evaluate the integral over conformal time. This is done easily by noticing that the substitution of $a^2$ for the de Sitter approximation cancels a factor of $\tau'^2$. Furthermore, there will be no contribution from the lower bound of the integral, as expected, due to the $-\infty(1+i\epsilon)$ in the exponent. Thus, the 3-point correlator for the first term in $\mathcal{L}_{\text{int}}$ becomes,
\begin{align}
\hspace{-0.5cm} \langle \zeta(\textbf{k}_1, 0) \zeta(\textbf{k}_2, 0) \zeta(\textbf{k}_3, 0) \rangle_1 = (2\pi)^3 \delta(\textbf{k}_1 + \textbf{k}_2 + \textbf{k}_3) \frac{H^4}{16\epsilon} \frac{1}{(k_1 k_2 k_3)^3} \left( \frac{(k_2 k_3)^2}{K} + \frac{k_1(k_2 k_3)^2}{K^2} + \text{sym.} \right). 
\end{align}

\begin{center}
\textbf{\textit{Term 2}}
\end{center}

The second term in $\mathcal{L}_{\text{int}}$, $a \epsilon^2 \zeta (\partial_i \zeta)^2$, can be correlated in similar fashion to the first. Without loss of generality, we can now choose to consider only $\zeta_{\textbf{k}_1} (\partial \zeta_{\textbf{k}_2}).(\partial \zeta_{\textbf{k}_3})$, where two additional terms will be carried through, by symmetry, under $1 \leftrightarrow 2$ and $1 \leftrightarrow 3$. To begin, one must notice that $(\partial_i \zeta)^2 \rightarrow (\textbf{k}_2.\textbf{k}_3) \zeta_{\textbf{k}_2} \zeta_{\textbf{k}_3}$ under a Fourier transform. This pre-factor to $\zeta$ can be expressed as a magnitude of $k$ by noting that,
\begin{align}
(\textbf{k}_1 + \textbf{k}_2 + \textbf{k}_3)^2 = 0 = k_1^2 + k_2^2 + k_3^2 + 2\textbf{k}_1.\textbf{k}_2 + 2\textbf{k}_2.\textbf{k}_3 + 2\textbf{k}_1.\textbf{k}_3,
\end{align}
by the triangle condition, hence,
\begin{align}
\textbf{k}_1.\textbf{k}_2 + \textbf{k}_2.\textbf{k}_3 + \textbf{k}_1.\textbf{k}_3 = -\frac{1}{2} (k_1^2 + k_2^2 + k_3^2).
\end{align}
Thus, the substitution $\textbf{k}_2.\textbf{k}_3 = -\frac{1}{2} k_1^2$ will be made in the subsequent work. Following the procedure above, only one term survives the Wick contraction\footnote{Technically, three terms survive the Wick contraction, however, as mentioned, the other two are being ignored for now by symmetry arguments.},
\begin{align}
\label{label1}
\langle \zeta(\textbf{k}_1, 0) \zeta(\textbf{k}_2, 0) \zeta(\textbf{k}_3, 0) \rangle_2  = \text{Re} \Bigg[ -4i v_{k_1}(0)& v_{k_2}(0) v_{k_3}(0) \epsilon^2 (2\pi)^3 \delta(\textbf{k}_1 + \textbf{k}_2 + \textbf{k}_3) \\ \nonumber \times & \int^{\tau}_{-\infty(1+i\epsilon)} d\tau' \text{ } a^2 v^*_{k_1}(\tau')  v^{*}_{k_2}(\tau')  v^{*}_{k_3}(\tau') \Bigg].
\end{align}
The pre-factor to the time integral, of course, remains the same as $\langle \zeta \zeta \zeta \rangle_1$; whereas the conformal time integral can be expanded as such,
\begin{align}
\hspace{-1cm} \int^{\tau}_{-\infty(1+i\epsilon)} d\tau' \text{ } a^2 v^*_{k_1}(\tau')  v^{*}_{k_2}(\tau')  v^{*}_{k_3}(\tau') = - \int^{\tau}_{-\infty(1+i\epsilon)} d\tau' \text{ } \frac{k_1^2}{2(H \tau')^2}  e^{-iK\tau} (1+ik_1\tau)(1+ik_2\tau)(1+ik_3\tau).
\end{align}
Evaluating this integral, taking the limit of $\tau \rightarrow 0$, and adding in the symmetric terms, yields,
\begin{align}
- \int^{\tau}_{-\infty(1+i\epsilon)} d\tau' \text{ }& \frac{k_1^2}{2(H \tau')^2}  e^{-iK\tau} (1+ik_1\tau)(1+ik_2\tau)(1+ik_3\tau)   \\ \nonumber
&= - \frac{i}{2H^2} \left(k_1^2 + k_2^2 + k_3^2 \right) \left[ -K + \frac{1}{K} (k_1 k_2 + k_1 k_3 + k_2 k_3) + \frac{1}{K^2} k_1 k_2 k_3 \right].
\end{align}
The correlator for the second term of $\mathcal{L}_{\text{int}}$ is therefore,
\begin{align}
\hspace{-0.5cm} \langle \zeta(\textbf{k}_1, 0) \zeta(\textbf{k}_2, 0) \zeta(\textbf{k}_3, 0) \rangle_2 = &(2\pi)^3 \delta(\textbf{k}_1 + \textbf{k}_2 + \textbf{k}_3) \frac{H^4}{32\epsilon^2} \frac{1}{(k_1 k_2 k_3)^3} \\ \nonumber
& \times \Bigg\{ \left(k_1^2 + k_2^2 + k_3^2 \right) \left[ -K + \frac{1}{K} (k_1 k_2 + k_1 k_3 + k_2 k_3) + \frac{1}{K^2} k_1 k_2 k_3 \right] \Bigg\}.
\end{align}

\begin{center}
\textbf{\textit{Term 3}}
\end{center}

Likewise, the evaluation of $\langle \zeta \zeta \zeta \rangle_3$ will begin by converting the final term, $2a^3 \epsilon^2 \dot{\zeta} (\partial_i \zeta) \partial_i ( \nabla^{-2} \dot{\zeta})$, to Fourier space. Firstly, it should be noted that, because there are three separate differential operators here, 3$!$ = 6 arrangements of the modes will survive the Wick contraction via symmetry (rather than three in the previous two evaluations). The general formula for this conversion to Fourier space becomes,
\begin{align}
\dot{\zeta} (\partial_i \zeta) \partial_i ( \nabla^{-2} \dot{\zeta}) \rightarrow \dot{\zeta}_{\textbf{k}} \frac{\textbf{k}'.\textbf{k}''}{k'^2} \dot{\zeta_{\textbf{k}'}} \zeta_{\textbf{k}''}.
\end{align}
Hence, the appropriate pre-factor to the time integral due to these six symmetric arrangements becomes,
\begin{align}
\hspace{-0.25cm} \frac{\textbf{k}_1.\textbf{k}_2}{k_1^2} + \frac{\textbf{k}_1.\textbf{k}_3}{k_1^2} + \frac{\textbf{k}_2.\textbf{k}_3}{k_2^2} + \frac{\textbf{k}_2.\textbf{k}_1}{k_2^2} +
\frac{\textbf{k}_3.\textbf{k}_1}{k_3^2} +
\frac{\textbf{k}_3.\textbf{k}_2}{k_3^2} = -\frac{1}{2} \left( \frac{k_3^2}{k_1^2} + \frac{k_3^2}{k_2^2} + \frac{k_2^2}{k_1^2} + \frac{k_2^2}{k_3^2} + \frac{k_1^2}{k_2^2} + \frac{k_1^2}{k_3^2} \right).
\end{align}
However, these pre-factors are grouped into three sets of two, due to the indistinguishable arrangements of $\zeta \dot{\zeta}  \dot{\zeta}$, leaving the following three time integrals,
\begin{align}
I_1 \defeq -\frac{1}{2} \left[ \frac{k_2^2}{k_1^2} +  \frac{k_1^2}{k_2^2} \right]  \int^{\tau}_{-\infty(1+i\epsilon)} d\tau' \text{ } a^2 v'_{k_1}(\tau') v'_{k_2}(\tau') v_{k_3}(\tau'), \\ \nonumber
I_2 \defeq -\frac{1}{2} \left[ \frac{k_3^2}{k_1^2} +  \frac{k_1^2}{k_3^2} \right]  \int^{\tau}_{-\infty(1+i\epsilon)} d\tau' \text{ } a^2 v'_{k_1}(\tau') v_{k_2}(\tau') v'_{k_3}(\tau'),  \\ \nonumber
I_3 \defeq -\frac{1}{2} \left[ \frac{k_3^2}{k_2^2} +  \frac{k_2^2}{k_3^2} \right]  \int^{\tau}_{-\infty(1+i\epsilon)} d\tau' \text{ } a^2 v_{k_1}(\tau') v'_{k_2}(\tau') v'_{k_3}(\tau'). 
\end{align}
These time integrals therefore take the form of the ones solved during the evaluation of the first term in $\mathcal{L}_{\text{int}}$ (\ref{mommm}). Hence, the momentum dependence is known as $\tau \rightarrow 0$, and these integrals reduce to,
\begin{align}
I_1 = -\frac{1}{2} \left[ \frac{1}{K} \left( \frac{k_2^4}{k_1} + \frac{k_1^4}{k_2} \right) + \frac{1}{K^2} \left( \frac{k_3 k_2^4}{k_1} + \frac{k_3 k_1^4}{k_2} \right) \right], \\ \nonumber
I_2 = -\frac{1}{2} \left[ \frac{1}{K} \left( \frac{k_3^4}{k_1} + \frac{k_1^4}{k_3} \right) + \frac{1}{K^2} \left( \frac{k_2 k_3^4}{k_1} + \frac{k_2 k_1^4}{k_3} \right) \right], \\ \nonumber
I_3 = -\frac{1}{2} \left[ \frac{1}{K} \left( \frac{k_2^4}{k_3} + \frac{k_3^4}{k_2} \right) + \frac{1}{K^2} \left( \frac{k_1 k_2^4}{k_3} + \frac{k_1 k_3^4}{k_2} \right) \right].
\end{align}
Therefore, combining these produces the 3-point correlator of the third term in $\mathcal{L}_{\text{int}}$, 
\begin{align}
\hspace{-0.5cm} \langle \zeta(\textbf{k}_1, 0) \zeta(\textbf{k}_2, 0) \zeta(\textbf{k}_3, 0) \rangle_3 = & -(2\pi)^3 \delta(\textbf{k}_1 + \textbf{k}_2 + \textbf{k}_3) \frac{H^4}{32\epsilon} \frac{1}{(k_1 k_2 k_3)^3} \\ \nonumber
& \times \Bigg\{ \left[ \frac{1}{K} \left( \frac{k_2^4}{k_1} + \frac{k_1^4}{k_2} \right) + \frac{1}{K^2} \left( \frac{k_3 k_2^4}{k_1} + \frac{k_3 k_1^4}{k_2} \right) \right] \\ \nonumber
& \times \left[ \frac{1}{K} \left( \frac{k_2^4}{k_1} + \frac{k_1^4}{k_2} \right) + \frac{1}{K^2} \left( \frac{k_3 k_2^4}{k_1} + \frac{k_3 k_1^4}{k_2} \right) \right] \\ \nonumber
& \times  \left[ \frac{1}{K} \left( \frac{k_2^4}{k_1} + \frac{k_1^4}{k_2} \right) + \frac{1}{K^2} \left( \frac{k_3 k_2^4}{k_1} + \frac{k_3 k_1^4}{k_2} \right) \right] \Bigg\}
\end{align} 

\begin{center}
\textbf{\textit{The Correction Term}}
\end{center}

Before obtaining the bispectrum for single field, slow roll inflation; the first order correction terms from the field redefinition (\ref{cor1}) must be computed. These corrections are simply three permutations of,
\begin{gather}
\label{cor111}
\hspace{-0.5cm} \langle \zeta (\textbf{x}_1) \zeta (\textbf{x}_2) \rangle \langle \zeta (\textbf{x}_1) \zeta (\textbf{x}_3) \rangle = \int \frac{d^3k_1}{(2\pi)^3}  \frac{d^3k_2}{(2\pi)^3} \text{ } e^{i\textbf{k}_1.(\textbf{x}_2 - \textbf{x}_1)}  e^{i\textbf{k}_2.(\textbf{x}_3 - \textbf{x}_1)} v_{\textbf{k}_1}(\tau) v^*_{\textbf{k}_1}(\tau)  v_{\textbf{k}_2}(\tau) v^*_{\textbf{k}_2}(\tau),
\end{gather}
under $1 \leftrightarrow 2$ and $1 \leftrightarrow 3$. As before, the mode functions are evaluated in the late time limit, $\tau \rightarrow 0$, where they become constant (\ref{consttt}). Thus, (\ref{cor111}) reduces to,
\begin{align}
\hspace{-0.5cm} \langle \zeta (\textbf{x}_1) \zeta (\textbf{x}_2) \rangle \langle \zeta (\textbf{x}_1) \zeta (\textbf{x}_3) \rangle = \int \frac{d^3k_1}{(2\pi)^3}  \frac{d^3k_2}{(2\pi)^3} \text{ } e^{i\textbf{k}_1.\textbf{x}_1} e^{i\textbf{k}_2.\textbf{x}_2} e^{-i\textbf{x}_3.(\textbf{k}_1 + \textbf{k}_2)} \frac{H^4}{16 \epsilon^2} \frac{1}{k_1^3 k_2^3}.
\end{align}
Finally, in order to be able to easily add the remaining two symmetric contributions; a trick can be made whereby the coordinates become symmetrised due to the substitution of the identity,
\begin{align}
 e^{-i\textbf{x}_3.(\textbf{k}_1 + \textbf{k}_2)} = \int d^3k_3 \text{ } \delta(\textbf{k}_1 + \textbf{k}_2 + \textbf{k}_3) e^{i\textbf{x}_3.\textbf{k}_3},
\end{align}
which yields,
\begin{align}
\hspace{-0.5cm} \langle \zeta (\textbf{x}_1) \zeta (\textbf{x}_2) \rangle \langle \zeta (\textbf{x}_1) \zeta (\textbf{x}_3) \rangle = \int & \frac{d^3k_1}{(2\pi)^3}  \frac{d^3k_2}{(2\pi)^3} \frac{d^3k_3}{(2\pi)^3} e^{i\textbf{k}_1.\textbf{x}_1} e^{i\textbf{k}_2.\textbf{x}_2} e^{i\textbf{k}_3.\textbf{x}_3} \\ \nonumber
& \times (2\pi)^3 \delta(\textbf{k}_1 + \textbf{k}_2 + \textbf{k}_3)  \frac{H^4}{16\epsilon^2} \frac{1}{(k_1 k_2 k_3)^3}k_3^3.
\end{align}
The two additional symmetry terms are now simply deduced via $k_3^3 \rightarrow (k_1^3 + k_2^3 + k_3^3)$. Hence, the correction term in its entirety, in Fourier space, reads,
\begin{align}
\frac{\eta}{4}( \langle \zeta_{\textbf{k}_1} \zeta_{\textbf{k}_2} \rangle \langle \zeta_{\textbf{k}_1} \zeta_{\textbf{k}_3} \rangle + \text{sym.}) = (2\pi)^3 \delta(\textbf{k}_1 + \textbf{k}_2 + \textbf{k}_3) \frac{H^4}{32 \epsilon^2} \frac{1}{(k_1 k_2 k_3)^3} \eta (k_1^3 + k_2^3 + k_3^2).
\end{align}

\subsection{The Bispectrum of Single Field, Slow Roll Inflation}

Finally, the three separate terms resulting from the substitution $\mathcal{L}_{\text{int}}$ into the \textit{in-in} formula, and the correction term, can be compiled to give the bispectrum of perturbations during single field, slow roll inflation. The pre-factors of all four shape functions remain (almost) the same; however, simplifying the $k$-dependence is no small task. This simplification process has been reviewed, in detail, in Ref$.$ \citep{wow}. Thus, upon performing this simplification, the bispectrum reads,
\begin{align}
B^{\text{mald}}(k_1, k_2, k_3) = \frac{H^4}{32 \epsilon^2} \frac{1}{(k_1 k_2 k_3)^3}& \Big[ (\eta - \epsilon)(k_1^3 + k_2^3 + k_3^3) \\ \nonumber
& + \epsilon (k_1^2 k_2 + k_2^2 k_1 + k_1^2 k_3 + k_3^2 k_1 + k_2^2 k_3 + k_3^2 k_2) \\ \nonumber
& + \frac{8 \epsilon}{K} (k_1^2 k_2^2 + k_1^2 k_3^2 + k_2^2 k_3^2) \Big],
\end{align}
which is the result of Maldacena's original derivation. In order to obtain the amplitude of the non-Gaussianity produced in this inflationary regime, one must first normalise with respect to the power spectrum, which removes a factor of $\frac{1}{\epsilon^2}$ in the shape function of $B^{\text{mald}}$. Thus, evaluating this appropriately normalised shape function, one finds it to be a superposition of the previously discussed local and equilateral shape functions,
\begin{align}
\label{121212}
S^{\text{mald}}(k_1, k_2, k_3) \approx (6\epsilon -2\eta) S^{\text{local}}(k_1, k_2, k_3) + \frac{5}{3} \epsilon S^{\text{equil}}(k_1, k_2, k_3).
\end{align}
However, while it is not immediately obvious by looking at this expression, the dominant shape function here is the local one; (\ref{121212}) has been shown to be $99.7\%$ correlated with $S^{\text{local}}$ \citep{bigreview}. Hence, the appropriate limit by which to evaluate this function is the \textit{squeezed limit}, $k_1 \ll k_2 \approx k_3$, thus yielding,
\begin{align}
\lim_{k_1 \rightarrow 0} S^{\text{mald}}(k_1, k_2, k_3) = (\eta + 2\epsilon)\frac{k_2}{k_1},\\ \nonumber
S^{\text{mald}} \approx \mathcal{O}(\epsilon, \eta) \rightarrow f_{\text{NL}}^{\text{mald}} \approx \mathcal{O}(\epsilon, \eta).
\end{align}
Therefore, the conclusion of Maldacena's seminal paper - that non-Gaussianity is slow roll suppressed in the squeezed limit - has been shown. As discussed, \textit{Planck} 2015 is probing $f_{\text{NL}} \approx \mathcal{O}(1)$, and hence this non-Gaussian amplitude is not within experimental bounds.



\chapter{Non-Gaussianity in Single-Field Models with Excited Initial States} 

\label{sec_4} 

\lhead{5. \emph{Non-Gaussianity in Single-Field Models with Excited Initial States}} 


The previous section detailed Maldacena's calculation of the single field, slow roll inflation bispectrum, whereby no observationally significant non-Gaussianity is generated. This section is concerned with how deviations to assumptions implicit within that calculation can generate an appreciable amplitude of non-Gaussianity. Specifically, the validity of the Bunch-Davis vacuum as the unique initial state of inflation will be investigated.

\section{Motivation}

The Big Bang is the most archetypical example of a necessarily trans-Planckian event in cosmology, and is ultimately responsible for the structure of the Universe we see today. Therefore, it is natural to seek to derive observational guidance for the nature of trans-Planckian physics from the Big Bang. Specifically, the question of whether the effects of such trans-Planckian physics manifest in the statistics of the cosmological observables has been the subject of much research \citep{tol,meer,chennonbd}. In the context of primordial non-Gaussianities, this investigation often reduces to treating the inflaton as a degree of freedom in an effective field theory (EFT) \citep{tol2}. Hence, the choice of an initial inflationary vacuum is largely a phenomenological matter. With this in mind, we can seek to derive the possibilities of excited (non-Bunch-Davis) initial states for producing an appreciable amplitude of primordial non-Gaussianity. The observation of these non-Gaussian signatures could thus act as a probe of unknown high-energy physics, and ultimately perhaps aid the development of UV-complete theories.

\section{Deviations From the Bunch-Davis Vacuum}

\subsection{Possible Vacua of a Scalar Field in de Sitter Space}

To begin with, this discussion must revisit the derivation of the Bunch-Davis vacuum in Section \ref{sec_331}. Namely, the explicit assumption that was made in order to arrive at this unique state was to choose the positive frequency mode,
\begin{align}
v_k(\tau) = \frac{1}{\sqrt{2k}} e^{-ik\tau},
\end{align}
at asymptotic past times ($\tau \rightarrow -\infty$), which recovers standard behaviour in Minkowski spacetime (as an initial condition). Importantly, this restriction yields the \textit{minimal excitation state}. It has been argued, therefore, that when treating the Bunch-Davis vacuum as the initial inflationary state, trans-Planckian effects can be ignored \citep{tol}, allowing all of the above calculations to appropriately take the limit $\tau_0 = -\infty$. The reasoning for this being that the Bunch-Davis vacuum is a state of minimal excitation, so there are no inflaton quanta to be subject to trans-Planckian effects. In order to consider a more general initial state, one must relax this constraint of minimal initial excitation. In doing so, the Bunch-Davis vacuum is found to be a single realisation of a one-parameter family of possible states, related by a Bogoliubov transformation. Formally, these states are invariant under the isometry group of de Sitter space, and are often referred to as $\alpha$-vacua \citep{alph1, alph2}. 

Following on from the mathematics in Section \ref{sec_331}, an arbitrarily exited (Gaussian) initial state is generated by action of the negative frequency operators on the vacuum,
\begin{align}
\label{exitstat}
| \textit{in} \rangle = \frac{1}{\sqrt{m! n! \dots}} \left[(a_{\textbf{k}_1}^{\dagger})^m (a_{\textbf{k}_2}^{\dagger})^n \dots \right] |0 \rangle.
\end{align}
The amount of excitation is parametrised by the Bogoliubov coefficient, $\beta_k$, satisfying,
\begin{align}
\label{excitgen}
u_k(\tau) = \alpha_k v_k(\tau) + \beta_k v_k^* (\tau),
\end{align}
which is normalised as such,
\begin{align}
|\alpha_k|^2 - |\beta_k|^2 = 1.
\end{align}
The Bogoliubov coefficient is related to the number density of `inflaton particles' with physical momentum $k$ by, $N_{\phi}(k) = |\beta_k|^2$. As expected, the Bunch-Davis vacuum is recovered with $\beta_k = 0$.  Deducing the amplitude of non-Gaussianity thus amounts to following $\beta_k$ through the mathematics of Section \ref{sec_3}, and noting its effect on the bispectrum. However, multiple complications arise by setting $\beta_k \neq 0$ that must be accounted for if one wishes to do this. These complications are detailed in the foundational paper, \textit{Enhanced Non-Gaussianity from Excited Initial States} by R. Holman and A. Tolley \citep{tol}, which the following work will now review. In this treatment, the initial state of inflation is not a vacuum state, therefore, it is no longer appropriate to simply take the limit of $\tau_0 \rightarrow -\infty$. Instead, a method must be developed which masks our ignorance of physics beyond a certain energy scale, $M$, at times $\tau< \tau_0$. This is done by treating the inflaton as a degree of freedom in an EFT. However, particular care must be taken in defining the cut-off scale, because unlike standard EFTs (for example, in particle physics), the expansion of the Universe can cause complications. To elaborate, the CMB is observed as a function of comoving momentum scales, $k$; however, at times sufficiently far in the past ($\tau_0$), these scales can become comparable to the cut-off, $k/a(\tau_0) \approx M$. Hence, $\tau_0$ is now chosen as the time by which all observable momentum scales in the CMB were below the cut-off (rendering the EFT valid for these momenta). Therefore, given a well defined cut-off, we can now begin to derive restrictions on $\beta_k$, which up to now has remained arbitrary.

\subsection{Restrictions on the Excitation Parameter} 

First, without reference to non-Gaussianity, constraints can be derived (due to the EFT treatment) from the primordial power spectrum. This is a well-researched topic \citep{psp1,psp2}, and corrections to $P_{\zeta}$ are typically of the order $\mathcal{O}(\frac{H}{M})$. Thus, because the power spectrum of the CMB is well described by standard inflation, any source of non-Gaussianity appearing as a function of $\frac{H}{M}$ is suppressed by this observation. Therefore, the inference $\frac{H}{M}\ll 1$ is often made, which is in agreement with the validity condition of the EFT. Furthermore, this implies $|k \tau_0 | \gg 1$; where $|k \tau_0 |$ is a factor that will appear frequently in the ensuing calculations. Secondly, an explicit constraint on $\beta_k$ is derived by enforcing that the excited states constructed in (\ref{exitstat}) are \textit{Hadamard}. That is to say, because the theory has been chosen to break down at energy scale $M$, no `inflaton particles' should have physical momenta $k>Ma(\tau_0)$. This condition manifests by demanding $\beta_k \rightarrow 0$ for $k>Ma(\tau_0)$ - in particular, $\beta_k$ must decay faster than $\frac{1}{k^2}$ \citep{tri}. Finally, and perhaps most importantly, $\beta_k$ is further restricted by considering the possible \textit{backreaction} from the energy density of the excited initial state on the dynamics of inflation. To quantify this backreaction, one must calculate the energy density due to inflaton quanta ($\beta_k$) using the energy-momentum tensor expectation value,
\begin{align}
\label{enerdens}
\rho = \langle \textit{in} | \hat{T}_{00} | \textit{in} \rangle =  \text{Re} \left[ \langle \textit{in} | -2i \hat{T}_{00} \int^{\tau}_{\tau_0} d\tau' \text{ } H^I_{\text{int}} |\textit{in} \rangle \right].
\end{align}
As expected, this expression must be appropriately renormalised due to UV-divergences; results quoted below achieved this using \textit{adiabatic regularisation} \citep{back1}. In fact, calculating backreactions from arbitrary modifications to the inflationary Lagrangian (higher derivative kinetic terms) reduces to calculating their contribution to $\hat{T}_{\mu \nu}$, and substituting into (\ref{enerdens}). A key result of Ref$.$ \citep{tol}, however, was to demonstrate that higher derivative terms do not appreciably affect the upper bound for $\beta_k$ derived from standard slow roll inflation.
Thus, this calculation of an upper bound on $\beta_k$ can be briefly outlined. One begins by appropriately assuming a (Hadamard condition satisfying) model for $\beta_k$ as $\beta_k = \beta_0 e^{-k^2 / (M a(\tau_0))^2}$.\footnote{In fact, most models which satisfy the Hadamard condition have been shown to lead to similar results up to factors $\mathcal{O}(1)$, thus, this model is assumed without much loss of generality.} The computation in (\ref{enerdens}) is then found to approximately reduce to\footnote{This expression was derived by noticing that the number density of inflaton quanta with momentum $k$ is $n \sim \int (2\pi a)^{-3}|\beta_k|^2 d^3k$, hence, the energy density is the first moment of this integral.},
\begin{align}
\rho = -\frac{1}{a(\tau)^4} \int \frac{d^3k}{(2\pi)^3} k | \beta_k |^2 = \frac{a(\tau_0)^4}{a(\tau)^4} |\beta_0|^2 M^4.
\end{align}
For a negligible backreaction, we require this energy density to be less than the background energy density provided by the first Friedmann equation of slow roll inflation, which is $\sim \epsilon M_{pl}^2 H^2$, implying,
\begin{align}
\frac{a(\tau_0)^4}{a(\tau)^4} |\beta_0|^2 M^4 < \epsilon M_{pl}^2 H^2 \rightarrow |\beta_0| < \sqrt{\epsilon}\frac{M_{pl} H}{M^2}.
\end{align}
The most stringent bound on $\beta_0$ is therefore obtained when $a(\tau_0)^4/a(\tau)^4 \sim \mathcal{O}(1)$.  A further constraint can be derived similarly using the second Friedmann equation, and these are combined to read,
\begin{align}
\label{brsup}
|\beta_0| < \sqrt{\epsilon}\frac{M_{pl} H}{M^2}, \\ \nonumber
|\beta_0| < \sqrt{\epsilon \eta}\frac{M_{pl} H}{M^2}.
\end{align}  
It will prove illuminating to now give a sensible order of magnitude estimate for $\beta_0$. Appropriately presuming a scenario whereby $M=10^{-2}M_{pl}$ \citep{mages}, and $\{\epsilon, \eta \} \sim 10^{-2}$, one finds,
\begin{align}
|\beta_0| \lesssim \frac{H}{M}.
\end{align}
Thus, $\beta_0$ is, in general, suppressed by the realm of validity of the EFT, $\frac{H}{M} < 1$. However, it can be tuned otherwise with a favourable choice of the cut-off scale. To summarise, assuming the Hadamard condition is satisfied, it has been deduced that the excitation of the initial state is suppressed via backreaction considerations. This is a relatively intuitive result, because increased numbers of inflaton quanta $N_{\phi} \sim |\beta_k|^2$, raise the initial energy density of the universe, and hence have the potential to stop accelerated expansion from occurring.

\subsection{The Excited Initial State Bispectrum}
\label{FILLLLL}
\begin{figure}[h]
\centering
\includegraphics[scale=0.42]{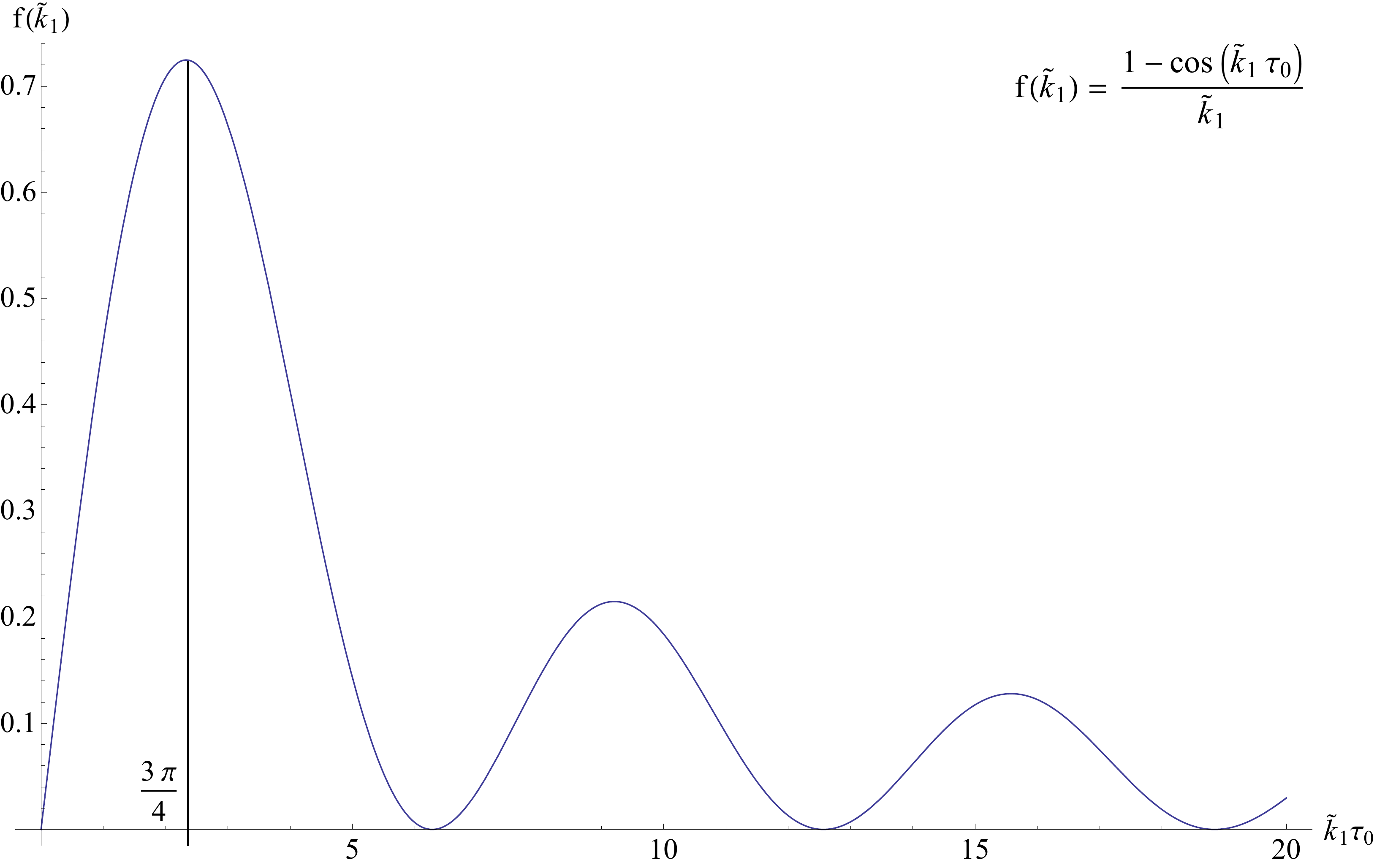}
\caption{A plot of the function $f(\tilde{k_1})$ determining the $k$ behaviour of the non-Bunch-Davis bispectrum (\ref{nbd1}). The arbitrary ordering $k_1 \geq k_2 \geq k_3$ has been used, and a clear peak is labelled at $\tilde{k}_1\tau_0 = 3\pi/4$.}
\label{peak}
\end{figure}

With these restrictions in mind, it is now possible to follow a similar process as Section \ref{sec_3} for calculating the quantum 3-point function. Instead, however, the mode functions that are substituted in place of interaction picture fields will take the more general form of (\ref{excitgen}), parametrising the degree of initial excitation. The EFT approach to inflation has been well reviewed (\textit{c.f.} \citep{eff}), and the relevant (third order) interaction Hamiltonian for an effective inflation action with a canonical kinetic term minimally coupled to gravity reads,
\begin{align}
H_{\text{int}} = -4 H \int d^3x \text{ } a^3 \epsilon^2 \zeta'^2 \partial^{-2} \zeta'.
\end{align}
Thus, this term can be substituted into the \textit{in-in} master formula, where (similarly to Section \ref{sec_3}) only three permutations survive the Wick contraction,
\begin{align}
\label{1wi}
\langle \zeta(\textbf{k}_1, 0) \zeta(\textbf{k}_2, 0) \zeta(\textbf{k}_3, 0) \rangle = & (2\pi)^3 \delta(\textbf{k}_1 + \textbf{k}_2 + \textbf{k}_3) 4 H \epsilon^2 \\ \nonumber
& \times \int^0_{\tau_0} d\tau' a^3 \frac{1}{k_1^2} u_{k_1}(0) u_{k_2}(0) u_{k_3}(0)u'^*_{k_1}(\tau')  u'^{*}_{k_2}(\tau')  u'^{*}_{k_3}(\tau') + \text{sym.},
\end{align}
where `sym$.$' denotes $1 \leftrightarrow 2$ and $1 \leftrightarrow 3$. Particularly, it is important to note that the lower bound of the integral, $\tau_0$, will not be set to $-\infty$, for the reasons discussed above. The effect of the non-Bunch-Davis initial state can now be brought into the calculation through the mode functions in (\ref{excitgen}). Specifically, these mode functions are now a superposition of the positive and negative frequency modes,
\begin{align}
u_{k_i}(\tau') = \alpha_k v_{k_i}(\tau') + \beta_k v_{k_i}^* (\tau'),
\end{align}
which yields three sets of Wightman functions,
\begin{align}
\label{513}
u_{k_i}(0)u^*_{k_i}(\tau') &= [\alpha_{k_i}v_{k_i}(0) + \beta_{k_i} v^*_{k_i}(0)][\alpha^*_{k_i}v^*_{k_i}(\tau') + \beta^*_{k_i} v_{k_i}(\tau')] \\ \nonumber
& = |\alpha_{k_i}|^2 v_{k_i}(0)v^*_{k_i}(\tau') + |\beta_{k_i}|^2 v^*_{k_i}(0)v_{k_i}(\tau') + ( \alpha_{k_i} \beta^*_{k_i} v_{k_i}(0) v_{k_i}(\tau') + \text{c.c.}),
\end{align}
for $i=1,2,3$, where c.c. denotes the complex conjugate. A number of simplifications can now be made for the sake of analytic transparency. Firstly, it was shown that $\beta_k$ is suppressed by a factor of $H/M$ (from the backreaction and observation of the CMB power spectrum), thus, the remainder of this calculation will only keep terms of linear order in $\beta_k$. Secondly, the contribution from the solely positive modes ($|\alpha_{k}|^2$) will simply add a small correction to the overall normalisation, and can hence be ignored. Within what remains lies the source of the defining feature of non-Bunch-Davis models, namely, that their bispectra peak in the \textit{flattened} limit. This can be seen by noticing that in (\ref{513}), terms linear order in $\beta_k$ are no longer absolute magnitudes of mode functions. This has the effect of \textit{mixing} the positive and negative frequency modes, which mathematically amounts to the $i^{th}$ Wightman function ($i=1,2,3$) in (\ref{1wi})  conjugating a singular $v_k(\tau')$ (whereas the other two remain unaltered). Hence, $k_i \rightarrow -k_i$ in the correlator for each $i$; where all correction terms of leading order in $\beta_k$ (now denoted $\Delta \langle \zeta \zeta \zeta \rangle$) are expressed as,
\begin{align}
\Delta \langle \zeta(\textbf{k}_1, 0) \zeta(\textbf{k}_2, 0) \zeta(\textbf{k}_3, 0) \rangle = & -i (2\pi)^3 \delta(\textbf{k}_1 + \textbf{k}_2 + \textbf{k}_3) \frac{H^4}{4 \epsilon} \frac{1}{(k_1k_2k_3)^3} \\ \nonumber 
& \times \int^0_{\tau_0} d\tau' \text{ } \sum_{i=1}^3 \left( \beta^*_{k_i} \frac{(k_1 k_2 k_3)^2}{k_i^2} e^{i \tilde{k}_i \tau'} + \text{c.c} \right).
\end{align}  
The notation, $\tilde{k}_i = K - 2k_i$, has been introduced as a result of the above analysis to account for $k_i \rightarrow -k_i$. For a given $i$, this integral evaluates to,
\begin{align}
-i \int^0_{\tau_0} d\tau' \text{ } e^{i \tilde{k}_i \tau} + \text{c.c.} = \frac{2(\cos(\tilde{k}_i \tau_0) -1)}{\tilde{k}_i},
\end{align}
leaving the final, non-Bunch-Davis (correction) 3-point correlator as,
\begin{align}
\label{nbd1}
\Delta \langle \zeta(\textbf{k}_1, 0) \zeta(\textbf{k}_2, 0) \zeta(\textbf{k}_3, 0) \rangle = & -(2\pi)^3 \delta(\textbf{k}_1 + \textbf{k}_2 + \textbf{k}_3) \frac{H^4}{2 \epsilon} \frac{1}{(k_1k_2k_3)^3} \\ \nonumber
& \times \sum^3_{i=1} \frac{k_1^2 k_2^2 k_3^2}{k_i^2 \tilde{k}_i} \text{Re}(\beta_{k_i})(\cos(\tilde{k}_i \tau_0) -1).
\end{align}
Clearly, a peak can now manifest due to the appearance of $\tilde{k}_i$ in the denominator. In fact, this peak is \textit{not} at $\tilde{k}_i = 0$ due to the oscillatory term decaying sufficiently fast as $\tilde{k}_i \rightarrow 0$. This oscillatory term has therefore regulated the possible divergence, and has a frequency controlled by the earliest time at which the momenta $k$ were below the EFT cut-off energy, $M$. Thus, $\tau_0$ is effectively a cut-off in and of itself, and is treated as such in \textit{Planck} \citep{Planck2013NG} where a cut-off momentum $k_c$ is defined via $k_c \approx 1/\tau_0$, which is used to regulate any divergences\footnote{Note that this cut-off can take a range of possible values, and is commonly $0.001 \leq k_c \leq 0.01$. Moreover, the speed of sound, $c_s$, typically makes an appearance in the definition of $k_c$, which will be detailed in a later section.}. The function determining the amplitude of this bispectrum can be seen in Fig$.$ \ref{peak}. Notably, a peak is displayed at $\tilde{k}_1 = 3\pi/4\tau_0 \sim 0$, meaning $k_1 \approx k_2 + k_3$ (under an arbitrary ordering of $k$) - i.e. flattened triangle configurations in $k$-space.
The bispectrum can thus be analytically evaluated in this limit, leading to,
\begin{align}
\lim_{\tilde{k}_1 \rightarrow 0} \Delta \langle \zeta \zeta \zeta \rangle \approx -(2\pi)^3 \delta(\textbf{k}_1 + \textbf{k}_2 + \textbf{k}_3) \frac{H^4}{2 \epsilon} \frac{1}{(k_1k_2k_3)^3}  (k_2^2 k_3^2) \text{Re}(\beta_{k_1}) \frac{1}{2} \tilde{k}_1 \tau_0^2.
\end{align}
Most importantly, however, the amplification with respect to the Bunch-Davis 3-point correlator $\langle \zeta \zeta \zeta \rangle^{\text{mald}}$ is calculated to be approximately,
\begin{align}
\frac{\Delta \langle \zeta \zeta \zeta \rangle}{\langle \zeta \zeta \zeta \rangle^{\text{mald}}} \Bigg|_{\tilde{k}_1 \rightarrow 0} \sim |\beta_{k_1} | |k_1 \tau_0|,
\end{align}
where $\text{Re}(\beta_{k_1}) = |\beta_{k_1}|$ has been assumed. Ref$.$ \citep{tol} now goes on to argue that, because the CMB data is a 2D projection, the amplification factor, $|k_1 \tau_0|$, will be lost in calculations of the angular bispectrum. However, a more in depth analysis of the detectability of this amplification is done in Ref$.$ \citep{meer}, which the ensuing work will now follow. First, following similar analysis techniques to Section \ref{sec_223}, it will be convenient to factorise out the $\frac{1}{k^6}$ scaling implicit in the non-Bunch-Davis bispectrum (\ref{nbd1}), to obtain the shape function,
\begin{align}
\label{NNNBBDD}
S^{\text{NBD}}(k_1, k_2, k_3) = \frac{k_2 k_3}{k_1 \tilde{k_1}} (1 - \cos(\tilde{k_1} \tau_0)) + \frac{k_1 k_3}{k_2 \tilde{k_2}} (1 - \cos(\tilde{k_2} \tau_0)) + \frac{k_1 k_2}{k_3 \tilde{k_3}} (1 - \cos(\tilde{k_3} \tau_0)).
\end{align}
Appropriately normalising this shape function leads to a non-Gaussian amplitude parameter of $f^{\text{NBD}}_{\text{NL}} \sim \epsilon |\beta_k|$ up to an $\mathcal{O}(1)$ factor. Thus, without reference to any amplification that may occur due to favourable triangle configurations, this amplitude of non-Gaussianity is undetectable. In fact, this amplitude is \textit{doubly} suppressed by both the slow roll parameter and the amount of initial excitation (\ref{brsup}). However, all is not lost, due to the non-Bunch-Davis shape function possessing an implicit $k$-scale dependence. This scale dependence becomes apparent when the shape function is factorised into the ratio parameters, $x_2$ and $x_3$, as such,
\begin{align}
\label{x1x2shape}
S^{\text{NBD}}(k_1, x_2, x_3) =& \frac{x_2 x_3}{-1 + x_2 + x_3}(1 - \cos[k_1 \tau_0 (-1 + x_2 + x_3)])\\ \nonumber
& + \frac{x_1 x_3}{1 - x_2 + x_3}(1 - \cos[k_1 \tau_0 (1 - x_2 + x_3)])\\ \nonumber
& + \frac{x_1 x_2}{1 + x_2 - x_3}(1 - \cos[k_1 \tau_0 (1 + x_2 - x_3)]).
\end{align}
Thus, the scale dependence is a function of the EFT cut-off, $k_1 \tau_0 = M/H \gg 1$. In order to deduce the effect this has on the observable amplitude of non-Gaussianity, one can correlate (see Section \ref{obsss}) (\ref{x1x2shape}) with itself. In doing so, one presumes a perfect (observationally constrained) template, which is sufficient (for now) to determine a maximum theoretical amplification of $f_{\text{NL}}$ as a result of this scale dependence. The correlation integral is evaluated in Ref$.$ \citep{meer} to be,
\begin{align}
F[S^{\text{NBD}},S^{\text{NBD}}] = \frac{\pi}{60} |k_1 \tau_0| + \frac{5}{4} \log  (|k_1 \tau_0| ) + 6.05.
\end{align}
Renormalising the amplitude will now lead to a non-Gaussianity parameter of,
\begin{align}
\label{NNN1}
f^{\text{NBD}}_{\text{NL}} \sim \epsilon |\beta_k|\sqrt{|k \tau_0|},
\end{align}
which is amplified by $\sqrt{|k \tau_0|}$ - the order of magnitude of which will be discussed shortly. In fact, one finds this amplification to only be present in shape functions well correlated with $S^{\text{NBD}}$. The above analysis can be extended to local and equilateral shape functions\footnote{Local and equilateral shape templates were the only experimentally constrained templates at the time of Ref$.$ \citep{meer}.}, resulting in,
\begin{align}
F[S^{\text{NBD}},S^{\text{local}}] \propto \log|k_1 \tau_0|,\\
F[S^{\text{NBD}},S^{\text{equil}}] \propto \text{const.},
\end{align}
where the constants of proportionality are $\mathcal{O}(1)$. Therefore, under a reasonable range of $|k_1 \tau_0|$, observational constraints on both $f^{\text{local}}_{\text{NL}}$ and $f^{\text{equil}}_{\text{NL}}$ are insufficient indicators of inflation with non-Bunch-Davis initial states due to the log damping (or complete cancellation) of the amplification factor. Ref$.$ \citep{meer} then goes on to propose a new, separable shape template which peaks for flattened triangle configurations, providing an (albeit small) upgrade over equilateral and local templates. This template was, unfortunately, still largely insufficient as it possesses the same problems discussed above. However, it became a precursor to the much improved orthogonal template (\ref{orthooo}) proposed by L. Senatore \textit{et al}. \citep{senatore} (which is now observationally constrained by \textit{Planck}).

The amplitude of non-Gaussianity for canonical, single field models with a non-Bunch-Davis initial state (\ref{NNN1}) (assuming a perfect template) can now be approximately determined. Firstly, substitution of the previously derived upper bound on $\beta_k$ (\ref{nbd1}) yields,
\begin{align}
\label{1010101022}
f^{\text{NBD}}_{\text{NL}} \sim \sqrt{\epsilon^3 \eta}\sqrt{\frac{M^2_{pl} H}{M^3}},
\end{align}
which is tunable via the unknown cut-off scale, $M$. We know $\{ \epsilon, \eta \} \sim 10^{-2}$, and $H/M_{pl} \sim 10^{-6}$ \citep{tol}, which reduces $f^{\text{NBD}}_{\text{NL}}$ to,
\begin{align}
f^{\text{NBD}}_{\text{NL}} \sim 10^{-7} \left( \frac{M_{pl}}{M} \right)^{\frac{3}{2}}.
\end{align}
Thus, in order to obtain $f^{\text{NBD}}_{\text{NL}} \sim \mathcal{O}(1)$, we require the cut-off scale to be $M \sim 10^{-\frac{14}{3}} M_{pl}$, which is within the realm of possibilities \cite{mages,eff} (for example, if $M \sim M_{\text{string}}$). However, this condition is on the very cusp of feasibly allowed values for $\beta_k$, as noted by Ref$.$ \citep{greene}, which derives a strict upper bound of $|\beta_k|<0.1$ from observations of the CMB power spectrum. It follows that, \textit{even with} amplification resulting from constraints derived using a maximally correlated shape template, significant non-Gaussianity from excited initial states alone is difficult to generate. This is the conclusion reached by a multitude of papers investigating general inflationary initial states in the context of non-Gaussianity \citep{tol,meer,chennonbd,sqcons,chen2,dey,tol2} (which are not all limited to flattened triangle configurations as the above work was\footnote{In fact, it has been noted \citep{sqcons} that non-Bunch-Davis models in the squeezed limit have the potential to break the consistency relation derived in Section \ref{sec_223}.}). If one wishes to further probe the possibility of non-Bunch-Davis initial inflationary states, a further element of the `no go' theorem (Section \ref{sec_223}) must be broken. In fact, there are many non-Bunch-Davis models for which this is the case; for example, certain \textit{feature} \citep{chen5} and multifield \citep{nbdmf} models. However, as we are already taking an EFT approach, it is natural to consider inflationary Lagrangians with a non-canonical kinetic term. Specifically, higher derivative interactions of the form ($\nabla \phi)^4$ are studied in the work this section began by reviewing \citep{tol}, amongst others \citep{meer,121212}. However, in order to eventually arrive at a model which is explicitly analysed by \textit{Planck}, the remainder of this work will adopt the approach of  X. Chen \textit{et al}. in \textit{Observational Signatures and Non-Gaussianities of General Single Field Inflation}, whereby a general $P(X,\phi)$ Lagrangian (\ref{PXPHI}) is considered \citep{chennonbd}. 

\subsection{Including a General Inflationary Lagrangian}

As detailed in Section \ref{sec_223}, a general inflationary Lagrangian takes the form,
\begin{align}
S = \int d^4x \sqrt{-g} \left[ \frac{M_{pl}^2}{2}\mathcal{R} - P(X,\phi) \right],
\end{align}
with the slow roll limit $P(X,\phi) = X - V(\phi)$. Varying this action yields the inflaton energy density,
\begin{align}
\rho = 2X \partial_X P - P,
\end{align}
by which the `sound speed' is then calculated from,
\begin{align}
c_s^2 \defeq \frac{dP}{d\rho} = \frac{\partial_X P}{\partial_X P + 2X \partial_X \partial_X P}.
\end{align}
Specifying $P$ thus amounts to explicitly choosing a model; we will however not restrict ourselves to a particular model. The aim is to therefore repeat Section \ref{sec_3} while keeping $P$ arbitrary, which is done in detail in Ref$.$ \citep{chennonbd}. Following this, it will prove convenient (for analytic transparency) to define the following parameters,
\begin{gather}
\Sigma \defeq \frac{H^2 \epsilon}{c_s^2}, \\
\lambda \defeq X^2 \partial_X \partial_X P + \frac{2}{3} X^3 \partial_X \partial_X \partial_X P, \\
s \defeq \frac{\dot{c}_s}{c_s H}.
\end{gather} 
In order to calculate the non-Bunch-Davis bispectrum in this regime, the appropriate interaction Hamiltonian is determined as, 
\begin{align}
H_{\text{int}} = - \int  d^3x \text{ } \Bigg\{ -& a^3 \left( \Sigma \left( 1 - \frac{1}{c_s^2} \right) + 2 \lambda \right) \frac{\dot{\zeta}^3}{H^3} + \frac{a^3 \epsilon}{c_s^4} \left( \epsilon - 3 + 3c_s^2 \right) \zeta \dot{\zeta}^2 \\ \nonumber
& + \frac{a \epsilon}{c_s^2} \left( \epsilon -2s +1 -c_s^2 \right) \zeta (\partial \zeta)^2 - \frac{2a \epsilon}{c_s^2} \dot{\zeta} (\partial \zeta) (\partial \chi) \Bigg\},
\end{align}
where $\chi$ is defined through $\partial^2 \chi = \frac{a^2 \epsilon}{c_s^2} \dot{\zeta}$. This is now substituted into the \textit{in-in} formula, and all four terms (plus the field redefinition correction) are evaluated separately, in similar fashion to Section \ref{Part4}. Note that the mode functions used in this calculation take the mixed form (\ref{513}), which will lead to the familiar polarity switch $k_i \rightarrow -k_i$ from the previous section. However, in contrast to that calculation, we will now take the lower limit of $\tau \rightarrow -\infty$ as an approximation. In doing so, a spurious divergence will appear, but can be regulated by a well motivated ansatz using the aforementioned cut-off, $k_c$.  

\begin{figure}[h]
\centering
\includegraphics[scale=0.42]{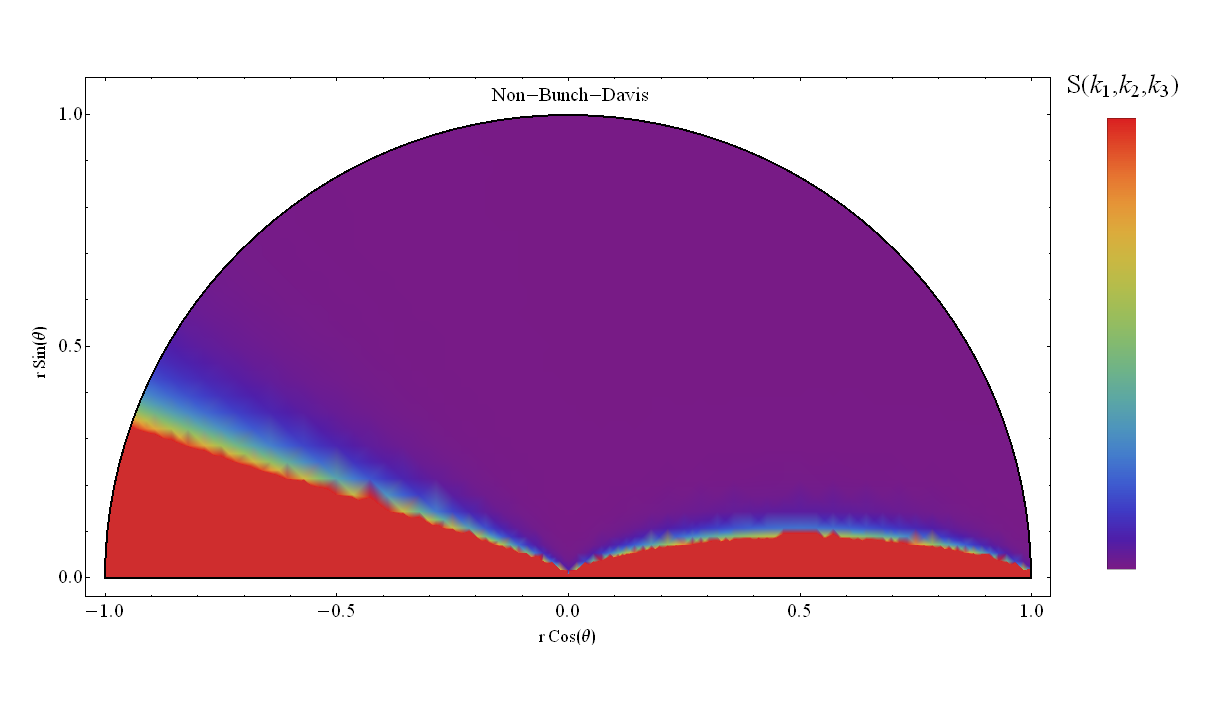}
\caption{A polar plot of the non-Bunch-Davis, general $P(X, \phi)$ shape function, $\tilde{S}_{\lambda}^{\text{ansatz}}$ (with $k_c = 0.001$). This function reveals sharp peaks along the regions $\theta \ll 1 \text{ } \forall \text{ } r$, and $\theta \approx \pi \text{ } \forall \text{ } r$, where the pole was cut-off for visual aid.}
\label{fig_flat2}
\end{figure}

Calculation of the 3-point correlator in this regime reveals a rather complicated $k$-dependence, expressed in full in (4.43) -- (4.49) of Ref$.$ \citep{chennonbd}. However, not all terms resulting from this calculation are of the same order in slow roll parameters. In fact, the `leading order' terms are \textit{not} slow roll suppressed, but would otherwise vanish in the slow roll limit ($c_s = 1$, $\frac{\lambda}{\Sigma} = 0$). Assuming $c_s < 1$ and $\frac{\lambda}{\Sigma} > 1$, the dominant terms in the bispectrum become,
\begin{align}
\label{aooaoaoaoaoaoa}
\langle \zeta_{\textbf{k}_1} \zeta_{\textbf{k}_2} \zeta_{\textbf{k}_3} \rangle = (2\pi)^3 \delta(\textbf{k}_1 + \textbf{k}_2 + \textbf{k}_3) \frac{H^4}{4 c_s^2 \epsilon^2} \frac{1}{k_1^3 k_2^3 k_3^2} (\tilde{\mathcal{A}}_{\lambda} + \tilde{\mathcal{A}}_{c}),
\end{align}
\begin{align}
\label{annn1}
\tilde{\mathcal{A}}_{\lambda} = k_1 k_2 k_3 \tilde{S}_{\lambda} = & \left( \frac{1}{c_s^2} - 1 - \frac{2 \lambda}{\Sigma} \right) \frac{3 (k_1 k_2 k_3)^2}{2} \\ \nonumber
& \times \left( \frac{|\beta_{k_1}|}{(-k_1+k_2+k_3)^3} + \frac{|\beta_{k_2}|}{(k_1-k_2+k_3)^3} + \frac{|\beta_{k_3}|}{(k_1+k_2-k_3)^3}  \right),
\end{align}
\begin{align}
\label{annn2}
\tilde{\mathcal{A}}_{c} = k_1 k_2 k_3 \tilde{S}_{c} = & \left( \frac{1}{c_s^2} - 1 \right) \\ \nonumber
& \times \sum_{p=1}^3 |\beta_{k_p}| \left( -\frac{1}{K} \sum_{i>j} k_i^2 k_j^2 + \frac{1}{2K^2} \sum_{i \neq j} k_i^2 k_j^3 + \frac{1}{8} \sum_i k_i^3 \right) \Bigg|_{k_p \rightarrow - k_p},
\end{align}
where the notation of Ref$.$ \cite{chennonbd} has been (mostly) adopted. It can be seen that the non-Bunch-Davis signature -  a peak in the limit of flattened triangle configurations - has been retained despite introducing a general $P(X, \phi)$ Lagrangian. As discussed, an unphysical pole is present in these shape functions at $\tilde{k}_i = 0$ due to taking the lower limit of $\tau = \tau_0 = -\infty$. This has the effect of removing the oscillatory nature of the bispectrum, which previously regulated the divergence in (\ref{nbd1}). In order to make up for this, a cut-off is introduced, $k_c \sim \frac{1}{c_s \tau_0}$, appearing in the following ansatz proposed by X. Chen \textit{et al}. \citep{chen2}, 
\begin{align}
\label{tototo}
\tilde{S}_{\lambda}^{\text{ansatz}} = k_1 k_2 k_3  \left( \frac{(-k_1+k_2+k_3)}{(-k_1+k_2+k_3 + k_c)^4}    +  \frac{(k_1-k_2+k_3)}{(k_1-k_2+k_3 + k_c)^4} +  \frac{(k_1+k_2-k_3)}{(k_1+k_2-k_3 + k_c)^4} \right),
\end{align}
which is (one of) the shape templates explicitly analysed by \textit{Planck} for non-Bunch-Davis  models\footnote{Note that $\tilde{\mathcal{A}}_{c}$ has very similar behaviour to $\tilde{\mathcal{A}}_{\lambda}$ but a slightly \textit{less} sharp peak due to the scaling of $1/\tilde{k}_i^2$ rather than $1/\tilde{k}_i^3$. Hence, it will be sufficient to only consider the regulated ansatz derived from $\tilde{\mathcal{A}}_{\lambda}$ hereafter.}. Plotting $\tilde{S}_{\lambda}^{\text{ansatz}}$ reveals a \textit{very} sharp peak for flattened triangles, and can be seen in Fig$.$ \ref{fig_flat2}.

Comparing this to the flattened template graphic in Fig$.$ \ref{fig_flat}, we see a very good match in general behaviour, however, Fig$.$ \ref{fig_flat2} possesses a much sharper peak. This results in a correlation of $\sim 90\%$ between $\tilde{S}_{\lambda}^{\text{ansatz}}$ and $S^{\text{flat}}$, as detailed by \textit{Planck} \citep{Planck2013NG,PlankNG}. It is worth noting that, at this point, very little `model-dependent' assumptions have been made. Namely, $P(X, \phi)$ has remained general, and $\beta_k$ is still bounded only by an approximate maximum through backreaction considerations (\ref{brsup}).  Thus, an order of magnitude estimate $f_{\text{NL}}$ can be derived for these general, $P(X, \phi)$, non-Bunch-Davis models. Clearly, the non-Gaussian amplitude of (\ref{annn1}) and (\ref{annn2}) will scale roughly as,
\begin{gather}
f_{\text{NL}}^{\lambda} \sim |\beta_k| \left( \frac{1}{c_s^2} - 1 - \frac{2\lambda}{\Sigma} \right), \\
f_{\text{NL}}^{c} \sim |\beta_k| \left( \frac{1}{c_s^2} - 1 \right).
\end{gather}
Therefore, allowing non-canonical kinetic terms sound speed has \textit{lifted} the slow roll suppression present in $f_{\text{NL}}^{\text{mald}}$. Moreover, the non-Gaussianity can be amplified greatly with $c_s^2 \ll 1$ or $ \lambda/\Sigma \gg 1$, whilst permitting $\beta_k$ to remain relatively suppressed, as it naturally is. Hence, models where inflation begins in a non-Bunch-Davis initial state \textit{and} possesses a low sound speed can produce an observationally significant amplitude of non-Gaussianity. Such behaviour (in $c_s$) is, as previously mentioned, displayed by (amongst others) K-inflation \citep{kinf} and DBI inflation \citep{DBI}, where the sound speeds discussed those references could reveal an amplification up to $f_{\text{NL}}^{\lambda/c} \sim \mathcal{O}(1)$ - $\mathcal{O}(100)$. Nevertheless, any detection of non-Gaussianity in the flattened limit would strongly favour non-Bunch-Davis initial states, regardless of any additional model-specific dependencies.  

\section{Observational Considerations}

\begin{figure}[h]
\centering
\includegraphics[scale=0.42]{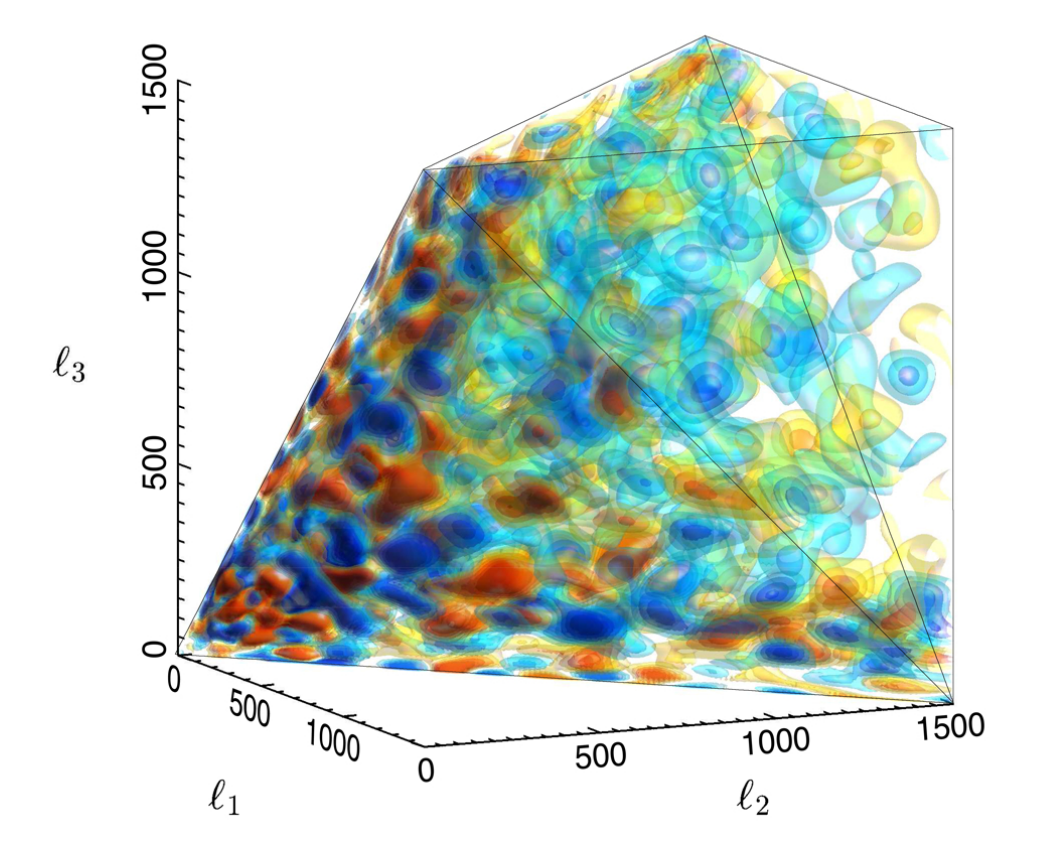}
\caption{Reconstruction of the CMB bispectrum from \textit{Planck} 2015 \texttt{SMICA} temperature only data. Plotted up to $l_{\text{max}} = 1500$, with the highest possible resolution ($n_{\text{max}}=2001$ polynomial modes in the modal expansion) \citep{PlankNG}.}
\label{CMB}
\end{figure}

A brief overview of the observational tools and underpinnings used in \textit{Planck} were provided in Section \ref{obsss}, hence, this section will remain short and only discuss particularly relevant results. First, it is possible for the full CMB bispectrum to be recovered from the \textit{Planck} 2015 data. Following techniques in Ref$.$ \citep{fs222}, and references therein, this bispectrum, reconstructed by means of a \textit{modal expansion}, can be seen depicted in Fig. \ref{CMB}. 

The diagrammatic methods involved in creating Fig$.$ \ref{CMB} are best explained in Ref$.$ \citep{fs223}, however, in brief: the CMB bispectrum is 3D function expressed in terms of multipole moments (\ref{wowowo}), which also adhere to a closed triangle condition. Thus, all possible $\ell$-configurations (where the bispectrum is non-zero) will fall within a \textit{tetrahedron} in $\ell$-space, and within this tetrahedron are contours depicting the amplitude of the bispectrum. Most important though are the numerical constraints on $f_{\text{NL}}$ that can be derived from the data, however, this visual reconstruction has been noted to display some `flattened characteristics' \citep{PlankNG}. Using the techniques detailed in Section \ref{obsss},\footnote{Specifically, \textit{Planck} uses the   \textit{Modal2} estimator in this case, which is one of four employed in the work.} with the `\texttt{SMICA}' dataset, constraints on the following primordial non-Gaussian amplitudes were derived,
\begin{gather}
f_{\text{NL}}^{\text{flat}} = 44 \pm 37, \label{nn1nn1} \\
f_{\text{NL}}^{\lambda + c} = 61 \pm 47, \\
f_{\text{NL}}^{\lambda, \text{ansatz}} = -8.7 \pm 5, \label{eeeeeeooeoeoeo}
\end{gather} 
using both temperature and polarisation data. These amplitudes correspond to the shape functions in (\ref{orthooo}), (\ref{annn1})+(\ref{annn2}), and (\ref{tototo}) respectively; where bounds are quoted at 68$\%$ CL. Although the mild deviation from zero of all three measurements is intriguing, it is too weak to constitute a detection. The best fit values fall within $\pm 2 \sigma$ of zero, and have $8\% \lesssim P \lesssim 30\%$ probability of occurring by chance alone\footnote{These probabilities are calculated by taking the number of standard deviations (denoted S/N in \textit{Planck}), and calculating $P(\text{by chance}) = 1 - \text{erf}(S/ \sqrt{2}N$).}, without implicating primordial non-Gaussianity. Despite no official detection of non-Gaussianity in this limit, these constraints are much improved over the previous \textit{Planck} 2013 and WMAP experiments. Thus, we can now seek to derive the constraints that these bounds have on parameters in the aforementioned models up to 68$\%$ CL. If we first entertain the crude approximation that $S^{\text{flat}}$ is perfectly correlated with $S^{\text{NBD}}$ (\ref{NNNBBDD}),\footnote{A further assumption will be implicit in this calculation, which is that the scale dependence of $S^{\text{NBD}}$ retains being well modelled by $S^{\text{flat}}$ over an appropriate range $|k \tau_0|$. Formally, the technique \textit{Planck} uses to treat scale dependence is to sample the parameter space (of the parameter which breaks the scale invariance), and `choose' the parameter with the best fit.} we can derive a (very approximate) restriction on $M$, and hence $\beta_k$ by direct relation. The theoretical $f^{\text{NBD}}_{\text{NL}}$ (for models where the \textit{only} modification is the non-Bunch-Davis initial state) was derived in (\ref{NNN1}), which, when substituting the upper bound for $\beta_k$ (\ref{brsup}), yields,
\begin{align}
f^{\text{NBD}}_{\text{NL}} \sim \sqrt{\epsilon^3 \eta}\sqrt{\frac{M^2_{pl} H}{M^3}},
\end{align}
which is (\ref{1010101022}). Assuming inflation occurs at an energy scale $H/M_{pl} \sim 10^{-6}$ (alongside the other assumptions made at the end of Section \ref{FILLLLL}), we find the maximum bound on $f^{\text{NBD}}_{\text{NL}}$ from (\ref{nn1nn1}) to be,
\begin{gather}
f^{\text{NBD}}_{\text{NL}} \lesssim 80 \\
10^{-7} \left( \frac{M_{pl}}{M} \right)^{\frac{3}{2}}  \lesssim 80. 
\end{gather}
Rearranging this expression, one finds the EFT scale $M$ at which `new physics' could occur is bounded by,
\begin{align}
\label{BRPP}
M < 8.6 \times 10^{-5} M_{pl}.
\end{align}
This is an extremely \textit{loose} bound, and does not reveal anything not already known, because many UV-complete theories function at energy scales which fall well within this constraint \cite{eff,greene,Baumann}. Furthermore, one can investigate the upper bound on the degree of initial excitation in a similar fashion. Substituting (\ref{BRPP}) into the formula for the backreaction constraints on $\beta_k$ (\ref{brsup}) one finds,
\begin{align}
\label{cococoo}
\beta_k < 7.4 \times 10^{3}.
\end{align}
As discussed, observational considerations involving the CMB power spectrum place an upper limit of $\beta_k < 0.1$. Clearly, the non-Gaussianity in the above scenario does \textit{not} further constrain any parameters of interest in solely non-Bunch-Davis models, due to the heavy suppression of $f^{\text{NBD}}_{\text{NL}}$. Moreover, fine tuning the ratio $H/M_{pl}$ yields only minor improvements over (\ref{BRPP}) and (\ref{cococoo}). 

We can now consider the model with a general inflationary Lagrangian and non-Bunch-Davis initial state (\ref{aooaoaoaoaoaoa}). The \textit{Planck} constraint for this model, assuming the separable ansatz (\ref{tototo}), can be seen in (\ref{eeeeeeooeoeoeo}). If, for the sake of this analysis, we apply these constraints directly to the shape $\tilde{\mathcal{A}}_c$, one finds the amplitude parameter is roughly bounded by,
\begin{align}
\label{111443}
\Bigg| \beta_k \left( \frac{1}{c_s^2} - 1 \right) \Bigg| < 3.7.
\end{align}
Relatively stringent bounds can be placed on sound speed from studies using equilateral template: $c_s \geq 0.087$ ($95\%$ CL). Therefore, favourably taking $c_s = 0.087$, we find (\ref{111443}) reduces to,
\begin{align}
\label{FINALLLL}
| \beta_k | \lesssim 0.03.
\end{align}
This is a somewhat tight bound on the Bogoliubov parameter, suggesting the degree of initial excitation in this model is low. Despite the very approximate treatment being provided here, this result highlights the possibilities of using non-Gaussianities to probe excited initial states. Furthermore, analysis techniques are constantly being improved upon, and this data could be more revealing in the near future (alongside a more proper investigation than what was provided here). Finally, it can be concluded that general non-Bunch-Davis models remain a particularly interesting prospect as they offer a unique signature shape (peaking for flattened configurations) and also possess the possibility of restricting parameters such as $\beta_k$ beyond what can be obtained from other observation \ref{FINALLLL}.



\chapter{Conclusion} 

\label{Chapter5} 

\lhead{6. \emph{Conclusion}} 


The work began by reviewing the background mathematics and physics required to study non-Gaussianities. This was a pedagogical introduction, covering classical statistics of random fields, the `standard' model of inflation, and cosmological perturbation theory. Most importantly, the gauge invariant quantity $\zeta$ was detailed, defining a convenient measure of curvature perturbations during and after inflation. In the comoving gauge, this variable was shown to be the \textit{only} primordial degree of freedom, and was used to keep track of perturbations in all ensuing calculations. However, before \textit{calculating} any (model specific) non-linearities generated during inflation, a phenomenological overview of non-Gaussianity was first provided. This began with a derivation of the local model of non-Gaussianity, which was able to demonstrate many of the essential foundations required to study non-Gaussianities, such as the bispectrum shape and amplitude. The three shape templates explicitly constrained by \textit{Planck} were then plotted and analysed, namely, the local, equilateral, and orthogonal templates. In each case, a brief review of the inflationary regimes which predict a such a bispectrum was given. Furthermore, the observational constraints derived from the \textit{Planck} 2015 data on these shapes was presented and discussed, along with a brief outline of general observational methodology for mapping the CMB.

A key result this analysis was that \textit{all} single field inflation models will possess a bispectrum suppressed by the spectral tilt for squeezed triangle configurations of Fourier modes. Thus, any detection of non-Gaussianity in the squeezed limit would favour multifield models. Moreover, bispectra peaking for equilateral configurations of Fourier modes were most commonly displayed by models with non-canonical kinetic terms in the Lagrangian; which opens up possibilities to probe certain models built from string theory \citep{DBI}. The remaining work from this point onward was concerned with calculating non-Gaussianity from inflationary models. To do this, a technique to calculate $n$-point correlation functions with time-dependent interacting states was reviewed - the \textit{in-in} formalism. Particularly, the resultant \textit{in-in} expression derived was to be evaluated at tree-level, which gives a first order account of any non-linearity. This formula was then used to calculate the non-Gaussianity generated by single field, slow roll inflation, following Maldacena's original work. The bispectrum from this calculation was found to be well correlated with the local template, therefore non-Gaussianity is generated on superhorizon scales. Furthermore, this non-Gaussianity is slow roll suppressed - in agreement with the previously derived consistency relation - meaning it is not detectable in the near future. 

Finally, the remaining work investigated inflationary regimes where a significant non-Gaussianity \textit{could} be generated, focusing primarily on models with non-Bunch-Davis initial states. These states are parametrised by a Bogoliubov coefficient, specifying the degree of initial excitation. Upon substitution into the \textit{in-in} formula, the resultant non-Bunch-Davis bispectra revealed a unique signature, namely, a peak for flattened triangle configurations of Fourier modes. However, non-Gaussianity from models with excited initial states alone were found to be largely undetectable due to stringent bounds on $\beta_k$ from backreaction considerations and observation of the CMB power spectrum suppressing $f^{\text{NBD}}_{\text{NL}}$. As such, the \textit{Planck} 2015 bounds for these models were found to be insufficient for constraining $\beta_k$ in a meaningful way, beyond what is already known. Thus, the possibility of non-canonical kinetic terms was also introduced via a general $P(X,\phi)$ Lagrangian. The bispectra from these models in fact retained the unique flattened signature, but were tunable as a function of the sound speed, $c_s$. Therefore, it was deduced that significant amplification of the non-Gaussian parameter in the flattened limit can occur in the if $c_s \ll 1$, which is the case for models such as K-inflation \citep{kinf}. A heuristic scenario using order of magnitude estimates for certain `known' parameters was then provided - i.e. favourably assuming $c_s \sim 0.087$ from $Planck$ data on the equilateral template, and $H/M \sim 10^{-6}$. This was was used to highlight the possibility that the magnitude of the Bogoliubov parameter could be constrained beyond current observational limits using non-Gaussianities, pending further investigation.



\addtocontents{toc}{\vspace{2em}} 

\appendix 




\addtocontents{toc}{\vspace{2em}} 

\backmatter



\label{Bibliography}

\lhead{\emph{References}} 

\bibliographystyle{unsrtnat} 

\bibliography{Bibliography} 

\end{document}